\documentclass[final,pdftex,twoside,3p]{elsarticle}

\usepackage[utf8]{inputenc}
\usepackage{amsmath}
\usepackage{amssymb}
\usepackage{bm}
\usepackage{graphicx}
\usepackage{color}
\usepackage{caption,subfig}
\usepackage{tikz}
\usetikzlibrary{decorations.markings}
\usepackage{lipsum}
\usepackage{url}
\usepackage{mathtools}

\font\FermiSmallfont=cmssq8 scaled 1200

\def\LANLppthead#1#2{
\null 
\begin{center}\vskip -1.7truein{\hbox to 6.5truein {
\hfill
\vbox to 1in {\vfill \FermiSmallfont
              \hbox{#1}
              \hbox{#2}
              \vfill}
}}\vskip-0.4truein\end{center}}%FNALppthead

\journal{Physics Reports}

\begin{document}
%------------------------------------------------------------------------------
% Title and abstract
%------------------------------------------------------------------------------
\title{Sterile neutrinos in cosmology}

\LANLppthead {UCI-TR-2017-03}{\ }

\begin{frontmatter}
\author{Kevork N.\ Abazajian}

\address{Center for Cosmology, Department of Physics \&
  Astronomy, University of California, Irvine, \\Irvine, California
  92697, USA}

\begin{abstract}
Sterile neutrinos are natural extensions to the standard model of
particle physics in neutrino mass generation mechanisms. If they are
relatively light, less than approximately 10 keV, they can alter
cosmology significantly, from the early Universe to the matter and
radiation energy density today. Here, we review the cosmological role such light sterile
neutrinos can play from the early Universe, including
production of keV-scale sterile neutrinos as dark matter candidates,
and dynamics of light eV-scale sterile neutrinos during the
weakly-coupled active neutrino era. We review proposed signatures of
light sterile neutrinos in cosmic microwave background and large scale
structure data. We also discuss keV-scale sterile neutrino dark matter
decay signatures in X-ray observations, including recent candidate
$\sim$3.5 keV X-ray line detections consistent with the decay of a $\sim$7 keV
sterile neutrino dark matter particle.
\end{abstract}

\end{frontmatter}

\tableofcontents

\section{Introduction}

The conclusive determination of neutrinos as massive approximately two
decades ago has opened a window into beyond the standard model of
particle physics regarding efforts to determine their mass generation
mechanism and the structure governing their
mixing~\cite{Olive:2016xmw}. On nearly the same timescale, the
standard model of cosmology has emerged with a robust framework, but
requiring two new components that are little understood: dark matter
and dark energy.  The simplest and earliest mechanisms proposed to
provide light neutrino masses invoke the presence of sterile neutrinos
via the ``seesaw'' mechanism
\cite{GellMann:1980vs,Yanagida:1979as,Mohapatra:1979ia}.

The Super-Kamiokande experiment has accomplished a precise measurement
of the zenith-angle dependence of the atmospheric neutrino flux,
leading to strong evidence of neutrino oscillations
\cite{Fukuda:1998mi}. The mixing was consistent with a neutrino mass
splitting of $|\delta m_{\rm atm}| \approx 50\rm~meV$ between mass
states associated largely with muon and tau flavor
neutrinos\footnote{Here, we describe mass splittings heuristically,
  as associated with their discovery method, but below the precise
  definition of a mass splitting is $\delta m_{ab}^2 = m_{\nu_a}^2 -
  m_{\nu_b}^2$ between vacuum mass eigenstates $\nu_a$ and $\nu_b$.}.
Evidence of the oscillations of solar neutrinos indicate the presence
of an additional splitting of $\delta m_\odot \approx 9\rm\ meV$
\cite{Cleveland:1998nv,Ahmad:2002jz}.  The atmospheric and solar mass
splittings are consistent with three flavor models that mirror the
flavor structure of the charged leptons and quarks. 

However, a persistence of anomalies in short baseline neutrino
experiments, starting with the Liquid Scintillator Neutrino Detector
(LSND) experiment, indicate a possibility of neutrino oscillations
with at least one mass splitting at the $|\delta m_{\rm SBL}| \sim
1\rm\ eV$ scale \cite{Aguilar:2001ty}. An consistent signal of
candidate flavor conversions of neutrinos at short baselines was seen
in the MiniBooNE experiment \cite{AguilarArevalo:2007it}. A related
signal exists in the discrepancy in reactor neutrino experiments that
could be indicative of this higher-scale mass splitting
\cite{Huber:2011wv,Kopp:2013vaa}. A third mass splitting would
require, at a minimum, a fourth mass eigenstate in the neutrino
sector, and the $Z_0$ width would require this fourth state to be
sterile \cite{Olive:2016xmw}. The experimental program and
theoretical background of sterile neutrinos related to short baseline
anomalies was reviewed in Refs.~\cite{Abazajian:2012ys,Lasserre:2014ita}.

There was a serendipitous start to both precision cosmology and
precision neutrino oscillation experiments in the late twentieth
century. This prompted a leading paper that the luminous red galaxy
(LRG) sample of the Sloan Digital Sky Survey would be highly sensitive
to the presence of massive neutrinos \cite{Hu:1997mj}. Though the
focus of that paper was massive active neutrinos, the physics also
applies to the cosmological presence of massive sterile neutrinos via
partial or complete thermalization \cite{Barger:1997yd}. The presence
of significant mixing between active and sterile neutrinos will also
partially or completely thermalize the sterile neutrinos
\cite{Langacker:1989sv} and contribute to both the mass density as
well as relativistic energy density at early times. A light sterile
neutrino mixing with active neutrinos could also lead to lepton number
generation \cite{foot:1995bm} and numerous matter-affected neutrino
mixing effects that distort neutrino spectra
\cite{abazajian:2004aj}. The effects of massive active and sterile
neutrinos and extra relativistic energy density on both large scale
structure (LSS) and the cosmic microwave background (CMB) were
reviewed in Ref.~\cite{Abazajian:2016hbv}. We discuss the effects of
massive eV-scale sterile neutrinos on cosmology in Section
\ref{sec:eV-scale}.

At a substantially higher mass scale, keV-scale mass sterile neutrinos
with proper mixing with active neutrinos can be produced as dark
matter via collisional processes in the early Universe in the
Dodelson-Widrow mechanism \cite{dodelson:1993je}. Couplings to other
particles can also produced in the early Universe at the abundance to
be part or all of the cosmological dark matter
\cite{Kusenko:2006rh}. Oscillation-based sterile neutrino dark matter
production can also occur via an enhanced Mikheev-Smirnov-Wolfenstein
(MSW) mechanism \cite{Wolfenstein:1977ue,Mikheev:1986gs} if there
exists a primordial lepton number, through the Shi-Fuller mechanism
\cite{dodelson:1993je}. Alternatively, the Dodelson-Widrow mechanism
is often dubbed ``non-resonant'' sterile neutrino dark matter while
the Shi-Fuller mechanism produces ``resonant'' sterile neutrino dark
matter. Both mechanisms produce sterile neutrino dark matter that can
behaves as ``warm'' dark matter, though resonant production could
produce dark matter indistinguishable from absolutely cold dark matter
(CDM) with current constraints. Non-oscillation production could also
produce sufficiently cold relic sterile neutrino dark
matter. Therefore, sterile neutrino dark matter can effectively be
CDM.

The non-resonant Dodelson-Widrow mechanism collisionally produces the
sterile neutrino dark matter in a non-equilibrium state, but with a
momentum distribution that reflects the active neutrinos'. However,
non-resonant sterile neutrino dark matter has a mapping between
particle mass-scale different from thermally produced warm dark matter
(WDM) \cite{blumenthal:1982mv} and therefore non-resonant sterile
neutrino dark matter's effect on structure formation is unique
\cite{colombi:1995ze}.  The resonant-production mechanism produces a
momentum distribution that is on average lower in average momentum at
a given temperature than non-resonant production due to the resonance
occurring first through the lower momenta modes. The resonant sterile
neutrino dark matter therefore is frequently called ``cooler'' than
non-resonantly produced sterile neutrino dark matter.  Neither
Dodelson-Widrow nor Shi-Fuller mechanisms for dark matter produce a
sterile neutrino in thermal equilibrium, therefore
temperature-based concepts like warm and cold are inaccurate to
apply. We discuss the connection between the thermal warm dark matter
case and oscillation and non-oscillation production mechanisms in
Section \ref{subsec:keVstructure}.

The value of the initial lepton asymmetry in the Universe, whether
trivial or substantial but within constraints, connects the
resonant-production mechanism with the non-resonant production
mechanism. Resonant production relies on a substantial but
observationally unconstrained primordial cosmological lepton
asymmetry. The exploration of the full parameter space of sterile
neutrino particle mass, mixing angle and lepton number was first
performed in Abazajian, Fuller \& Patel~\cite{abazajian:2001nj}, as
well as an exploration of constraints on sterile neutrino dark matter
from primordial nucleosynthesis, the CMB, structure formation, and
diffuse X-ray to gamma-ray backgrounds. Ref.~\cite{dolgov:2000ew} also
discussed diffuse X-ray constraints on non-resonantly produced sterile
neutrino dark matter. Abazajian, Fuller \&
Patel~\cite{abazajian:2001nj} were the first to point out the
serendipitous connection between the allowable parameter space for
sterile neutrino dark matter and the detectability of the radiative
decay line of sterile neutrino dark matter with observations of
dark-matter rich objects like galaxies and galaxy clusters by the
then-recently launched {\em Chandra} and {\em XMM-Newton X-ray Space
  Telescopes}. The follow-up paper Ref.~\cite{abazajian:2001vt} was
the first to study X-ray constraints from {\em XMM-Newton}
observations of the Virgo Cluster, and also made forecasts for future
observations, including those from proposed large-exposure and
large-effective-area X-ray microcalorimeter spectrometer missions like
{\em Constellation-X}. Ref.~\cite{abazajian:2001vt} also discussed
that increased sensitivity could also be attained from large exposures
via a stacking analysis of the spectra of a number of similar
clusters.

For a considerable amount of time, only upper limits to decay rates
were found from searches for the X-ray line. In 2014, Bulbul et
al.~\cite{bulbul:2014sua} reported a high-significance,
$4\sigma-5\sigma$, detection in stacked observations of 73 clusters
with the MOS and PN spectrometers aboard {\em XMM-Newton}, as well as
a consistent signal from the Perseus cluster of galaxies observed with
the {\em Chandra} telescope. Boyarsky et al.~\cite{Boyarsky:2014jta}
found a consistent signal from the Andromeda galaxy as well as Perseus
using data from the {\em XMM-Newton} satellite. There has been
significant scrutiny of the results as well as follow-up observations
that see commensurate signals in other astronomical observations. We
review these signals and challenges in its interpretation in
Section \ref{sec:X-ray}.

\section{Neutrino Oscillation, Mass Generation and Sterile Neutrinos}

\subsection{Neutrino Oscillation and Massive Neutrinos}

A quantum state created in some atomic, nuclear or particle
interaction need not be created in an energy eigenstate that is
stationary---that is, a state that does not evolve---in the vacuum.  A
``textbook'' example of such a problem is the spin precession of a
spin $1/2$ system in an external magnetic field $\bf B$ (e.g.,
see Ref.~\cite{Sakurai1995}).  A spin $1/2$ particle that is, for
example, created in a spin state aligned with the positive $x$-axis
($|S_x +\rangle$), and enters a region with a magnetic field aligned
along the positive $z$-axis, will evolve or ``precess.'' In the region
with a magnetic field along the $z$-axis, the energy eigenstates are
spin $1/2$ particles with their spins aligned either in the
positive or negative $z$-axis ($|S_z \pm\rangle$).  The state $|S_x +,
t=0 \rangle$ will have a probability at some later time, $t^\prime$ to
be either in a $|S_x +\rangle$ or $|S_x -\rangle$ state, since $|S_x
+\rangle$ is not a stationary state of the system.  The probability of
finding it in a $|S_x +\rangle$ or $|S_x -\rangle$ state has an {\it
  oscillatory} form.

The case of the massive neutrino, $\nu$, propagating in vacuum is very
similar to spin precession.  The neutrino interacts via the Weak
Interaction and displays the three family ($e,\mu,\tau$) symmetry of
the lepton sector of the standard model of particle physics.  The
simplifying assumption of the standard model is that the neutrino is
massless, which automatically makes a neutrino of a particular flavor
($e,\mu,\tau$) an energy eigenstate.    If neutrinos have
mass, theory does not require that the mass states are identical with
the flavor states ($|\nu_\alpha\rangle$, where $\alpha=e,\mu,\tau$).
Most generally, Weak Interaction flavor and vacuum mass states are not
identical.  A flavor state $|\nu_\alpha\rangle$ , is, most generally,
a superposition of mass eigenstates $|\nu_a\rangle (a=1,2,3...)$.
Since the time evolution operator ${\cal U}(t) = \exp(-iHt)$ in a
field/force free vacuum is ${\cal U}(t) = \exp(-iE_i
t)=\exp(-i\sqrt{m_i^2 + p^2}t) \approx \exp(-ix m_i^2/2p)$, the mass
states are energy eigenstates in vacuum (we have identified $x=t$).

If we make the often-applied approximation that there are only two
neutrino flavors and mass states important to consider, the two
flavors ($\alpha,\beta$) are superpositions of the two mass states
({\it i.e.}, ``mixed''):
\begin{align}
|\nu_\alpha \rangle &= \phantom{-}\cos \theta |\nu_1\rangle + \sin \theta | \nu_2
 \rangle \nonumber \cr
|\nu_\beta \rangle &= -\sin \theta |\nu_1\rangle + \cos \theta | \nu_2
 \rangle,
\end{align}
and the analogy to the spin $1/2$ precession system is
apparent.  A neutrino created in some Weak Interaction process as a
flavor state (via flavor conservation), will evolve and the
probability of finding it later in a state $\alpha$ will have an
{\it oscillatory} form.  

One of the most interesting features of neutrino physics is the
coherent effect of the medium through which the neutrino is
propagating.  The effective mass and mixing angles can be very
different for neutrinos propagating through dense environments. Such
coherent effects in the sun's interior can produce a coherent level
crossing and transformation of neutrino flavor in the sun---the
Mikheev-Smirnov-Wolfenstein (MSW) mechanism
\cite{Wolfenstein:1977ue,Mikheev:1986gs}.  Resonant
transformations can also occur in the the dense environment of
core-collapse supernovae, and can induce effects
on neutrino heating behind the shock \cite{Fuller1992}.  
Resonant neutrino transformations can also occur in the early
Universe, which we explore further below
\cite{Savage:1990by,Foot:1996qk,Shi:1996ic}.

Considerable evidence suggests the existence of small neutrino masses
and mixings, and that neutrino mass states that have different masses.
The first indications of anomalies in the neutrino sector came from
the measurement of the flux of electron neutrinos coming from the Sun
by the Davis Experiment \cite{Davis:1994jw}, which found that flux to
be about a third of that predicted flux from solar models
\cite{Bahcall:1989ks,Haxton:1998sd,Balantekin:1999re}. Observations of
the neutrino flux from the sun are found to be far below expected
rates, and indicate neutrino flavor conversion via the
MSW effect with relatively large mixing angles
and at a mass splitting scale of $|\delta m_\odot| \sim 8\ \rm
meV$. The flavor conversion was confirmed by the Sudbury Neutrino
Observatory which sees the full flux in flavor-independent neutral
current interactions, while also seeing the deficit of electron
neutrinos in charge-current interactions \cite{Ahmad:2002jz}. This
flavor conversion was confirmed in the KamLAND laboratory reactor
neutrino experiment, which sees the same flavor conversion but with
reactor antineutrinos \cite{Eguchi:2002dm}. The Super-Kamiokande
atmospheric neutrino experiment sees a zenith angle dependence of the
atmospheric neutrino flux that is fit most simply with an model of
neutrino masses for the $\mu$ and $\tau$ flavored neutrinos, $\nu_\mu$
and $\nu_\tau$, split by a mass difference of $|\delta m_\mathrm{atm}|
\sim 50\rm\ meV$ \cite{Fukuda:1998mi}, and this has been verified and
refined by long-baseline neutrino laboratory experiments
\cite{Ahn:2002up,Michael:2006rx}.  

The splittings of the mass eigenstates are becoming more precisely
measured in neutrino oscillation experiments \cite{Capozzi:2016rtj},
while determining the value of the absolute mass in the laboratory
requires the accurate measurement of the endpoint of the spectrum of
beta-decay neutrinos \cite{Beck:2010zzb}.  Neutrinoless double-beta
decay experiments could simultaneously determine whether neutrinos
are Majorana particles ({\it i.e.}, their own antiparticles) as well as
their absolute mass scale \cite{Avignone:2007fu}. In addition, long
baseline neutrino oscillation experiment will likely determine whether
the neutrino eigenstates are ``normal ordered'' with $m_1 < m_2 < m_3$
or ``inverted order'' $m_3 < m_1 < m_2$. A measure of this ordering in
concert with a measurement of the sum of neutrino masses can reveal
the nature of the neutrino mass-generation mechanism and therefore
open a window into high-energy scale physics \cite{Mohapatra:2006gs}.

The Los Alamos Liquid Scintillator Neutrino Detector (LSND) is a
short-baseline beam-dump experiment that finds a flavor excess that
may be consistent with neutrino oscillations at a eV-scale mass
splittings, $|\delta m_\mathrm{SBL}| \sim 1\, \mathrm{eV}$
\cite{Athanassopoulos:1997pv}. The MiniBooNE Experiment sees an
anomaly in its antineutrino channel that may be consistent with the
LSND signal \cite{Aguilar-Arevalo:2013pmq}. In total, there are three
mass-splitting scales of $|\delta m_\odot| \ll |\delta m_\mathrm{atm}|
\ll |\delta m_\mathrm{SBL}|$, which indicates the need for at least
four mass eigenstates to accommodate three disparate mass
splittings. It is difficult to include all atmospheric, solar and
short baseline oscillations, and constraints on short baseline
oscillations, in a single neutrino mass and mixing model of four or
more neutrino states \cite{Gariazzo:2017fdh}.

\subsection{Generating Neutrino Mass Beyond the Standard Model}
In the canonical Standard Model (SM) of particle physics, neutrinos,
combined with the charged leptons, form left-handed electroweak
$SU(2)$ doublets, $L_\alpha$ with family $\alpha$. This is associated
via CPT with a right handed antiparticle state $L^c_{\alpha,R}$, which
transform as\footnote{Here, we adopt the notation and reflect some of
  the discussion from Ref.~\cite{Langacker:1998fq}}.
\begin{equation}
\left(\begin{matrix}\nu_e\\ e^-\end{matrix}\right)_L
  \xleftrightarrow{\mathrm{CPT}} \left(\begin{matrix} e^+\\\nu_e^c \end{matrix}\right)_R.
\end{equation}
The $L$ and $R$ refer to left and right chiral projections, which in
the case of zero mass also correspond to helicity states. 
The SM has that the neutrinos have no electric charge, no color, as
well as the ansatz of masslessness with no inclusion of a
``right-handed'' $N_R$ that would be required to produce mass. $N_R$
would be $SU(2)$-singlets, with no weak interactions except those
induced by mixing with the active neutrinos. Such ``sterile
neutrinos'' can be added to the SM, and are predicted in most
extensions. For sterile neutrinos, the $R$ state is the particle and
the $L$ state is the anti-particle,
\begin{equation}
N_R \xleftrightarrow{\mathrm{CPT}} N_L^c.
\end{equation}
Below, after this discussion of neutrino mass generation, we will
refer to sterile neutrinos as $\nu_s$: $\nu_s\equiv N_R$. 

Neutrinos can have masses through the same mechanism that provides the
charged leptons masses, via a Dirac mass term, where there are two
distinct neutrinos $\nu_L$ and $N_R$:
\begin{equation}
-\mathcal{L}_\mathrm{Dirac} = m_D\left(\bar\nu_L N_R + \bar N_R\nu_L\right) = m_D\bar\nu\nu
\end{equation}
and the Dirac field is defined as $\nu \equiv \nu_L+N_R$. This mass
term conserves lepton number $L = L_\nu+L_N$. For three or more
families (flavors) of neutrinos, this mechanism is easily generalized,
and the mass terms become matrices. The charged current interactions
then involve a leptonic mixing matrix analogous to the
Cabibbo-Kobayashi-Maskawa (CKM) quark mixing matrix, with the leptonic
one dubbed the Pontecorvo-Maki-Nakagawa-Sakata (PMNS) matrix
\cite{Pontecorvo:1957cp,Pontecorvo:1957qd,Maki:1962mu}. This allows
for oscillations between the light neutrinos.

The generation of mass requires a $SU(2)$ breaking generated by a
Yukawa coupling
\begin{equation}
-\mathcal{L}_\mathrm{Yukawa} = h_\nu \left(\bar\nu_e\bar e\right)_L
\left(\begin{matrix}\varphi^0\\ \varphi^-\end{matrix}\right) N_R + H.c. \label{diracmass}
\end{equation}
This is gives the Dirac mass for the neutrinos in the same way as for
charged leptons with $m_D = h_\nu v/\sqrt{2}$. The vacuum
expectation value (VEV) of the Higgs doublet is
$v=\sqrt{2}\langle\varphi^0\rangle=\left(\sqrt{2} G_F\right)^{-1/2} =
246\ \mathrm{GeV}$, and the Yukawa coupling is $h_\nu$. The question
in this mass generation mechanism is why $h_\nu$ is so small, with
$h_\nu \lesssim 10^{-11}$ for $m_{\nu_e} < 2\ \mathrm{eV}$
\cite{Kraus:2004zw,Aseev:2011dq}.  

Neutrinos may also have masses via Majorana mass interaction, which
involves the right-handed antineutrino, $\nu_R^c$, instead of a
separate Weyl neutrino. That is, it is a transition from an
antineutrino to a neutrino, or equivalently as a creation or
annihilation of two neutrinos. If present, this would lead to
neutrinoless double beta decay. The Majorana mass term is
\begin{align}
-\mathcal{L}_\mathrm{Majorana} &= \frac{1}{2} m_T\left(\bar\nu_L \nu_R^c + \bar
\nu_R^c\nu_L\right) = \frac{1}{2} m_T\bar\nu\nu\\
&=  \frac{1}{2} m_T\left(\bar\nu_L C \bar\nu_L^T + H.c.\right), \label{activemajorana}
\end{align}
where $\nu = \nu_L + \nu_R^c$ is a two-component state that is
self-conjugate, satisfying $\nu = \nu^c = C \bar\nu^T$; $C$ is the
charge conjugation matrix. There could also be a Majorana mass term
among the sterile neutrinos alone,
\begin{equation}
-\mathcal{L}_\mathrm{Majorana, sterile} = \frac{1}{2} M_N\left(\bar N_L^c N_R
+ \bar N_R N_L^c\right).\label{sterilemajorana}
\end{equation}

\subsubsection{The Seesaw Mechanism and Related Models}
When $\nu_L$ is an active neutrino, isospin is violated by one unit by
the Majorana term, and $m^T$ must be generated by an elementary Higgs
triplet or by an effective operator from two Higgs doublets arranged
to act as a Higgs triplet. The latter case is
\begin{equation}
\frac{\lambda_{ij}}{M_N}\left(L_i \varphi \right)^T \left(L_j \varphi\right),\quad i,j=e,\mu,\tau,
\end{equation}
where $\varphi$ is the Higgs doublet, $\lambda_{ij}$ are dimensionless
couplings, and $M_N$ is a cutoff mass scale above electroweak symmetry
breaking. This generates the Majorana masses 
\begin{equation}
m_{ij} = \frac{\lambda_{ij}\langle H\rangle^2}{M_N} \sim
\frac{m_D^2}{M_N}.\label{seesaw}
\end{equation}
For $M_N$ large, $m_{ij} \ll m_D$, the high-scale completion of this
mechanism determines the type of seesaw mechanism
\cite{Mohapatra:2006gs}. If $M_N$ is at the Planck scale, $M_N \sim
M_\mathrm{Planck} \sim 10^{19}\ \mathrm{GeV}$, and $\lambda_{ij}\sim
1$. the neutrino mass scales are smaller than the observed splittings,
$m_{ij} \sim 10^{-5}$. This intriguingly indicates new physics below
the Planck scale.

The candidate mass insertions of
Eqs.~\eqref{diracmass},\eqref{activemajorana},\eqref{sterilemajorana}
can be distilled into a generalized form of new terms to enter the SM
to generate neutrino mass, dubbed the ``new Standard Model'' or
``neutrino Standard Model'' ($\nu$SM) \cite{deGouvea:2005er} or the
``neutrino Minimal Standard Model'' ($\nu$MSM) \cite{asaka:2005an},
\begin{equation}
\mathcal{L} \supset -h_{\alpha i} L_\alpha N_i \varphi - \frac{1}{2}
M_{ij}N_i N_j + H.c.,
\end{equation}
where $h_{\alpha i}$ are the Yukawa couplings for the flavor states
$\alpha = e,\mu,\tau$ and $M_{ij}=M_{ji}$ ($i,j=1,2,...$) are the
Majorana masses.  

There is quite a bit of freedom as to the values of sterile neutrino
masses $M_N$ and Yukawa couplings $h_{\alpha,i}$ allowed
\cite{deGouvea:2007hks}. Experimental data limits currently places a
wide range of values for these: $10^{-12}\lesssim h_{\alpha i}
\lesssim 1$ while $M_{ij} = 0$ for all $i,j$, where Majorana neutrinos
``fuse'' to Dirac neutrinos, to $M_{ij} \sim
10^{16}\ \mathrm{GeV}$. There is a relatively small range at
$0.001\ \mathrm{eV}\lesssim M_{ij}\lesssim 1\ \mathrm{eV}$, which is
disfavored since it induces large sterile neutrino mixing which is
disfavored by oscillation data. The seesaw mechanism is simply the
case where $M_i \gg h_{\alpha i} v$, and the three sterile neutrinos
are split from the mass states associated with with active neutrinos.

Truly only two of the heavy sterile neutrinos are needed to produce
the solar and atmospheric mass scales, and the third is free to play
the role of either a largely sterile mass eigenstate that leads to the
short baseline anomalies or could play the role of dark matter
\cite{deGouvea:2005er,asaka:2005an}. Being open about the sterile
neutrino numbers reflecting the family structure of the standard model
allows for three sterile neutrino states and therefore one of these to
be a relative light state: a sterile neutrino associated with short
baseline oscillations or a dark matter sterile neutrinos. It is often
mistakenly taken that the short baseline anomalies indication of
potential light eV-scale neutrinos are motivation for dark matter
sterile neutrinos at the keV scale, but simplicity argues for only one
of the two sterile neutrino mass states to be likely, either eV or keV
in scale.

The mass scale of the active neutrinos is given by Eq.~\eqref{seesaw},
$m_\alpha \sim h^2 v^2/M_N$ while the active-sterile mixing angles
scale as 
\begin{equation}
\theta \sim \frac{h v}{M} \sim \sqrt{\frac{m_\alpha}{M}}.
\end{equation}
This allows for a predominantly sterile mass state to have a mass $m_s
\equiv M \sim \mathrm{keV}$ that has arbitrarily small (or large)
mixings as long as the ``active'' mass scale $m_\alpha$ at the
appropriate scale. Given that the lightest active neutrino mass
eigenstate is not bounded from below, neither is the mixing
angle. This is a natural mechanism for a sterile neutrino to have the
properties like that of a dark matter sterile neutrino, or an eV-scale
short baseline neutrino. However, this mechanism cannot provide both
eV-scale and keV-scale sterile neutrinos within a three family
structure. The origin of the splitting of the lower scale, eV or keV,
sterile neutrinos from the higher scale sterile neutrinos ($\gg
\mathrm{keV}$) must be come from a correction to zero mass for the
light sterile, or from a suppression relative to the higher
scale. Several mechanisms regarding both methods were explored in
Refs.~\cite{Merle:2013gea,Adhikari:2016bei}.

\subsubsection{Other Mass Generation Mechanisms}

There are numerous mechanisms for active sterile neutrino mass
generation that are related to the original seesaw described above, or
independent. As mentioned above, a triplet Majorana mass $m_T$ can be
generated by a Higgs triplet VEV $v_T$, in what is dubbed a Type-II
seesaw mechanism
\cite{Lazarides:1980nt,Magg:1980ut,Mohapatra:1980yp,Schechter:1980gr}. In
that case, $m_T = h_T v_T$, where $h_T$ is the
Yukawa coupling. The smallness of $m_T$ would be provided by a small
$v_T$. A combination of type I terms and type II is called a ``mixed
seesaw.'' Mohapatra \& Smirnov \cite{Mohapatra:2006gs} provide an
excellent review of seesaw and related neutrino mass generation
mechanisms. 

One case where additional singlets contributing to the neutrino mass
generation mechanism provided by a high-energy theory is dubbed an
inverse seesaw \cite{Mohapatra:1986bd}.  In
Merle~\cite{Merle:2013gea}, mass generation mechanisms for keV-scale
(and for that matter eV-scale) masses were usefully categorized as
those that arise from ``bottom-up'' models, where the natural sterile
neutrino scale is zero, with has a model-specific correction to shift
it to non-zero, or arise from ``top-down'' mechanisms, where the
natural scale is high ($\gg \mathrm{keV}$), but split downward to low
mass scales by suppression mechanism. Examples of bottom-up mechanisms
include two based on the flavor symmetries of $L_e-L_\mu-L_\tau$
\cite{Shaposhnikov:2006nn,Lindner:2010wr} and $Q_6$
\cite{Grossman:1999ra}. In other bottom-up models, neutrino masses are
only produced via higher order loops \cite{Zee:1980ai}.

One top-down model uses an extra dimension compactified in a spherical
orbifold geometry on $S_1/Z_2$, with the Yukawa couplings and right
handed Majorana masses on a standard model brane, while the right
handed neutrino wave functions exponentially localize on a hidden
brane. This allows splitting of the scales of the seesaw, with two
right handed neutrinos at high scale and one at a keV scale, in what
is called a split seesaw mechanism \cite{Kusenko:2010ik}. This was
extended to a flavor symmetry model in an $A_4$ extended split seesaw
\cite{Adulpravitchai:2011rq}. A version of the split seesaw has varied
localizations of both the right handed Majorana masses on the hidden
brane, and is dubbed the separate seesaw
\cite{Takahashi:2013eva}. Another set of top-down models  are
based on the Froggat-Nielsen mechanism
\cite{Chen:2006hn,Kamikado:2008jx,Merle:2011yv}.

Another class of models include the extended seesaw mechanism, where a
singlet fermion from a supersymmetric model can produce light sterile
neutrinos \cite{Chun:1995js,Barry:2011wb,Zhang:2011vh,Dev:2012bd}. In
Ref.~\cite{chun:1999cq} the axino plays the role as the singlet in
$R$-parity violating supersymmetry models. Another class of models
involve mirror sector \cite{Foot:1995pa,Berezhiani:1995yi} or
left-right symmetric models \cite{Bezrukov:2009th,Nemevsek:2012cd}
which in some cases can provide light eV-scale sterile neutrinos as
well as the keV scale dark matter sterile neutrino
\cite{Borah:2016lrl}. Arguably just as natural as the seesaw mechanism
is the possibility of Dirac neutrino masses,
e.g. Ref.~\cite{Chen:2012jg}. There are many further models of light
sterile neutrino mass generation, which are well reviewed in
Refs.~\cite{Abazajian:2012ys,Merle:2013gea,Adhikari:2016bei}.

\section{Thermal History of the Universe and Sterile Neutrino Production}

The cosmic microwave background (CMB) is now a very well-studied
thermal black body with small anisotropies that carry information from
the surface of last scattering in the early Universe. That information
includes the mass-energy content of the Universe and primordial
perturbation spectrum \cite{Hu:2001bc}, as well as the background of
cosmic neutrinos \cite{Abazajian:2016hbv}. The CMB has a current
photon temperature of $T_{\gamma,0} = 2.72548\pm 0.00057\rm\ K$
\cite{Fixsen:2009ug} and number density of $n_{\gamma,0} = 410.7
(T_{\gamma,0}/2.72548\rm\ K)^3$. The photon background had higher
temperature in the past, increasing inversely with the scale factor
$T\propto a^{-1}$, during phases where the degrees of freedom coupled
to the photons is constant. The photon number density correspondingly
increased as $n_{\gamma} \propto a^{-3}$, which implies a hot, dense
early phase. This led to the ionization of the primordial gas and
Thompson scattering coupling of the photon-electron fluid at early
times, prior to the CMB surface of last scattering.

At even earlier times, the photon background had sufficient energy to
pair produce an electron-positron background $\gamma +
\gamma\leftrightarrow e^+ + e^-$ at $T_\gamma \gtrsim
0.1\ \mathrm{MeV}$, with a neutrino background being created at higher
temperatures $T\gtrsim 1\ \mathrm{MeV}$: $e^+ + e^- \leftrightarrow
\nu_\alpha + \bar\nu_\alpha$, and, above respective energy thresholds
the entire standard model of particle physics catalog of particles is
created in the thermal bath, along with any non-standard particles
with sufficiently strong interactions \cite{KolbTurner1990}.  

In the neutrino-coupled era at $T\gtrsim 1\ \mathrm{MeV}$, the effects
of neutrino scattering take place in an environment where neutrino
oscillations would necessarily also need to be taken into account. In
the standard flavor and $CPT$ symmetric neutrino background, there are
not significant effects to due active neutrino oscillations, but
asymmetries can produce interesting collective effects
\cite{dolgov:2002ab,wong:2002fa,abazajian:2002qx}. These collective
oscillations constrain the total asymmetry of the Universe and are
tied to the mixing angle to which solar neutrino oscillations are
largely sensitive, $\theta_{12}$ \cite{abazajian:2002qx}.  In the
standard thermal history, active neutrinos are clearly created in
their flavor states but then freely propagate as mass eigenstates that
are symmetric except for a the slight heating that occurs from
partial coupling of the neutrinos during electron-positron
annihilation \cite{Dicus:1982bz,Mangano:2005cc}.

\subsection{Quantum Statistical Mechanics of Neutrinos}

Most of the topics in this report reflect neutrino mass and mixing in
the early Universe, where conditions are {\it very} different from an
ambient vacuum.  To approach the evolution of neutrinos in the early
Universe, a quantum statistical formulation must be adopted.  We
follow here the descriptions of the statistical formulation of a
quantum system through the density matrix as described in
Refs.\ \cite{Landau:qm,Sakurai1995}. The density operator, $\rho$, or
its matrix is the most general form for the quantum description of a
statistical system, and contains all physical information of the
system that we consider.  This seems ambitious, but it is simply the
nature of the density matrix by definition.  A general quantum system
here is described by a wave function $|\Psi(x,q)\rangle$, where $x$
are the coordinates or properties of the system that we are interested
in and $q$ are the coordinates we are the rest of the coordinates of
the system.  The average expectation value of some observable
$\hat{f}$, $\langle f\rangle$ in wave-function integral form is
\begin{equation}
\langle f\rangle =
\int{\int{\Psi^{\ast}(x,q)\hat{f}\Psi(x,q)\,dq\,dx}}.
\label{averagef}
\end{equation}

We can reduce the general system by averaging over all other
coordinates $q$, so that there remains only a weight $w_x$ for the
system to be characterized by $|\Psi(x)\rangle$.  The ensemble average
of an observable $\hat{f}$ is
\begin{equation}
\langle f\rangle = \sum_x \langle \Psi(x)|\hat{f}|\Psi(x) \rangle .
\label{averagef2}
\end{equation}
We can introduce the orthogonal complete bases $|a^\prime\rangle$ and
$|a^{\prime\prime}\rangle$ that may be of interest in our description
of the quantum system.  The average (\ref{averagef2}) can be written,
using completeness, as
\begin{align}
\begin{split}
\langle f\rangle &= \sum_x w_x \sum_{a^\prime} \sum_{a^{\prime\prime}}
\langle \Psi(x)| a^\prime\rangle \langle a^\prime | \hat{f} |
a^{\prime\prime}\rangle \langle a^{\prime\prime} | \Psi(x)\rangle\\
&= \sum_{a^\prime} \sum_{a^{\prime\prime}} \left({\sum_x w_x \langle
a^{\prime\prime} | \Psi(x)\rangle \langle \Psi(x)|
a^\prime\rangle}\right) \langle a^\prime | \hat{f} |
a^{\prime\prime}\rangle.
\label{averagef3}
\end{split}
\end{align}
In order to simplify the average over the coordinates of interest, we
define the density operator as
\begin{equation}
\hat\rho  = \sum_x{w_x |\Psi(x)\rangle\langle\Psi(x)|}\, ,
\end{equation}
which can be projected into a matrix in the basis of a set of states
$|a^\prime\rangle$ via
\begin{equation}
\langle a^{\prime\prime}|\hat\rho|a^\prime\rangle = \sum_x{ w_x
\langle a^{\prime\prime} | \Psi(x)\rangle \langle\Psi(x) |
a^\prime\rangle}.
\end{equation}
Therefore, the ensemble average Eq.\ (\ref{averagef3}) can be written
simply as 
\begin{align}
\begin{split}
\langle f \rangle &= \sum_{a^\prime} \sum_{a^{\prime\prime}} \langle
a^{\prime\prime} |\hat\rho | a^\prime\rangle \langle a^\prime | \hat{f}
|a^{\prime\prime}\rangle\\
&= \sum_{a^{\prime\prime}} \langle
a^{\prime\prime} |\hat\rho  \hat{f} |a^{\prime\prime}\rangle\\
&= {\rm tr}(\hat\rho \hat{f}).
\end{split}
\end{align}
Since this trace is independent of the basis representation, any
convenient basis is used.

The diagonal elements of the density matrix describe the
probability distribution of the system in a specific basis state
$|a^\prime\rangle$:
\begin{equation}
\langle f \rangle\big|_{a^\prime} = \langle
a^{\prime} |\hat\rho  \hat{f} |a^{\prime}\rangle.
\label{diag}
\end{equation}
In this way, the average value of an observable in a state
$|a^\prime\rangle$ can be found readily. Therefore, the analysis of
the density matrix of a system is a powerful formulation.  The
diagonal elements, Eq.~\eqref{diag}, serve as the statistical
distribution functions for the system.

For the early Universe, spatial coordinates are unnecessary under the
assumption of spatial homogeneity, which is the case in the standard
picture.  For neutrinos in this environment, we may project the
density matrix into a mass eigenstate basis or a flavor basis.  Now we
have the density matrix as a a tool for describing the statistical
properties of a quantum mechanical system, but in order to find how
neutrinos {\it evolve} in the early Universe, we need to be able to
evaluate the time evolution of the relativistic quantum statistical
ensemble.

The weak interactions of mixed active-sterile neutrinos have finite
width to produce sterile states, and serve to couple sterile density
operator amplitude to the thermal environment. The time evolution of
the density operator can be described by the Heisenberg equation of
motion:
\begin{equation}
\label{hberg2}
i \frac{\partial \rho}{\partial t} = [\rho,H],
\end{equation}
with an analogous equation for the antineutrino density matrix, $\bar\rho$.
Now, the problem lies in the choice of the Hamiltonian of the neutrino
ensemble $H$.  The first application of the evolution of the system
through a Heisenberg equation of the density matrix was done by Dolgov
\cite{Dolgov:1980cq}, who took the matrix elements of elastic ($\nu
e\leftrightharpoons\nu e$) and inelastic ($\nu\nu\leftrightharpoons
e^+ e^-$) to write a simple singlet neutrino production formulation.

Another way of looking at the time evolution of the density operator
is through a scattering matrix approach such that the differential
evolution of the density operator is given by \cite{McKellar:1992ja}
\begin{equation}
\label{scatteringrho}
\hat{\rho}^{(f)} = \hat{S} \hat{\rho}^{(i)}\hat{S}^\dagger
\end{equation}
which can be used to derive the scattering matrix elements for population
of singlet neutrinos through two-body elastic and inelastic processes,
and the time rate of change of the density operator.

The integration over contributing scattering matrix elements in the
right hand side of the density operator evolution equations or
``quantum kinetic equations'' \eqref{hberg2} \& \eqref{scatteringrho}
lead to the integrated and drastically less cumbersome ``quantum rate
equations.''  The derivation of the rate equations from the kinetic
equations has been described by McKellar \& Thomson
\cite{McKellar:1992ja}, Sigl \& Raffelt \cite{Sigl:1992fn}, and Bell
{\it et al.} \cite{Bell:1998ds}. 

One approach to solutions of the evolution of the sterile neutrino
states approaches the solution of Eq.~\eqref{hberg2} directly. This is
separately coupled to an additional relation governing the expansion of
the Universe, the Friedmann equation, with approximations to the
subsequent time-temperature evolution and scattering rates
\cite{Asaka:2006rw,Asaka:2006nq,Ghiglieri:2015jua}.
The most significant effects that enter the evolution of neutrinos in
the early Universe are the self energies produced the asymmetry and
thermal potentials. These self energies are from propagating active
neutrinos' interactions with the plasma
\cite{notzold:1987ik}. There are three contributions to the neutrino
self energy: (a) an imaginary part proportional to the net neutrino
scattering rate or opacity, (b) a real part due to finite weak gauge
boson masses ($V^{\rm th}$), and (c) a real part proportional to
asymmetries in weakly interacting particles ($V^{\rm L}$). This
exposition follows the treatment in Venumadhav et
al.\ \cite{Venumadhav:2015pla} that employs some of the definitions
from Ref.~\cite{notzold:1987ik}. We will specify to different regimes
of applicability later.

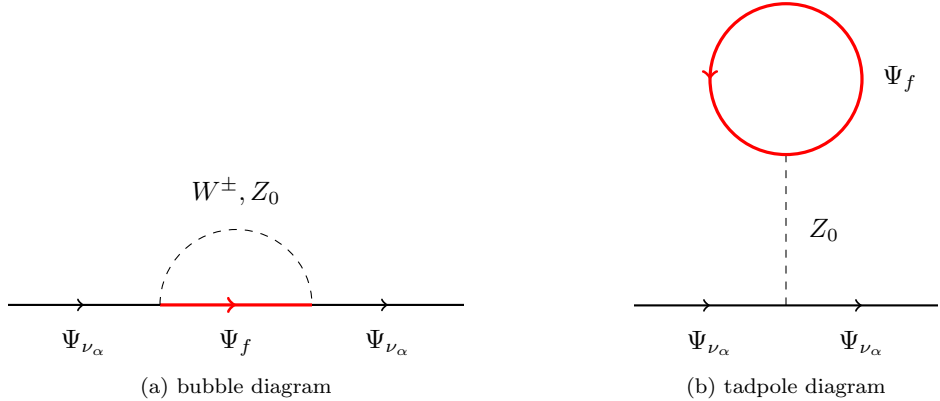
\begin{figure}[t]
  \hspace*{\fill}%
  \subfloat[\label{fig:bubble}bubble diagram]{
    \begin{tikzpicture}[scale=1.]
      \begin{scope}[decoration={
	markings,
	mark=at position 0.5 with {\arrow{>}} }
      ]
      \draw[black, thick, postaction={decorate}] (-3, 0) -- (-1, 0);
      \node at (-2,-0.5) {$\Psi_{\nu_\alpha}$};
      \draw[red, very thick, postaction={decorate}] (-1, 0) -- (1, 0);
      \draw[black, thick, postaction={decorate}] (1, 0) -- (3, 0);
      \node at (2,-0.5) {$\Psi_{\nu_\alpha}$};
      \draw[black, dashed] (-1, 0) arc (180:0:1cm);
      \node at (0, 1.5) {$W^\pm,Z_0$};
      \node at (0, -0.5) {$\Psi_f$};
      \end{scope}
    \end{tikzpicture}
  }  
  \hfill 
  \subfloat[\label{fig:tadpole}tadpole diagram]{
    \begin{tikzpicture}[scale=1.]
    \begin{scope}[decoration={
	markings,
	mark=at position 0.5 with {\arrow{>}} }
      ]
      \draw[black, thick, postaction={decorate}] (-2, 0) -- (0, 0);
      \node at (-1,-0.5) {$\Psi_{\nu_\alpha}$};
      \draw[black, thick, postaction={decorate}] (0, 0) -- (2, 0);
      \node at (1,-0.5) {$\Psi_{\nu_\alpha}$};
      \draw[black, dashed] (0, 0) -- (0, 2);
      \node at (0.5, 1) {$Z_0$};
      \draw[red, very thick, postaction={decorate}] (0, 3) circle (1cm);
      \node at (1.5, 3) {$\Psi_f$};
    \end{scope}
    \end{tikzpicture}
  }
   \hspace*{\fill}%

  \caption{\label{fig:oneloop}These are the lowest order contributions to a
    propagating active neutrino's self energy. Red lines are thermal
    propagators. In (a), $f$ is any species with weak charge. In (b),
    $f = \nu_\alpha, \alpha^-$.}
\end{figure}

Figure \ref{fig:oneloop} shows the lowest-order contributions to the
neutrinos' self energy. Thick red lines are thermal propagators of
weakly charged species in the background plasma. There are two
contributions: bubbles and tadpoles, shown in Fig.~\ref{fig:bubble} and
\ref{fig:tadpole} respectively. The background fermion is a lepton of
the same flavor in the former, and any weakly charged species in the
latter.

A massless active neutrino's propagator is
\begin{align}
  G_{\nu_\alpha}^{-1}(p_{\nu_\alpha}) & = {\not}{p_{\nu_\alpha}} -
  b_{\nu_\alpha}(p_{\nu_\alpha}) {\not}u \left( 1 - \gamma_5 \right)/2
  \mbox{,} \label{eq:propcorrect} \\ b_{\nu_\alpha}(p_{\nu_\alpha}) &
  = b_{\nu_\alpha}^{(0)} + b_{\nu_\alpha}^{(1)} \omega_{\nu_\alpha}
  \mbox{,} \qquad \omega_{\nu_\alpha} = - p_{\nu_\alpha} \cdot u
  \mbox{,} \label{eq:vsplit}
\end{align}
where $p_{\nu_\alpha}$ and $u$ are the neutrino and
plasma's four-momenta, ${\not}v$ is shorthand for 
$\gamma^{\mu} v_{\mu}$, and $b_{\nu_\alpha}$ is the left handed neutrino's self
energy. The relation Eq.~\eqref{eq:vsplit} divides the self energy into separate
contributions that affect the particle and anti-particles in
Eq.~\eqref{eq:propcorrect} differently. 

The finite density potential in the early universe can be dominated by
asymmetries in the lepton number. Therefore it is often referred to as the
``lepton potential'' $V^L$.  It takes the form
\begin{equation}
\label{vl}
V^L = \frac{2 \sqrt{2} \zeta (3)}{\pi^2}\,G_{\rm F} T^3 \left({\cal
L}^\alpha \pm \frac{\eta}{4}\right),
\end{equation}
where we take \lq\lq $+$\rq\rq\ for $\alpha=e$ and \lq\lq$-$\rq\rq\ for
$\alpha=\mu,\tau$.
Here we define the lepton number ${\cal{L}^\alpha}$ in terms
of the lepton numbers in each active neutrino species as
\begin{equation}
\label{L}
{\cal{L}^\alpha} \equiv 2 L_{\nu_\alpha} +
\sum_{\beta\neq\alpha}{L_{\nu_\beta}},
\end{equation}
with the final sum over active neutrino flavors other than
$\nu_\alpha$. Here $\eta\equiv n_{\rm b}/n_\gamma$ is the baryon to
photon ratio.
An $\alpha$-type neutrino asymmetry is defined as 
\begin{equation}
L_{\nu_{\alpha}}\equiv\frac{n_{\nu_{\alpha}}-n_{\bar{\nu}_{\alpha}}}{n_\gamma},\label{leptonnumber}
\end{equation}
where the photon number density $n_{\gamma}=2\zeta(3)T^{3}/\pi^{2}$.

There are a  large number of scattering processes that contribute to the
neutrino opacity (i.e.~the imaginary part of the self-energy) at
temperatures $10$ MeV $\leq T \leq 10$ GeV. Accurate neutrino
opacities are needed since they control the production of of sterile
neutrinos as well as the damping of the active neutrinos'
oscillations. Quantum damping, also described as the quantum Zeno
effect, is the suppression of conversions due to the interaction
length of neutrinos being much smaller than the oscillation
length. That is, there is no appreciable conversion possible unless
neutrinos are able to sufficiently propagate to have a non-trivial
probability to convert to sterile neutrinos.  

Early work on neutrino interactions in the early
Universe proposed that neutrinos largely scattered off relativistic
particles and therefore scaled their cross-sections with the
center-of-mass (CM) energy \cite{notzold:1987ik, McKellar:1992ja,
  abazajian:2001nj}.  In addition, these early calculations neglected
the effects of particle statistics. Under these two simplifying
assumptions, the opacity $\Gamma(E_{\nu_\alpha})$ for an input
neutrino of energy $E_{\nu_\alpha}$ is of the form
\begin{equation}
  \Gamma(E_{\nu_\alpha}) = \lambda(T) G_{\rm F}^2 T^4 E_{\nu_\alpha} \mbox{,} \label{eq:pscaling}
\end{equation}
where $G_{\rm F}$ is the Fermi coupling constant, and $\lambda(T)$ is
a constant that depends on the number and type of available
relativistic species in the cosmic plasma. References
\cite{Asaka:2006rw,Asaka:2006nq} subsequently developed a framework to
include particle masses, loop corrections, and particle statistics in
the neutrino opacity calculation. These were also included in
Venumadhav et al.\ \cite{Venumadhav:2015pla}, where
previously-neglected contributions to the scattering rates such as
two- and three-body fusion reactions were included,
also. Ref.\ \cite{Venumadhav:2015pla} was first in using chiral
perturbation theory to compute the hadronic contribution to the
scattering opacity below the quark-hadron transition, with significant
qualitative and quantitative modifications to the form of
Eq.~\eqref{eq:pscaling}.

\subsubsection{Partial to Full Thermalization in Symmetric Thermal
  Backgrounds}

Oscillations induced by the mass generation mechanism between light
eV-scale to keV-scale sterile neutrinos allows for them to be
populated by scattering processes in the cosmological background. The
initial work on this mechanism used a quasi-classical treatment that
included the effects of quantum damping and thermal
self-energy. Langacker \cite{Langacker:1989sv} showed that quantum
damping suppresses active-sterile neutrino oscillations at the highest
temperatures in the early Universe, and that active-sterile mass
splittings of $\delta m^2 \gtrsim 10^{-7} \ \mathrm{eV}^2$ will fully
thermalize at maximal mixing, while $\delta m_{\alpha s}^2 \sim 1
\ \mathrm{eV}^2$ will thermalize with mixings of $\sin^2 2\theta >
10^{-4}$. Subsequent work found that finite temperature effects
further suppressed the mixing and conversion of active to sterile
neutrinos, with similar regions of parameter space where sterile
neutrinos would be thermalized, notably the parameter space associated
with the large mixing angle solution to the solar neutrino problem as
well as the short baseline anomalies
\cite{Barbieri:1989ti,Barbieri:1990vx,Enqvist:1990dq,Enqvist:1990ek,Enqvist:1991qj,Cline:1991zb}.

The essential discoveries of early work are in the matter and thermal
affected contributions to the neutrino self-energy. They lead to an
effective mixing between the active and sterile neutrinos that
suppresses mixing at high temperatures:
\begin{equation}
\sin^2 2\theta_m = \frac{\Delta^2 (p) \sin^2 2\theta}{\Delta^2 (p)
\sin^2 2\theta + \left[\Delta (p) \cos 2\theta - V^L -
V^T(p)\right]^2}.\label{matterangle}
\end{equation}
Combining this with quantum damping, the effective production rate of
sterile neutrinos is
\begin{equation}
\Gamma(\nu_\alpha\rightarrow \nu_s; p,t) \approx
\frac{\Gamma_\alpha}{2} \langle P_m(\nu_\alpha \rightarrow
\nu_s;p,t)\rangle.
\label{conversrate}
\end{equation}
This probability $P_m$ depends on the amplitude of the matter mixing
angle and the quantum damping rate
\begin{align}
  D(p)&=\Gamma_\alpha(p)/2 \label{damping} \\
  \bar D(p)&=\bar\Gamma_\alpha(p)/2, \nonumber
\end{align}
for neutrinos and antineutrinos
\cite{Stodolsky:1986dx,Bell:1998ds},
\begin{align}
\langle P_m(\nu_\alpha\rightarrow\nu_s; p,t) \rangle &\approx \frac{1}{2}
\frac{\Delta(p)^2 \sin^2
2\theta}{\Delta(p)^2\sin^2 2\theta + D^2(p) +
[\Delta(p) \cos 2\theta - V^L - V^T(p)]^2} 
\label{avgprob}\\
\langle P_m(\bar{\nu}_\alpha\rightarrow\bar{\nu}_s; p,t) \rangle &\approx
\frac{1}{2} \frac{\Delta(p)^2 \sin^2 2\theta}{\Delta(p)^2\sin^2
2\theta + \bar{D}^2(p) + [\Delta(p) \cos 2\theta + V^L - V^T(p)]^2}.
\label{avgprobnubar}
\end{align}
There are a number of consequences of this collision-dominated
conversion process:
\begin{enumerate}
\item With appreciable mixing and specific mass differences, sterile
  neutrinos will be fully thermalized by the oscillations via collision
  dominated production, when no other new physics is included. This
  makes sterile neutrinos that thermalize be subject to classic bounds
  from primordial nucleosynthesis and over-closure as well as new
  bounds on relativistic density and hot dark matter from the CMB and
  LSS. These effects are predominantly present for active-sterile mass
  splittings of $\delta m \gtrsim 10^{-7} \ \mathrm{eV}^2$ to $\delta
  m_{\alpha s} \sim 100 \ \mathrm{eV}^2$.
\item At keV-scales, there is a level of mixing at which the
  sterile neutrino is produced at the proper amount to be all or part
  of the dark matter, is mildly relativistic to nonrelativistic at
  matter-radiation equality and CMB decoupling, but can act as hot to
  ``warm'' dark matter since it has a significant free streaming
  scale. This is the Dodelson-Widrow mechanism. There are constraints
  on sufficiently large free streaming scales from observations of
  cosmological and even Milky Way Local Group structure. We discuss
  these constraints in Section \ref{subsec:keVstructure} below.
\item Even during collision-domination, the system can have a
  resonances between the terms in brackets in the denominators of
  Eqs.\ \eqref{matterangle},\eqref{avgprob},\eqref{avgprobnubar}. This
  allows for the resonant {\it creation or destruction} of lepton
  number with associated production or destruction of sterile neutrino
  density. The multitude of potential processes that that can engender
  is described in the following section. The case of conversion of
  primordial lepton number to sterile neutrinos via this resonance is
  the Shi-Fuller mechanism.
\end{enumerate}

\subsubsection{Creation of Lepton Number}

Since the sterile neutrino states are a bath of particles that are
typically initially unpopulated and are placed in thermal contact with
the active neutrinos via matter-affected oscillations, the system was
found to be analogous to two thermal state condensed matter systems
\cite{Harris:1980zi,Stodolsky:1986dx}. In such cases a two-state
neutrino system is projected into a polarization vector description
\begin{align}
\rho(p) &= \frac{1}{2}P_0(p)\left[1+\mathbf{P}(p)\cdot\bm{\sigma}\right],\cr
\bar\rho(p) &= \frac{1}{2}\bar P_0(p)\left[1+\mathbf{\bar P}(p)\cdot\bm{\sigma}\right],
\end{align}
where
\begin{equation}
\mathbf{P}(p)\equiv P_x(p)\mathbf{\hat x}+ P_y \mathbf{\hat y} + P_z(p) \mathbf{\hat z},
\end{equation}
where $P_0$ and $\mathbf{P}$ are convenient representations of the
magnitude of the elements of the density matrix $\rho$ in terms of the
Pauli matrices $\bm\sigma$. $\mathbf{P}$ is often called the
polarization of the neutrino system, is dependent on time, and
effectively represents the presence of asymmetry, such as the presence
of active neutrinos over sterile neutrinos. Recall that diagonal
elements of $\rho$ and $\bar\rho$ are the relative number density
distributions or phase space distributions (PSDs) of $\nu_\alpha$ or
$\bar\nu_\alpha$ and $\nu_s$ or $\bar\nu_s$.

The time evolution of the polarization vector $\mathbf{P}$ and
magnitude $P_0$ is obtained from evolving the density matrix forward
in time including the presence of thermal, matter and scattering
effects \cite{Bell:1998ds}. The evolution of $\mathbf{P}(p)$,
$P_{0}(p)$, $\bar{\mathbf{P}}(p)$ and $\bar{P}_{0}(p)$ are given by
the following equations (e.g., \cite{Kainulainen:2001cb}),
\begin{align}
  \frac{d\mathbf{P}}{dt} & =  \mathbf{V}(p) \times \mathbf{P}(p) - 
     D(p)[P_{x}(p)\hat{\mathbf{x}} + 
     P_{y}(p)]\hat{\mathbf{y}}] + 
     \frac{d P_{0}}{dt}\hat{\mathbf{z}}\nonumber \\
  \frac{dP_{0}}{dt} & \simeq 
  \Gamma(p)\left[\frac{f_{eq}(p)}{f_{0}(p)}-\frac{1}{2}(P_{0}(p)+P_{z}(p))\right].
\label{eq:qke}
\end{align}
The equations for the anti-particles are given by the substitutions
$\mathbf{P} \rightarrow \bar{\mathbf{P}}$,
$P_{0}\rightarrow\bar{P}_{0}$,
$\mathbf{V}(p)\rightarrow\bar{\mathbf{V}}(p)$, $f_{eq}(p) \rightarrow
\bar{f}_{eq}(p)$.  $\bar{\mathbf{V}}(p)$ is obtained by replacing
$L^{(\alpha)}$ by $-L^{(\alpha)}$. The total collision rates of the
flavor neutrino and anti-neutrino are approximately equal,
$\Gamma(p)\simeq\bar{\Gamma}(p)$.  These 8 equations form the quantum
kinetic equations for active-sterile oscillation.

The damping coefficient is given by Eq.~\eqref{damping}.
The rotation vector $\mathbf{V}(p)$ has the following components
\begin{align}
  V_{x}(p) & =  \frac{\delta m^{2}}{2p}\sin2\theta_{0}\nonumber \\
  V_{y}(p) & =  0\nonumber \\
  V_{z}(p) & =  V_{0}(p)+V_{L}(p),
\end{align}
where
\begin{align}
  V_{0}(p)&= -\frac{\delta m^{2}}{2p}\cos2\theta_{0}+V_{1}\nonumber \\
  V_{1}(p)&= -\frac{7\sqrt{2}}{2}\frac{\zeta(4)}{\zeta(3)} \frac{G_{F}}{M_{z}^{2}}n_{\gamma}pT[n_{\nu_{\alpha}}+n_{\nu_{\bar{\alpha}}}]\nonumber \\
  V_{L}(p)&= \sqrt{2}G_{F}n_{\gamma}L^{(\alpha)}.\nonumber
\end{align}
Here $n_{\gamma}$ is the photon equilibrium number density,
$n_{\nu_{\alpha}}\text{ and }n_{\nu_{\bar{\alpha}}}$ are normalized to
unity and $f_{eq}(p)$ is the Fermi-Dirac distribution with a chemical
potential $\mu_{\alpha}$. 

It was discovered that this active-sterile neutrino mixing system
could generate a lepton number under the proper conditions, namely
that the mass eigenstate more closely associated with the sterile
neutrino is lighter than that of the predominantly active neutrino
with which it mixes, $m_4 < m_1$ \cite{foot:1995bm}, and furthermore
it was found that this process could be chaotic in nature with the
potential indeterminate sign in the lepton number \cite{Shi:1996ic}.

\begin{figure}[t]
\begin{center}
   \includegraphics[height=10cm]{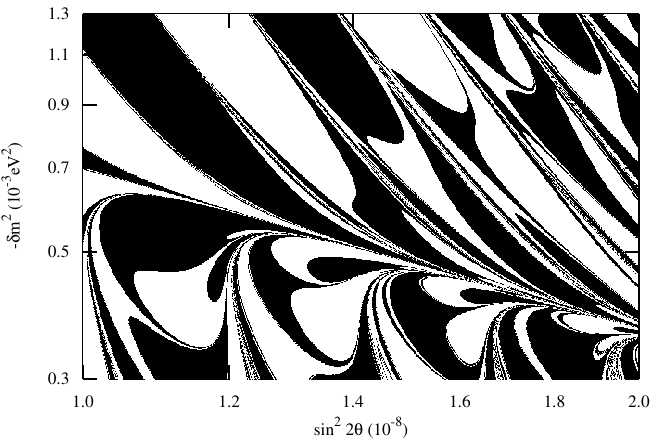}
\caption{Shown is the final sign of the lepton asymmetry as a function
  of mass splitting and mixing angle in the case of
  $\nu_\tau\leftrightarrow\nu_s$ mixing for a specific range where
  arbitrary experimental accuracy could be unable to predict the sign
  of the lepton number of the Universe. White indicates a positive
  sign and black negative, from
  Ref.~\cite{Abazajian:2008dz}.\label{fractal}}
\end{center}
\end{figure}

There is one significant aspect of the fact that the chaos is
introduced by the system: it leads to the sign of the lepton number
not being predictable for certain parts of the active-sterile neutrino
mixing parameter space. Furthermore, it has a fractal nature in parts
of the active-sterile neutrino mixing parameter space, leading to the
potential indeterminacy in the sign of any lepton number generated in
the early Universe. In such a case, no matter the precision and
accuracy of the parameters of any discovered light scale sterile
neutrino, we would be unable to predict the sign of the lepton number
generated, and it would be necessary to determine other ways, such as
its effects on the weak rates setting the stage for primordial
nucleosynthesis \cite{Abazajian:2008dz}. A view of the sign of the
lepton number as a function of a sterile neutrino's position in
parameter space is shown in Fig.~\ref{fractal}.

\section{Light eV-scale Sterile Neutrinos: Thermalization Conditions
  and Early Universe Constraints}
\label{sec:eV-scale}
If a sterile neutrino is present, its effects on the early Universe
may include lepton number generation, destruction, thermalization and
potentially all of the above. The contingencies of what the sequence
of what would occur in the presence of arbitrary numbers of sterile
neutrinos present with arbitrary parameters has not been completely
explored. There have been a number of papers that have studied the
evolution of a four neutrino system in the case of only one sterile
neutrino with appreciable mixing, consistent with short baseline
anomalies
\cite{Bell:1998sr,Abazajian:1999tc,dibari:2001ua,abazajian:2002bj}.

A synthesis of the possibilities was in Abazajian et
al.~\cite{abazajian:2004aj}. There, it was shown that the presence of
a lepton number in the early Universe---in models with four or more
sterile neutrinos--- almost certainly produces nonthermal active and
sterile neutrino spectra, regardless of the origin of the lepton
asymmetry. It was also shown that a standard MSW resonance with smooth
transformation until lepton number is depleted is impossible, but must
be stochastic given the nature of the resonance evolution.  Mixing
among the active neutrinos almost certainly transfers some of the
distortions and asymmetry to the electron neutrino sector
\cite{dolgov:2002ab,abazajian:2002qx,wong:2002fa}, which will directly
affect the weak rates an light element production in the early
Universe.

A number of the possible four neutrino scenarios were studied in
Ref.~\cite{Smith:2006uw}. Much of the parameter space for
active-sterile mixing is ruled out by light element abundances from
primordial nucleosynthesis, plus hot dark matter ($\Omega_{\nu_s}$)
constraints from the CMB plus LSS, but not all. The full quantum
kinetic treatment and implications remain to be explored for 3 active
neutrino---with all respective known mixings---and $N$ sterile
neutrinos and their implications for light elements, the CMB and LSS.

\section{Sterile Neutrino keV-scale Dark Matter}

\subsection{Oscillation-based Production}

In the case of production that primarily takes place via oscillations,
the parameters governing production are closely tied to the
detectability of sterile neutrino dark matter via X-ray observations
and effects on structure formation.  The primordial initial lepton
number of the Universe, Eq.~\eqref{leptonnumber}, is a single
parameter that connects the resonant-production Shi-Fuller mechanism
with the non-resonant production Dodelson-Widrow mechanism.

For the non-resonant Dodelson Widrow mechanism the production rate at
high temperatures decreases with increasing temperature due to finite
temperature and damping effects, so that the rate of production to the
Hubble expansion rate is $\Gamma/H \propto T^3$, while at lower
temperatures the collisions fade rapidly and $\Gamma/H \propto
T^{-9}$. The peak rate of sterile neutrino production occurs at
temperature \cite{Barbieri:1989ti,dodelson:1993je}
\begin{equation}
T_{\rm max} \approx 133{\rm\,
MeV}\left(\frac{m_s}{1{\rm\,keV}}\right)^{1/3},
\label{tpeak}
\end{equation}
where $m_s$ is the mass eigenstate that is predominantly sterile.
Therefore, for keV-scale sterile neutrino dark matter, sterile
neutrinos with sufficiently high masses---that avoid structure formation
bounds---are produced above temperatures $T \gtrsim 100$ MeV.

For Dodelson-Widrow production the proper mixing required for a given
sterile neutrino particle mass scale is best determined now by
Ref.~\cite{abazajian:2005gj}, which gives a fitting form for the
proper level of production of dark matter at the proper density
\begin{align}
m_s \approx & \ 3.40\ {\rm keV}\ \left(\frac{\sin^2
  2\theta}{10^{-8}}\right)^{-0.615}\ \left(\frac{\Omega_{\rm DM}
  }{0.26}\right)^{0.5}\cr
&\times\ \left\{ 0.527\ {\rm erfc}\left[ -1.15 \left(\frac{T_{\rm
        QCD}}{170\rm\ MeV}\right)^{2.15}\right]\right\},
\label{prediction}
\end{align}
where $\sin^2 2\theta$ is the mixing between the predominantly active
and sterile mass states; the dark matter density relative to critical
is $\Omega_\mathrm{DM} \equiv \rho_\mathrm{DM}/\rho_\mathrm{crit}$,
$\rho_\mathrm{crit} \approx 1.05375 \times 10^{-5} h^2\ \mathrm{GeV\,
  cm^{-3}}$; the Hubble constant $H_0$ is scaled as $h \equiv
H_0/(100\ \mathrm{km\, s^{-1}\, Mpc^{-1}})$; and, $T_\mathrm{QCD}$ is
the temperature of the cross-over quark-hadron transition.

For resonant production in universes with nontrivial primordial lepton
numbers, the production temperature depends on the mass, mixing angle
and the primordial lepton number, and has not been condensed into a
three parameter fit. A sense of the production temperature scale is
achieved by looking at where the resonance, which moves from low to
high momenta, would be appreciably through the active neutrino
momentum distribution and therefore have depleted the primordial
lepton number and created the dark matter. The position of the
resonance is given by \cite{shi:1998km}
\begin{align}
\label{eres}
\epsilon_{\rm res} &\approx {\frac{\delta m^2}{
{\left( 8\sqrt{2}\zeta(3)/\pi^2\right)} G_{\rm F} T^4 {{L}} }}\\
&\approx 3.65 {\left({\frac{\delta m^2}{(7\,{\rm
keV})^2}}\right)} {\left({\frac{{10}^{-3}}{{{L}}}}\right)}
{\left({\frac{170\,{\rm MeV}}{T}}\right)}^4 \nonumber\ ,
\end{align}
where $\epsilon_{\rm res} \equiv p/T|_{\rm res}$ is the position of
the resonance. 

The most accurate calculations for resonant production are given by
Venumadhav et al.~\cite{Venumadhav:2015pla}\footnote{The code for the
  numerical calculation of the production is available at
  \texttt{https://github.com/ntveem/sterile-dm}}.  For lepton numbers
and sterile neutrino particle masses consistent with constraints the
production largely occurs at $T\gtrsim 100\ \mathrm{MeV}$, where the
process is collisionally dominated, i.e.~the interaction contribution
dominates the vacuum oscillations. In this regime, the evolution of
the density matrix separates out and yields a quasi-classical
Boltzmann transport equation for the diagonal terms, which are the
momentum space distributions of the active and sterile components
\cite{Bell:1998ds,Volkas:2000ei,Lee:2000ej}. The quasi-classical
Boltzmann equation that governs the incomplete coming to equilibrium
of the sterile neutrino momentum space distribution function is
\begin{align}
  &\frac{\partial}{\partial{t}}f_{\nu_{\rm s}}(p,t) - H\,p\,
  \frac{\partial}{\partial p }f_{\nu_{\rm s}}(p,t) = \nonumber\\
&\quad \sum_{\nu_x + a + \dots \rightarrow i + \dots} \int{\frac{d^3 p_a}{(2\pi)^3 2 E_a} \dots \frac{d^3 p_i}{(2\pi)^3 2 E_i} \dots (2\pi)^4 \delta^4(p+p_a + \dots - p_i - \dots)}\nonumber\\
&\qquad\qquad\qquad \times \frac12 \left[ \langle P_{\rm m}(\nu_\alpha \rightarrow \nu_{\rm s}; p,t) \rangle \left(1-f_{\nu_{\rm s}}\right) \sum |\mathcal{M}|^2_{i + \dots \rightarrow a + \nu_\alpha + \dots} f_i \dots \left(1 \mp f_a \right) \left( 1 - f_{\nu_\alpha} \right) \dots \right. \nonumber\\
& \qquad\qquad\qquad\qquad ~ \left. - \langle P_{\rm m}(\nu_{\rm s} \rightarrow \nu_\alpha; p,t) \rangle f_{\nu_{\rm s}}\left( 1 - f_{\nu_\alpha} \right) \sum |\mathcal{M}|^2_{\nu_\alpha + a + \dots \rightarrow i + \dots} f_a \dots \left( 1 \mp f_i \right) \dots \right] \mbox{.}
\label{eq:classicalboltz}
\end{align}
There is an analogous equation for the antineutrinos. Here, the $f(p)$
are momentum space distribution functions for particles with momentum
$p$ and energy $E$, and $H$ is the Hubble expansion rate at that
time. The sum on the right hand side is over all reactions that
consume or produce an active neutrino. The symbol $\sum |\mathcal
M|^2$ denotes the squared and spin-summed matrix element for the
reaction, and the factors of $(1 \mp f)$ implement Pauli blocking/Bose
enhancement, respectively. The factor of $1/2$ accounts for the fact
that only one active neutrino state in the two-state system interacts
\cite{Harris:1980zi,Stodolsky:1986dx,Raffelt:1991ck}. The $P_{\rm m}$
are matter-affected active--sterile oscillation probabilities, which
depend on the vacuum mixing angle $\theta$, and are modified by
interactions with the medium as described above. In terms of these
quantities, the oscillation probabilities are
\cite{Volkas:2000ei,Lee:2000ej}
\begin{equation}
  \langle P_{\rm m}(\nu_\alpha \leftrightarrow \nu_{\rm s}; p,t) \rangle 
   = \, (1/2) \Delta^2 (p) \sin^2{2\theta} 
   \Bigl\{ \Delta^2 (p) \sin^2{2\theta} + D^2(p) 
   + \left[\Delta (p) \cos 2\theta - V^{\rm L} - V^{\rm th}(p)\right]^2 \Bigr\}^{-1} \mbox{,} \label{eq:probabilities}
\end{equation}
where $D(p)$ is the quantum damping rate described above. Here,
$\Delta(p)$ is the momentum scaled vacuum oscillation rate, $\Delta(p)
\equiv \delta m^2_{\nu_\alpha,\nu_{\rm s}}/2p$. The neutrino self
energy is split into the lepton asymmetry potential $V^{\rm L}$, and the
thermal potential $V^{\rm th}$ (the asymmetry contribution enters with
the opposite sign in the version of Eq.~\eqref{eq:probabilities} for
antineutrinos). The net interaction rate for an active neutrino is
\begin{align}
%  ~~~ & \!\!\!\!
  \Gamma_{\nu_\alpha}(p) 
   &= \sum_{\nu_x + a + \dots \rightarrow i + \dots} \!\! \int \frac{d^3 p_a}{(2\pi)^3 2 E_a} \dots \frac{d^3 p_i}{(2\pi)^3 2 E_i} \dots
  (2\pi)^4 \delta^4(p+p_a + \dots - p_i - \dots) \nonumber \\
  &\qquad\qquad\qquad\qquad\qquad \times \sum |\mathcal{M}|^2_{\nu_\alpha + a + \dots \rightarrow i + \dots} f_a \dots \left( 1 \mp f_i \right) \dots \label{eq:scatteringrate}
\end{align}
The resulting Boltzmann equation for quantum-damped,
collisionally-driven sterile neutrino production is \cite{abazajian:2001nj}
\begin{align}
  ~ ~ ~ & \!\!\!\!
  \frac{\partial}{\partial{t}}f_{\nu_{\rm s}}(p,t) - H\, p\,
\frac{\partial}{\partial p }f_{\nu_{\rm s}}(p,t) \nonumber \\
& \approx
 \frac{\Gamma_{\nu_\alpha}(p)}{2} \langle P_m(\nu_\alpha
\leftrightarrow \nu_{\rm s}; p, t) \rangle
\left[f_{\nu_\alpha}(p,t) - f_{\nu_{\rm s}}(p,t)\right] \mbox{,}
\label{eq:productionmaster}
\end{align}
and there is a related equation for antineutrinos. There have been some
questions as to the subtleties regarding the effects of
quantum-damping in the case of resonance \cite{boyanovsky:2006it}, but
tests with the full density matrix formalism find that the
quasi-classical treatment is appropriate \cite{kishimoto:2008ic}.

\begin{figure}[t]
\begin{center}
   \includegraphics[height=10cm]{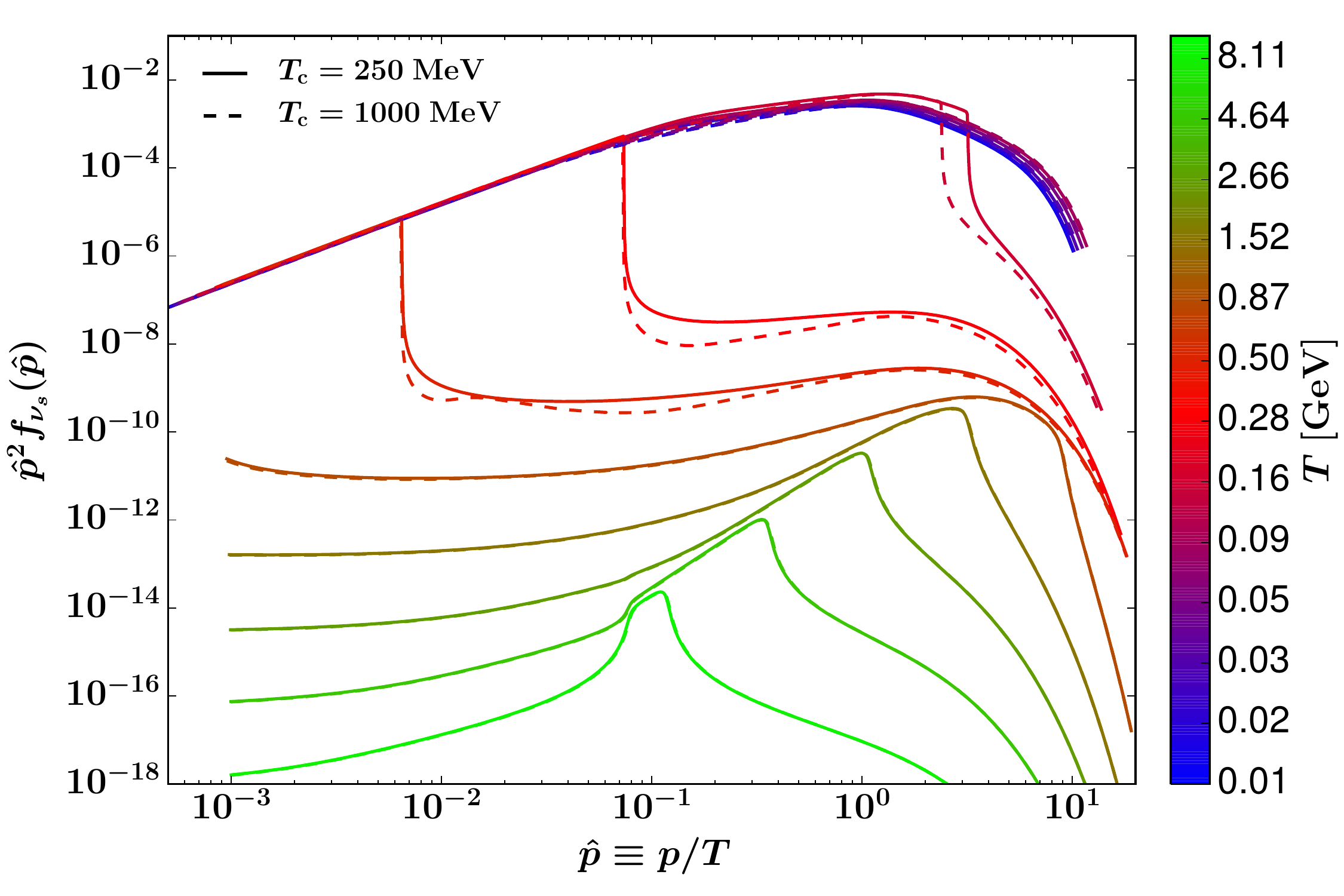}
\end{center}
\caption{We illustrate the temperature-evolution of the sterile
  neutrino's momentum space distributions for the example case of the
  central model of Fig.~\ref{fig:parspace} with $(m_{\rm s},
  \sin^2{2\theta}) = (7.1 \ {\rm keV}, 4 \times 10^{-11})$. Solid and
  dashed lines distinguish results with varying neutrino
  opacities. This figure is from Ref.~\cite{Venumadhav:2015pla}.}
\label{fig:loge2fe_temp}
\end{figure}

\begin{figure}[t]
\begin{center}    
\includegraphics[height=10cm]{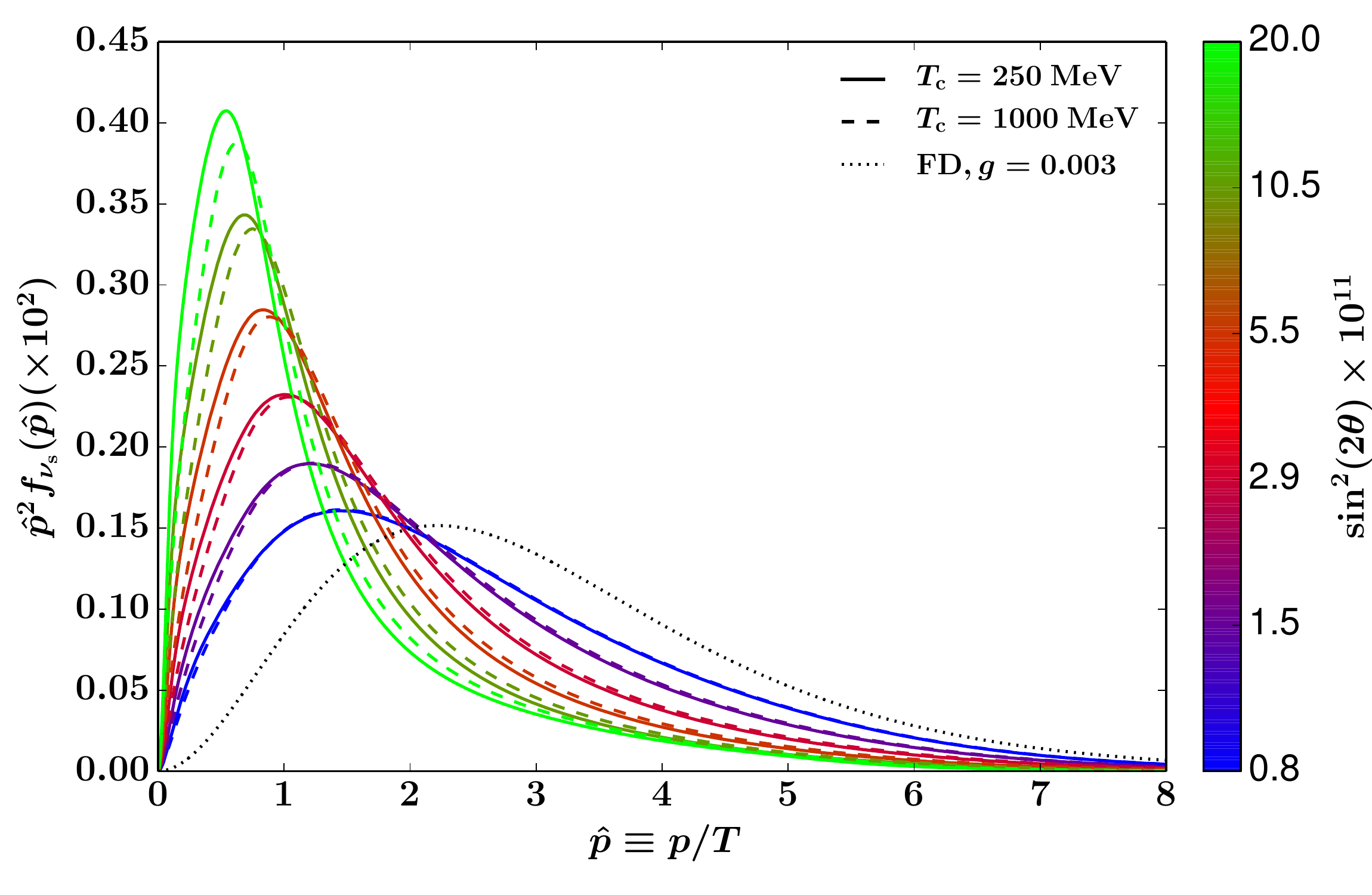}
\end{center}
\caption{Shown here are sterile neutrino momentum space distributions
  at close of production at $T=10$ MeV for parameters consistent with
  the candidate signal at 3.5 keV. For reference we show the momentum
  space distribution for the active neutrinos, as the dotted line. The
  dark matter has a ``colder'' distribution with $\langle
  p_\mathrm{sterile}\rangle < \langle
  p_\mathrm{active}\rangle$. Parameters here are shown as stars in
  Fig.~\ref{fig:parspace}. This figure is from
  Ref.~\cite{Venumadhav:2015pla}. \label{fig:psds}}
\end{figure}

An example of how resonant sterile neutrino dark matter production
occurs is shown in Fig.~\ref{fig:loge2fe_temp}, where one can see that
there are two resonances, but one is dominant, proceeding from low to
high momenta. An example of final momentum space distributions are
shown in Fig.~\ref{fig:psds}. These are for the the region of
parameter space for the 7 keV resonantly-produced sterile neutrino
decaying dark matter consistent with detections of an unidentified
line at $\approx 3.5\,\mathrm{keV}$, shown as stars in
Fig.~\ref{fig:parspace}. When compared to the Fermi-Dirac distribution
of the active neutrinos, the sterile neutrino dark matter distribution
is ``cooler'' (though explicitly non-thermal), with $\langle
p_\mathrm{sterile}\rangle < \langle p_\mathrm{active}\rangle$. This
means that for a given particle mass, the free streaming scale is
smaller, and specifically compared to Dodelson-Widrow active
neutrinos, which have a momentum space distribution closer to that of
the active neutrinos, yet still nonthermal
\cite{abazajian:2005gj}. This means that for a fixed particle mass,
resonantly produced dark matter acts more like CDM than that particle
mass Dodelson-Widrow mechanism produced sterile neutrino. We discuss
implications for structure formation calculations in the following
section.

\subsection{Low Reheating Temperature Universe and the Visible Sterile Neutrino}

\begin{figure}[t!]
\begin{center}    \includegraphics[width=6in]{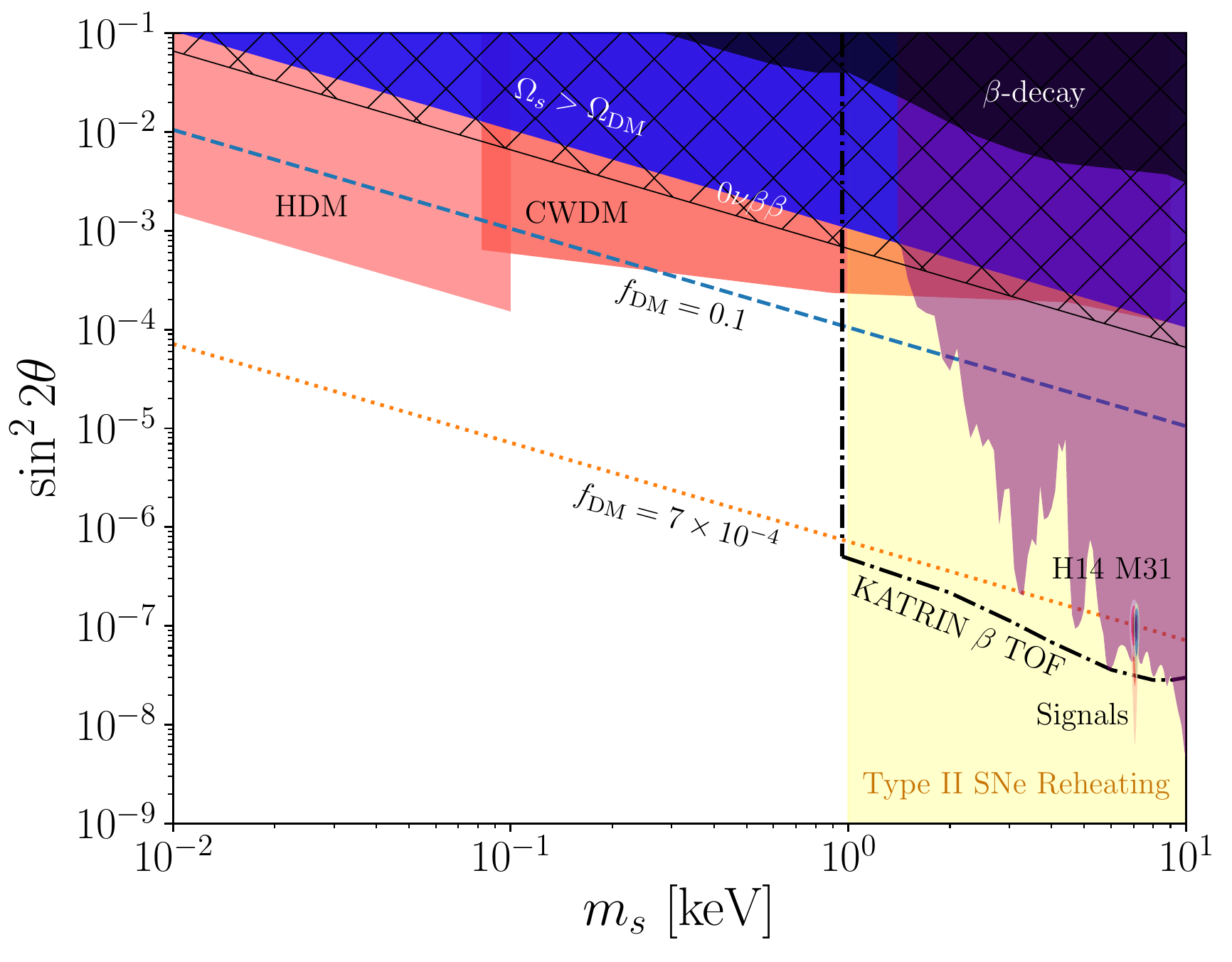}
\end{center}
\caption{Shown here is the parameter space for a possible low
  reheating temperature universe with $T_R = 5\,\mathrm{MeV}$, for the
  case of $\nu_s\leftrightarrow\nu_e$ mixing. The stacked galaxy
  cluster \cite{bulbul:2014sua} and M31 \cite{Boyarsky:2014jta}
  candidate signals are shown near 7 keV and $\sin^2 2\theta\approx
  10^{-7}$.  The contours for fraction of dark matter production
  $f_\mathrm{DM}$ at this reheating are shown, with the lower one
  $f_\mathrm{DM} = 7\times 10^{-4}$ required to match that needed for
  the candidate signals. The black solid regions are constrained by
  laboratory experiments, independent of any astrophysical or
  cosmological models: from neutrinoless double-beta decay searches in
  the hatched region labeled 0$\nu\beta\beta$
  \cite{KamLAND-Zen:2016pfg}, or by nuclear beta decay kink searches
  in the solid black region labeled $\beta$-decay
  \cite{Olive:2016xmw}. The black dot-dashed line is the forecast
  1$\sigma$ sensitivity of time-of-flight measures from the KATRIN
  $\beta$-decay experiment \cite{Steinbrink:2017ung}. We also show
  constraints from large scale structure limits on the presence of hot
  dark matter (HDM) in light pink \cite{Ade:2015xua}, from mixed cold
  and warm dark matter models \cite{anderhalden:2012jc}, as well as
  local group dwarf galaxy count constraints on WDM (labeled CWDM)
  in red \cite{horiuchi:2013noa}. The blue region
  $\Omega_s>\Omega_\mathrm{DM}$ overproduces the dark matter.  The
  constraint from M31 observations for the same fraction of dark
  matter as the signals is shown in purple, labeled H14 M31
  \cite{horiuchi:2013noa}. The yellow region is where sterile
  neutrinos deplete energy in the core of a Type II supernova
  \cite{abazajian:2001nj,Hidaka:2006sg,Arguelles:2016uwb}, though
  portions of this region may also be responsible for supernova shock
  enhancement \cite{Hidaka:2006sg} or the origination of pulsar kicks
  \cite{Kusenko:2006rh}. \label{lowreheat}}
\end{figure}

One possibility in the case of oscillation-based production is that
the Universe never heated up to the full peak-production temperatures
of the given production mechanism. In this case, large mixing angles
can produce the proper production levels because of the reduction of
the level of a strong thermal bath. This possibility was proposed in
Gelmini et al.~\cite{Gelmini:2004ah}.  Such low reheating scenarios
can under-produce the cosmological relativistic energy density measured
as the effective number of neutrinos, $N_\mathrm{eff}$, but for
reheating temperatures of $T_R = 5\ \mathrm{MeV}$, $N_\mathrm{eff}$
can be within 10\% of its canonical value, which remains consistent
with current measures of $N_\mathrm{eff}=3.15\pm 0.23\ (1\sigma)$
\cite{Ade:2015xua}. Future CMB experiments are forecast place stronger
limits on the relativistic energy density, as $N_\mathrm{eff}$, to
within $\sim$1\% \cite{Abazajian:2016yjj}.  Though low reheating
temperature universes are an unusual cosmology, they are possible,
with baryogenesis occurring at low scale via an Affleck-Dine mechanism
\cite{affleck:1984fy}. Perhaps most interestingly is that this
plausible scenario marries quite disparate fields of cosmological dark
matter production, cosmological large scale structure constraints on
hot dark matter (HDM), galaxy formation limits on WDM and mixed WDM
plus CDM, Type II supernova reheating models, as well as
laboratory-based nuclear beta-decay searches and neutrinoless double
beta decay searches.

For the case of a $T_R = 5\,\mathrm{MeV}$ reheating temperature, the
production of sterile neutrino dark matter proceeds via partial
thermalization due to low temperatures and low mixing angles combined,
allowing larger mixing angles that the standard thermal history to
produce the requisite dark matter density. Here, the $\nu_s$
distribution function turns out to be
\begin{equation}
f_s(E,T) \approx 3.2
d_\alpha\left(\frac{T_R}{5\,\mathrm{MeV}}\right)^3\sin^2 2\theta
\left(\frac{E}{T}\right) f_\alpha(E,T),
\end{equation}
where $d_\alpha = 1.13$ for $\nu_\alpha = \nu_e$, and $d_\alpha=0.79$
for $\nu_\alpha = \nu_{\mu,\tau}$. The fraction of the sterile
neutrino distribution produced is
\begin{equation}
  f\equiv\frac{n_{\nu_s}}{n_{\nu_\alpha} \approx 10 d_\alpha\sin^2 2\theta\left(\frac{T_R}{5\,\mathrm{MeV}}\right)^3}.
\label{fractionsterile}
\end{equation}
In such a model, one must match the requisite decay rate with the
fraction of dark matter that is sterile neutrino dark matter via the
production calculation at $T_R=5\,\mathrm{MeV}$. In such a case, the
signals at $3.55\,\mathrm{keV}$ \cite{bulbul:2014sua,Boyarsky:2014jta}
are at a relatively much higher mixing angle, $\sin^2 2\theta \approx
10^{-7}$, as shown in Fig.~\ref{lowreheat}, where they are within
reach of laboratory experiments. To match the decay rate, the
production makes sterile neutrino dark matter that is a small fraction
of the total dark matter $f_\mathrm{DM} \equiv
\Omega_s/\Omega_\mathrm{DM} \approx 7\times 10^{-4}$ for the case of
the signal from stacked clusters \cite{bulbul:2014sua}.

The parameter space of low-reheating temperature models of
$T_R=5\,\mathrm{MeV}$, for the case of $\nu_s\leftrightarrow\nu_e$
mixing, is shown in Fig.~\ref{lowreheat}. We also show regions that are
constrained by laboratory experiments, independent of any
astrophysical or cosmological models. Kink searches in nuclear
beta-decay are direct kinematic searches
\cite{Olive:2016xmw}. Neutrinoless double-beta decay searches
(0$\nu\beta\beta$) are sensitive to $\langle m\rangle_s = m_s
\sin^2\theta e^{i\beta_s}$, where $\beta_S$ is the Majorana
$CP$-violating phase. The best 0$\nu\beta\beta$ limits are currently
$|\langle m\rangle| < 61 - 165\,\mathrm{meV}$, which correspond to
$m_s\sin^2 2\theta < 0.66\,\mathrm{eV}$
\cite{KamLAND-Zen:2016pfg}. Kinematic reconstruction by time-of-flight
in beta decay in the KATRIN $\beta$-decay experiment has sensitivity
to this region \cite{Steinbrink:2017ung}. Even partial thermalization
of the sterile neutrino in cases where its free streaming scale is
large gives constraints from the CMB and large scale structure,
$\sigma m_i + f m_s<0.23\,\mathrm{eV}$ \cite{Ade:2015xua}, where $f$
is the fraction of the active neutrino density in sterile neutrinos
(Eq.~\eqref{fractionsterile}). From affecting galaxy counts, there are
constraints on specific ratios of mixed cold and warm dark matter
models. Namely, mixed CWDM models constrain 5\% WDM with equivalent
$m_s = 0.08\,\mathrm{keV}$, 20\% fraction WDM with equivalent $m_s =
0.90\,\mathrm{keV}$, 80\% fraction WDM with equivalent $m_s =
4.5\,\mathrm{keV}$ \cite{anderhalden:2012jc}, plus local group dwarf
galaxy count constraints on WDM at 100\% fraction limits at $m_s
\approx 9\,\mathrm{keV}$
\cite{horiuchi:2013noa,Cherry:2017dwu}. Constraints from M31
observations apply, but move commensurately up for the same fraction
of dark matter as the signals shown \cite{horiuchi:2013noa}. Sterile
neutrinos deplete energy in the core of a Type II supernova
\cite{Kainulainen:1990bn,abazajian:2001nj,Hidaka:2006sg,Arguelles:2016uwb},
though portions of this parameter space can also be responsible for
supernova shock enhancement \cite{Hidaka:2006sg} or the origination of
pulsar kicks in portions of that region \cite{Kusenko:2006rh}.

The sterile neutrino dark matter's large mixing angle in low-reheating
scenarios is much more accessible to laboratory experiments like
$\beta$-decay
\cite{Mertens:2014nha,Mertens:2014osa,Steinbrink:2017ung}.  There are
also efforts at testing these couplings via K-capture experiments
\cite{Finocchiaro:1992hy,Hindi:1998ym} including novel methods
employing atom traps that search for the massive admixed neutrino in
nuclear beta decay via complete kinematic reconstruction of the final
state \cite{Finocchiaro:1992hy,Smith:2016vku}.

  \subsection{Non-oscillation Production: Particle Decays}
In addition to the non-resonant and resonant oscillation production
models, a few other production mechanisms have been proposed via
particle decays. For most cases, a generic scalar $S$-particle is
introduced with an interaction Lagrangian
\begin{equation}
\mathcal{L}_\mathrm{int} =
\frac{y}{2}\overline{\left(\nu_R\right)^c}\nu_R S+ h.c.
\end{equation}
$S$-particles are created in some process in the early Universe and
could decay into sterile neutrinos at some later point. Clearly, the
abundance is now independent of the active-sterile mixing angle, but
in general, if this mechanism is responsible for the bulk of
production, the mixing angle must be below that produced by
non-resonant thermal production. In addition, depending on the nature
of the mechanism, the scalar-decay sterile neutrino dark matter could
be ``warmer'' or ``colder'' than in oscillation production. A thorough
review of these production mechanisms is given in
Ref.~\cite{Adhikari:2016bei}, and we provide a brief overview here. 

Regardless of the mechanism, if the active-sterile mixing is large
enough for Dodelson-Widrow or Shi-Fuller production to be significant,
that has to be included. In the various particle-decay mechanisms, the
``parent'' $S$-particle can itself be in or out of thermal equilibrium
at the time the sterile neutrinos are produced. The decay into the
sterile neutrino dark matter has been considered via several scalar
particles as a single particle decay channel or via one part of
multiple possible decay channels
\cite{shaposhnikov:2006xi,Kusenko:2006rh,Petraki:2007gq,Bezrukov:2009yw,Kusenko:2012ch,Merle:2015oja}.
In scalar decay cases, the production of sterile neutrinos is governed
by a Boltzmann kinetic equation describing their distribution
$f(p,t)$ as a function of momentum $p$ and time $t$ \cite{shaposhnikov:2006xi}
\begin{equation}
  \frac{\partial f}{\partial t} - H p \frac{\partial f}{\partial p} =
  \frac{2m_S \Gamma}{p^2}\int_{p+m^2_s/4p}^\infty{f(E)\ dE}
\end{equation}
for a scalar particle of mass $m_S$. $\Gamma$ is the partial width for
scalar decay into sterile neutrinos. If the decays occur when the
number of degrees of freedom in the background plasma is constant,
then this kinetic equation is analytically integrable to find the late
time number density of sterile neutrinos,
\begin{equation}
n_0 \approx \frac{3\Gamma M_\mathrm{Pl} \zeta(5)}{3.32\pi m_S^2
  \sqrt{g_\ast}}\ T^3, 
\end{equation}
where $g_\ast$ is the number of statistical degrees of freedom during
the production epoch, $M_\mathrm{Pl}$ is the Planck mass, and
$\zeta(5)$ is the zeta function. The average momentum of the sterile
neutrinos is $\langle p\rangle = \pi^6/(378\zeta(5))T \approx 2.45 T$,
which is lower than thermal $\langle p\rangle \approx 3.15 T$, but not
overwhelmingly colder.

There
have been explorations where the parent particle is not a scalar, but
rather a vector \cite{Shuve:2014doa} or fermion
\cite{Abada:2014zra}. In these decay-production cases, or even the
oscillation-production cases, further decays of particles, including
related more massive sterile neutrinos, can further ``cool'' the
sterile neutrino dark matter relative to the thermal bath
\cite{Asaka:2006ek,Petraki:2008ef,Boyanovsky:2008nc,Patwardhan:2015kga}. They
often cannot be not arbitrarily cooled, and could still be constrained
by measures of its effects on structure formation
\cite{abazajian:2006yn,Menci:2017nsr}.

\section{Sterile Neutrinos and Cosmological Structure Formation}

The effects of massive sterile neutrinos that are completely or
partially thermalized are like that of the active neutrinos, except
that the particle mass scales and therefore structure scales can vary
more considerably. The primary physical effect for all mass scales is
the damping of structure on scales below the free streaming scale of
the particle in the early Universe before structure starts to grow
appreciably, at matter-radiation equality,
\begin{equation}
\lambda_\mathrm{FS} = \int_0^\mathrm{EQ} \frac{v(t)dt}{a(t)} \approx
1.2\,\mathrm{Mpc}\left(\frac{1\,\mathrm{keV}}{m_\nu}\right)\left(\frac{\langle
p/T\rangle}{3.15}\right)\mbox{,}\label{lambdaFS}
\end{equation}
where $\langle p/T\rangle$ is the average momentum over temperature of
the particle species. For a thermalized Fermi gas with small chemical
potential, $\langle p/T\rangle = 7\pi^4/180\zeta(3) \approx 3.15$. In
order to produce structure below galaxy cluster scales of
$\sim$10~Mpc, the sterile neutrino can either not be the entire dark
matter for the eV-scale since it would be hot dark matter, or it must
be at the keV-scale such that it is not fully thermalized in its
production process. In order to avoid the
Cowsik-McLelland/Gershtein-Zeldovich bound, the particle mass must be
$M < 94h^2\,\mathrm{eV}$ so as to not over-close the Universe
\cite{KolbTurner1990}.

Because sterile neutrinos in many cases are nonthermal, including the
case of dark matter, then the free-streaming scale change due to
``cooler'' $\langle p/T\rangle$ must be taken into account in
Eq.~\eqref{lambdaFS}. This is accurately performed by calculating the
free streaming modification of the transfer function of the linear
total matter power spectrum by employing numerical Boltzmann solvers
such as \verb|CAMB| \cite{lewis02}. \verb|CAMB| now has the ability to
include extra massive sterile neutrinos, and modified versions are
employed to solve for the case of sterile neutrino warm dark matter
with nonthermal momentum space distributions
\cite{abazajian:2005gj,abazajian:2014gza,Venumadhav:2015pla}.

\subsection{eV-Scale Sterile Neutrinos}
\label{subsec:eVstructure}
For the case of eV-scale sterile neutrinos that are partially or
completely thermalized, the constraints on their modification of the
linear matter power spectrum as measured in large scale structure
(LSS) are similar to the constraints on massive active neutrinos,
except for the fact that a fourth neutrino species must be included in
the calculation. That is, thermalizing a sterile neutrino clearly
alters the total radiation energy density at early times, when they
are relativistic, but then the massive neutrino modifies the matter
density. Therefore, a partially or fully thermalized sterile neutrino
is subject to constraints from big bang nucleosynthesis (BBN)
\cite{Steigman:1977kc} as well as LSS. Note that massive sterile
neutrinos are sometimes thought to contribute to the relativistic
energy density as quantified by the effective number of neutrinos,
which is specifically defined as the relativistic energy density
relative to a single {\it massless} active neutrino density,
\begin{equation}
N_\mathrm{eff} \equiv \rho_r/n_\nu.
\label{neff}
\end{equation}
Therefore, when the sterile neutrino is nonrelativistic, as in the era
of structure formation with photon temperatures $T_\gamma \lesssim
1\,\mathrm{eV}$, $N_\mathrm{eff}$ is not an accurate descriptor.
$N_\mathrm{eff}$ is an accurate characterization of eV-scale neutrinos
at the BBN epoch at $T_\gamma \approx 1\,\mathrm{MeV}$, where
relativistic energy density modifies weak freeze-out, the
neutron-to-proton ratio entering BBN, and therefore the light element
abundances, which constrain the number of thermalized sterile
neutrinos in the early Universe \cite{Cyburt:2015mya}. Note that
keV-scale dark matter sterile neutrinos do not affect BBN since they
are, by definition, produced at the abundance of dark matter, which is
very small in relative density to radiation in the radiation-dominated
BBN epoch.

The effects of massive sterile neutrinos on LSS observables are
described in detail in the review by Abazajian \& Kaplinghat
\cite{Abazajian:2016hbv}. The effect of neutrinos on large scale
structure is reflected on their effects on the primordial power
spectrum of perturbations in the early Universe. The evolution of the
inhomogeneities are linear in early times. The solution of the evolution of
inhomogeneities from inflation through the radiation dominated, matter
dominated and vacuum dominated eras are delineated clearly in recent
texts, e.g. Ref.~\cite{2003moco.book.....D}. In short, the
gravitational potential, $\Phi$, drives the linear growth of
perturbations. The primordial potential $\Phi_\mathrm{P}$ is modified
by the physics of the perturbations in the early epoch, described by the
scale dependent transfer function. Growth of inhomogeneities further
modifies the linear potential to give that which may be observed at late
times, $\Phi(k,a)$, where $k$ is the wavenumber of the potential and
$a$ is the scale factor of the Universe:
\begin{equation}
  \Phi(k,a) = \Phi_\mathrm{P}\times \left\{\text{transfer
    function}(k)\right\} \times \left\{\text{growth
    function}(k,a)\right\}\ .
\end{equation}
Neutrinos affect these processes by not participating in gravitational
collapse until they become nonrelativistic. Instead, neutrinos
free-stream out of the gravitational potential wells that are
populated by dark matter and baryons. Massive eV-scale sterile
neutrinos produce the same effects as active neutrinos since they are
both decoupled from baryons (via charged leptons) very early ($T\sim
1\ \mathrm{MeV}$).  All forms of massive neutrinos do participate as
matter when they become nonrelativistic. This occurs at a scale when
$k_B T\nu(a) \sim m_{\nu} c^2$, which corresponds to a redshift of
$z_\mathrm{nr} \sim
300\ \left(m_{\nu}/50\ \mathrm{meV}\right)$. Primordial fluctuations
in neutrinos are streamed away at scales below the horizon at
$z_\mathrm{nr}$. In comoving space, this corresponds to a wavenumber
\begin{equation}
  k_\mathrm{nr} \equiv a_\mathrm{nr} H\left(a_\mathrm{nr}\right)
  \approx 0.04\, a^2\sqrt{\Omega_m a^{-3} +
    \Omega_\Lambda}\left(\frac{m_\nu}{50\rm\, meV}\right) h/\mathrm{Mpc}
\end{equation}
where $\Omega_m \equiv \rho_m/\rho_\mathrm{crit}$ is the fraction of
critical density in all matter. Larger than this scale,
$k_\mathrm{NR}$, adiabatic perturbations in neutrinos, dark matter and
baryons are coherent and are sufficiently described by a single
perturbation $\delta_m = \delta\rho_m/\rho_m$.
However, even when the neutrinos are nonrelativistic, they have finite
velocity dispersion that prevents clustering and that free-streaming
scale is what is described in Eq.~\eqref{lambdaFS}.

The detailed effects of massive eV-scale sterile neutrino free
streaming below the scale defined by Eq.~\eqref{lambdaFS} is
quantified accurately by Boltzmann solvers of perturbation
evolution. The effects suppress structure below
$\lambda_\mathrm{FS}$ in the linear power spectrum $P(k)$ by an amount
at small scales by
\begin{equation}
\frac{\Delta P_\delta}{P_\delta} \simeq -8
\frac{\Omega_\nu}{\Omega_{\rm M}}=-8\frac{\Sigma m_\nu}{94
  \ \Omega_{\rm M}h^2\ {\rm eV}}\,,
\end{equation}
where $\Omega_{\nu}=\rho_{\nu,0}/\rho_\mathrm{crit,0}$ is the present
critical density in neutrinos and $\Sigma m_\nu$ is the total neutrino
mass in active and sterile neutrinos. The amplitude of the matter
power spectrum at small scales is typically parameterized by
$\sigma_8$, the rms amplitude of fluctuations integrated over a scale
of $8\ h^{-1}\rm\, Mpc$. There are a number of LSS data sets that
indicate suppressed power at small scales, the so-called ``$\sigma_8$
problem.''  These include weak lensing, cluster abundance, and
Lyman-$\alpha$ forest measures of the small scale amplitude
\cite{Battye:2013xqa,Wyman:2013lza,Dvorkin:2014lea,Beutler:2014yhv}. These
could indicate a non-trivial non-zero neutrino mass, possibly in the
form of a sub-eV-scale active or sterile neutrino. However, for this
to be in the form of a sterile neutrino, the extra relativistic energy
density it would impose at early times is constrained by BBN
\cite{Giusarma:2014zza}, though partial thermalization could
accommodate both \cite{Jacques:2013xr}. The tension between $H_0$
inferred from the CMB and the $H_0$ measured directly in the local
Universe could also indicate extra radiation density. However,
evidence suggests that other new physics may be responsible such as
features in the primordial power spectrum \cite{Canac:2016smv} or
non-standard dark energy \cite{DiValentino:2016hlg}.

\subsection{keV-Scale Neutrinos: Warm to Cold Dark Matter}
\label{subsec:keVstructure}
For many simple models of warm dark matter particle production, there
is a unique relationship between the particle mass and its effect on
the free streaming scale, $\lambda_\mathrm{FS}$,
Eq.~\eqref{lambdaFS}. The topic of what particle-mass constraints
arise from structure formation is often fraught with
misunderstanding. Often constraints in the literature are provided on
an ill-defined ``warm dark matter'' particle mass, but sometimes more
carefully on ``thermal warm dark matter'' particle mass. The case of
truly thermal warm dark matter is well defined, as well as the
standard Dodelson-Widrow production mechanism. These two cases were
originally explored in Colombi, Dodelson \&
Widrow~\cite{colombi:1995ze}, which we follow here. Other cases
typically have no clear mapping between particle mass and measurable
structure formation effects $\lambda_\mathrm{FS}$, but are certainly
quantifiable, which we also discuss below.

For the case of a purely thermal warm dark matter ``$x$'' particle, the models
can be interpolated between hot dark matter (HDM) and CDM with
the single function  
\begin{equation}
f_x(v) = \frac{\beta}{e^{p/\alpha T_\gamma}+1},
\label{eq:warmPDF}
\end{equation}
where $T_\gamma$ is the photon temperature, $v=p/(p^2+m^2_x)^{1/2}$;
$m_x$ is the particle's mass. Clearly, there are three parameters that
define this distribution function, $\alpha$, $\beta$ and $m_x$. For the
case of standard fermionic hot dark matter, $\alpha = (4/11)^{1/3}$,
$\beta = 1$. The particle mass is then set to provide the requisite
fraction of critical density $\Omega_\mathrm{HDM}$. Therefore, there
is no freedom in the shape of the distribution function and, for pure
HDM, a fixed distribution function for a given density. For the case
of {\it pure} CDM, the velocity dispersion is negligible and the shape
or form of the distribution function is not important. For the case of
the parameterization in Eq.~\eqref{eq:warmPDF}, CDM corresponds to the
limiting case of $\alpha = \rm constant$, $\beta \rightarrow 0$ and
$m_x \rightarrow \infty$ (equivalently, $\alpha\rightarrow 0$ and
$m_x\rightarrow \infty$). 

For what generic thermally-produced WDM like both sterile neutrinos
and gravitinos, $\alpha$ and $\beta$ can be different from the HDM
values. Fixing the density of particles to provide all or a fraction
of the cosmological dark matter gives a constraint
\begin{equation}
\Omega_x h^2 = \beta
\left(\frac{\alpha^3}{4/11}\right)\left(\frac{m_x}{94\,\mathrm{eV}}\right).
\end{equation}
We are left with a two free parameters, which are typically chosen to
be $m_x/\alpha$ (proportional to $m_x/T_x$) and $\alpha$. The
parameter $m_x/\alpha$ (or $m_x/T_x$) defines the shape of the dark
matter power spectrum and therefore the free streaming scale. A value
of $\alpha$ corresponds to early-decoupled dark matter---like
gravitino dark matter---and another to the simplest production models
of Dodelson-Widrow sterile neutrino dark matter. In the class of
models where Eq.~\eqref{eq:warmPDF} describes the distribution
function for sterile neutrino dark matter and early-decoupled thermal
WDM, the particle masses are related as \cite{viel:2005qj}
\begin{equation}
m_\text{DW,ideal} \approx
4.46\,\mathrm{keV}\,\left(\frac{m_\text{thermal}}{1\,\mathrm{keV}}\right)\,\left(\frac{0.12}{\Omega_x
h^2}\right).\label{idealDW}
\end{equation}
However, as discussed in Colombi, Dodelson \&
Widrow~\cite{colombi:1995ze} and detailed in Abazajian
\cite{abazajian:2005gj}, this relation is altered if the number of
degrees of freedom in the plasma is changing during the production
epoch, which is often a dramatic because peak production typically
occurs near the quark-hadron transition in the early Universe
\cite{abazajian:2002yz}, where the time-temperature evolution is
significantly altered, and the number of weak-scatterers drastically
changes. These effects combine to ``cool'' Dodelson-Widrow produced
sterile neutrino dark matter \cite{abazajian:2005gj}, which alters the
relation for Eq.~\eqref{idealDW} to be different by 20\% for typical
parameters, and alters the constraints for Dodelson-Widrow dark matter
when these effects are included \cite{abazajian:2005xn}. The full
momentum space distributions even for the Dodelson-Widrow case must be
included into cosmological Boltzmann solvers in order to arrive at
accurate constraints, though Eq.~\eqref{idealDW} is approximately
correct. Given that pure Dodelson-Widrow sterile neutrino dark matter
is excluded by Local Group constraints \cite{horiuchi:2013noa}, this
is not a significant issue any longer, but pertains to accurate
characterizations of WDM particle mass limits in general. For the case
of resonantly-produced Shi-Fuller sterile neutrino dark matter the
quasi-thermal momentum space distribution of Eq.~\eqref{eq:warmPDF} is
invalid, with the momentum space distribution highly nonthermal and
``cooler'' than thermal (Fig.~\ref{fig:psds}). Therefore, the full
momentum space distribution must be employed in its effects on
structure formation \cite{Venumadhav:2015pla} and that then used for
structure formation constraints
\cite{Schneider:2016uqi,Cherry:2017dwu}.

One of the most potentially powerful constraints on the matter power
spectrum at small scales affected by WDM is the clustering of gas as
measured along the line of sight to distant quasars in the
Lyman-$\alpha$ forest
\cite{viel:2005qj,abazajian:2005xn,Schneider:2016uqi,croft:1999mm,mcdonald:2004xn,seljak:2006qw,viel:2006kd,viel:2007mv,viel:2013fqw,Baur:2015jsy,Irsic:2017ixq}. The
claimed level of the most recent constraints by Ir\v{s}i\v{c} et
al.~\cite{Irsic:2017ixq} place an approximate limit on the ideal
Dodelson-Widrow mass (Eq.~\eqref{idealDW}) when mapped from the
thermal WDM limits of $m_\text{thermal} \ge 5.3\,\mathrm{keV}$ to be
$m_\text{DW,ideal} \ge 41\,\mathrm{keV}$ (95\% CL). When combined with
X-ray limits, they strongly exclude Dodelson-Widrow sterile neutrino
dark matter from being all of the dark matter. These are considerably
weakened when sterile neutrinos are only partially the dark matter
\cite{Boyarsky:2008xj}. The Lyman-$\alpha$ forest is a potentially
very powerful tool to measure the small-scale matter power
spectrum. It relies on mapping the clustering of neutral gas in one
dimension to the 3-dimensional full matter power spectrum. For some
time, this has been known to have potential systematic problems in
entangling the thermal history of the intergalactic medium---via the
thermal broadening and Jeans pressure-support of the gas---with the
underlying matter power spectrum \cite{gnedin:2001wg}. This has become
more apparent with very high resolution hydrodynamic simulations
like those in Kulkarni et al.~\cite{Kulkarni:2015fga}, where pressure
support in the gas was shown to greatly affect the flux power spectrum
at high redshift, and the recovered nonlinear flux power spectrum at
late time varied greatly from the linear theory methods typically used
in cosmological analyses like that done to probe WDM. Kulkarni et
al. also show the temperature density relation has a dispersion that
is highly non-Gaussian and that temperature-density relation should be
augmented with a third pressure smoothing scale parameter
$\lambda_F$. The Lyman-$\alpha$ forest is argued further to be best
used as a probe of the epoch of cosmological reionization
\cite{Onorbe:2017ftn}.

A potentially strong probe of WDM vs.\ CDM is the formation of
structure at high redshift probed by galaxy number counts
\cite{schultz:2014eia} as well as reionization
\cite{Barkana:2001gra}. The limits from reionization from the optical
depth to the cosmic microwave background as measured by Planck are at
the level of $m_\text{thermal} \gtrsim 1.3\,\mathrm{keV}$
\cite{schultz:2014eia}, while recent limits from the luminosity
function of high redshift galaxies are at the level of
$m_\text{thermal} \ge 2.5\,\mathrm{keV}$ at 2$\sigma$
\cite{Menci:2017nsr}. The sensitivity of the {\it James Webb Space
  Telescope} (JWST) to galaxy counts at even higher redshift, which
are even more sensitive to WDM suppression of structure formation,
will push to sensitivities to even higher thermal WDM particle masses
\cite{schultz:2014eia}. JWST is planned to launch in October
2018\footnote{\texttt{https://jwst.nasa.gov}}. Detailed observations
of reionization, the Lyman-$\alpha$ forest and high-redshift galaxy
counts could differentiate between the variations of the shape of the
suppression scales of different WDM production scenarios
\cite{Bozek:2015bdo}.

In studies of the formation of the small scale structure as probed in
the Local Group of galaxies and the cores and central densities of
galaxies, there remain too-low of a central density profile compared
to that expected in CDM of dwarf galaxies that are satellites as well
as in the field---i.e., not gravitationally bound to another galaxy
\cite{BullockARAA}. This has been dubbed the too-big-to-fail
problem, and it can be alleviated by WDM of the proper free streaming
scale, at approximately the free streaming scale provided by
$m_\text{thermal} \approx 2\,\mathrm{keV}$
\cite{lovell:2011rd,anderhalden:2012jc}. Very significantly, it was
pointed out that this free streaming scale was matched by sterile
neutrino dark matter in the region of parameter space consistent with
the 3.5 keV candidate dark matter decay signal in the X-ray, discussed
below \cite{abazajian:2014gza,Horiuchi:2015qri}. Resonantly-produced Shi-Fuller 
sterile neutrino dark matter in the 7 keV, $\sin^2 2\theta \sim
10^{-10}$ region produce a range of cutoff scale consistent with $1.5
\,\mathrm{keV} \lesssim m_\text{thermal} \lesssim 3.0\,\mathrm{keV}$
\cite{abazajian:2014gza,Venumadhav:2015pla}.

The Milky Way's Local Group satellite galaxy counts can also provide
limits on the free streaming scale since the free streaming suppresses
dwarf galaxy formation
\cite{bode:2000gq,polisensky:2010rw,Horiuchi:2015qri,Schneider:2016uqi}. As
dwarf galaxies are discovered by deep all-sky observations, the limits
have increased to place tension with WDM suppression scales in the
region consistent with resonantly-produced 7.1 keV sterile neutrino
dark matter in the region of the 3.55 keV signal
\cite{Cherry:2017dwu}. This tension may make more attractive the
scenario where 10\% to 20\% of the dark matter is Dodelson-Widrow
sterile neutrino dark matter, with the rest being some other form.  A
10\% to 20\% fraction of Dodelson-Widrow sterile neutrinos can produce
the 3.55 keV signal, with a mixing angle that is commensurately
approximately five to then times larger, in order to continue to match
the observed flux in the signal with a smaller sterile neutrino dark
matter mass in the field of view \cite{bulbul:2014sua}. This case,
where there is a mixed cold plus warm dark matter, escapes constraints
from galaxy counts and the Lyman-$\alpha$ forest
\cite{Boyarsky:2008xj}, and may still alleviate small-scale structure
challenges \cite{anderhalden:2012jc}.

\section{keV Sterile Neutrino Dark Matter Detectability in X-ray Observations}
\label{sec:X-ray}

\subsection{Methods \& Current Results}

The fact that a light, neutral lepton, like a sterile neutrino, would
have a radiative decay mode was first pointed out and calculated by
Shrock~\cite{Shrock:1974nd} and independently by Pal \&
Wolfenstein~\cite{Pal:1981rm}. For the Majorana
neutrino case, the decay rate is 
\begin{equation}
\Gamma_\gamma(m_s,\sin^2 2\theta) \approx 1.36 \times 10^{-30}\,{\rm
s^{-1}}\ \left(\frac{\sin^2 2\theta}{10^{-7}}\right)
\left(\frac{m_s}{1\,\rm keV}\right)^5,
\end{equation}
where $m_s$ is the mass eigenstate most closely associated with the
sterile neutrino, and $\theta$ is the mixing angle between the sterile
and active neutrino.  For the case of a Dirac sterile neutrino, the
decay rate is reduced by a factor of two. The decay of a
nonrelativistic sterile neutrino into two (nearly) massless particles
produces a line at energy $E_\gamma = m_s/2$.

The radiative decay time is orders of magnitude greater than the age
of the universe (which is necessary for a viable dark matter
candidate) but the number of particles in the field of view of the
{\it Chandra} or {\it XMM-Newton} observatories is approximately $\sim
10^{78}$, which makes this decay for the candidate signal we will
discuss below to be at the level of $10^{48}\rm\,s^{-1}$ for a
$10^{15}\,M_\odot$ cluster of galaxies halo.

The radiative decay of sterile neutrino dark matter was found to be
constrained by the diffuse X-ray background by Drees \&
Wright~\cite{Drees:2000qr}, and this was reiterated in
Ref.~\cite{dolgov:2000ew}. The first proposals of the estimated
sensitivity of looking for the radiative decay toward massive dark
matter halos like galaxy clusters and galaxies were presented in
Abazajian, Fuller \& Patel~\cite{abazajian:2001nj}. A detailed
signal-to-background analysis of contemporary X-ray telescopes as
well as forecasts for future telescopes was performed in Abazajian,
Fuller \& Tucker~\cite{abazajian:2001vt} (AFT). That work found that
the sensitivity provided by microcalorimeter spectroscopy and much
greater effective area of the at-the-time proposed {\em
  Constellation-X} telescope made it particularly sensitive to the
radiative decay signal. In AFT, it was also proposed as a hedge that ``An
exposure equivalent to [that of {\em Constellation-X}] could be
obtained by a stacking analysis of the spectra of a number of similar
clusters.''

\begin{figure}[t!]
\begin{center}    \includegraphics[width=6in]{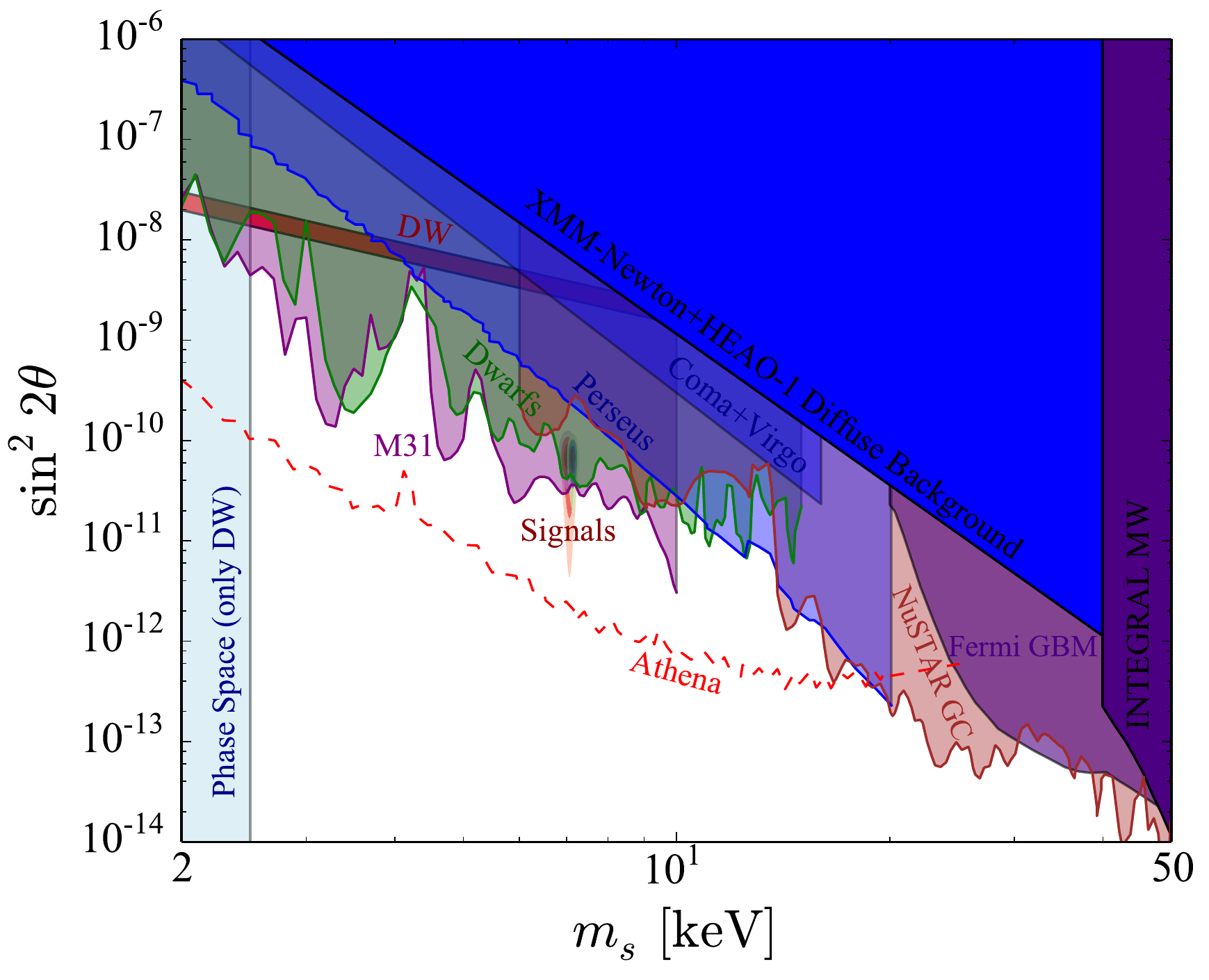}
\end{center}
\caption{The full parameter space for sterile neutrino dark matter,
  when it comprises all of the dark matter, is shown. Among the most
  stringent constraints at low energies and masses are constraints
  from X-ray observations M31 Horiuchi et al.~\cite{horiuchi:2013noa},
  as well as stacked dwarfs \cite{malyshev:2014xqa}. Also shown are
  constraints from the diffuse X-ray background
  \cite{Boyarsky:2005us}, and individual clusters ``Coma+Virgo''
  \cite{Boyarsky:2006zi}. At higher masses and energies, we show the
  limits from Fermi GBM \cite{Ng:2015gfa} and INTEGRAL
  \cite{Boyarsky:2007ge}. The signals near 3.55 keV from M31 and
  stacked clusters are also shown
  \cite{bulbul:2014sua,Boyarsky:2014jta}. The vertical mass constraint
  only directly applies to the Dodelson-Widrow model being all of the
  dark matter, labeled ``DW,'' which is now excluded as all of the
  dark matter. The Dodelson-Widrow model could still produce sterile
  neutrinos as a fraction of the dark matter. We also show forecast
  sensitivity of the planned {\it Athena X-ray Telescope}
  \cite{Neronov:2015kca}. \label{fig:fullparams}}
\end{figure}

Since the initial proposal of the search of the radiative decay line
in clusters of galaxies, field galaxies, and the cosmic X-ray
background by AFT, many groups have conducted follow-up searches for
the signature decay in the cosmic X-ray background
\cite{Boyarsky:2005us}, clusters of galaxies \cite{Watson:2006qb},
individual dwarf galaxies
\cite{Boyarsky:2006fg,loewenstein:2008yi,Loewenstein:2009cm,Loewenstein:2012px},
the Andromeda galaxy \cite{Watson:2006qb,horiuchi:2013noa} and the
Milky Way \cite{Riemer-Sorensen:2006fh}.  Among the best current
constraints is that from an analysis of {\it Chandra} X-ray
observations of the Andromeda galaxy by Horiuchi et
al. \cite{horiuchi:2013noa}.  Constraints from stacked dwarf galaxy
observations are comparable in strength~\cite{malyshev:2014xqa}. There
are also constraints from Perseus observations from {\it Suzaku} which
extend to higher energies and masses \cite{tamura:2014mta}. At the
highest energies, constraints exist from the {\it Fermi Gamma-Ray
  Space Telescope} Gamma-Ray Burst Monitor (GBM) \cite{Ng:2015gfa} and
INTEGRAL observations of the Milky Way halo
\cite{Boyarsky:2007ge}. Several of these constraints are shown in
Fig.~\ref{fig:fullparams}. For a considerable amount of time since
these methods were proposed, no significant detections of unidentified
candidate dark matter lines had been found, with only upper limits to
the decay flux.

\begin{figure}[t!]
\begin{center}
\includegraphics[width=6in]{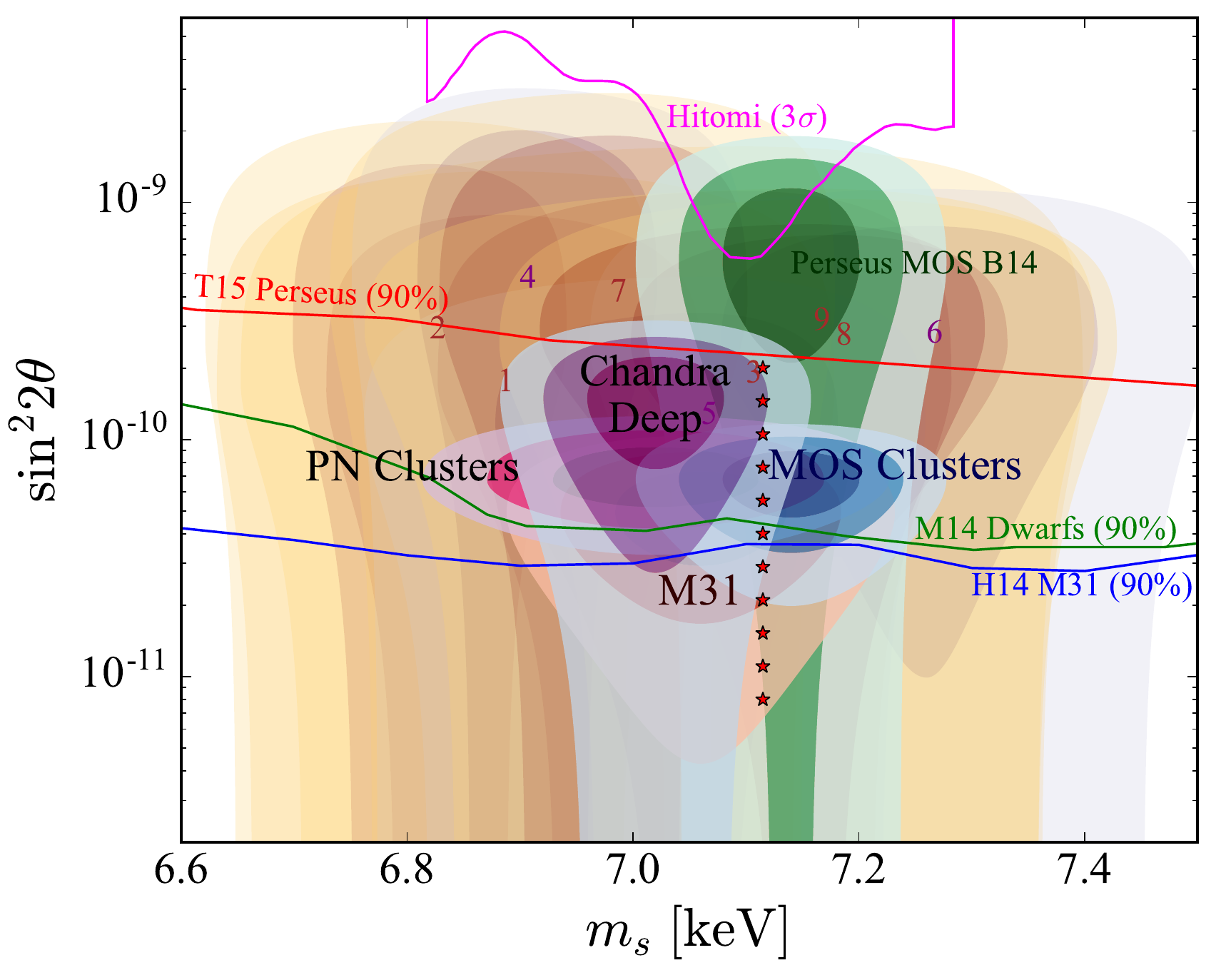}
\end{center}
\caption{X-ray line detections consistent with sterile neutrino dark
  matter are shown here. The dark colored regions are $1,2$ and $3$
  $\sigma$ from the MOS (blue) and PN (red) stacked clusters by Bulbul
  et al.~\cite{bulbul:2014sua}, the Bulbul et al. core-removed Perseus
  cluster (green), and M31 (orange) from Boyarsky et
  al.~\cite{Boyarsky:2014jta}. Also shown are the 1 and 2 $\sigma$
  regions of the detection in the Galactic Center (GC)
  \cite{Boyarsky:2014ska} as well as the $>\! 2\sigma$ line detections
  in 1.\ Abell 85; 2.\ Abell 2199; 3.\ Abell 496 (MOS); 4.\ Abell 496
  (PN); 5.\ Abell 3266; 6.\ Abell S805; 7.\ Coma; 8.\ Abell 2319;
  9.\ Perseus by Iakubovskyi et
  al.~\cite{Iakubovskyi:2015dna}. Numbers in the plot mark the
  centroid of the regions, with MOS detections in orange and PN in
  purple.  We also show, in purple, the region consistent with the
  signal in {\em Chandra} Deep Field observations, with errors given
  by the flux uncertainty, {\it i.e.}, not including dark matter
  profile uncertainties \cite{Cappelluti:2017ywp}. The lines show
  constraints at the $90\%$ level from {\it Chandra} observations of
  M31 (14) \cite{horiuchi:2013noa}, stacked dwarf galaxies (M14)
  \cite{malyshev:2014xqa}, and Suzaku observations of Perseus (T15)
  \cite{tamura:2014mta}. Stars mark the models shown in
  Fig.~\ref{fig:psds}. \label{fig:parspace}}
\end{figure}

In early 2014, Bulbul et al.~\cite{bulbul:2014sua} used stacked
cluster observations totaling over 6 Ms in exposure time and detected
an unidentified line near 3.55 keV in energy at high significance, 4
to 5$\sigma$, using both the PN and MOS CCDs aboard {\it
  XMM-Newton}. The signal was also detected in {\it Chandra}
observations of the Perseus cluster, at 2.2$\sigma$ in that work. As
seen in Fig.~\ref{fig:parspace}, the signal immediately straddled the
robust constraints from M31 \cite{horiuchi:2013noa}, which used {\it
  Chandra} data and were available even at that time. The Bulbul et
al.\ analysis was quite thorough in studying the possible atomic and
instrumental sources of the line, and anticipated much of the followup
work. They showed that potassium lines are far too low in emissivity
and relative abundance to account for the line, and that limits on
partner lines of Ar {\sc XVII} and Cl {\sc XVII} with stronger
emissivity strongly constrain those lines near the 3.55 keV
feature. Bulbul et al. also proposed that the planned {\it ASTRO-H}
X-ray Space Telescope would be sensitive to the Doppler broadened
energy of the dark matter line due to its velocity dispersion, which
turned into one of the key analyses of the mission prior to its
failure.\footnote{This velocity broadening occurs for the plasma and
  dark matter virialized system because $m_s \approx 7\,\mathrm{keV}
  \ll m_\mathrm{Ar}\sim m_\mathrm{K}\sim 40\,\mathrm{GeV}$, leaving
  the dark matter particle to have a much higher velocity dispersion
  for a matched kinetic energy distribution. Note that ions such as Ar
  and K will have turbulent velocity broadening as well, at a level of
  120 km/s relative to 100 km/s thermal broadening at
  $T=4\,\mathrm{keV}$, though still very low relative to the dark
  matter.} The position of the signal in the parameter space relative
to other detections and constraints near the signals is shown in
Fig.~\ref{fig:parspace}.

Shortly after the Bulbul et al. result, Boyarsky et
al.~\cite{Boyarsky:2014jta} reported the signal as being seen at
3$\sigma$ in {\it XMM-Newton} observations of M31, which had more
exposure time at that point than that available from {\it Chandra}, as
well as at 2.3$\sigma$ from {\it XMM-Newton} observations of the
outskirts of the Perseus Cluster. They reported a combined statistical
significance of 4.4$\sigma$. The M31 signal parameters are shown in
Fig.~\ref{fig:parspace}.

Significant observational work followed. An early result
was the detection of the signal in the Milky Way's Galactic Center
\cite{Boyarsky:2014ska,Jeltema:2014qfa}. Urban et
al.~\cite{Urban:2014yda} detected the signal toward the Perseus
Cluster using {\it Suzaku} data at the level of 7.4$\sigma$ (reported
as a $C$-statistic). Urban et al. reported that the radial flux
profile was inconsistent with that expected from a dark matter halo,
but a subsequent analysis of the same data found that profile to be
consistent with a halo and with dark matter decay
\cite{Franse:2016dln}. The line had weak indications at
$\sim$2$\sigma$ in stacked {\it Suzaku} cluster observations
\cite{Bulbul:2016yop}. The 3.5 keV line was seen at $>$2$\sigma$
significance in eight new clusters by Iakubovskyi et
al.~\cite{Iakubovskyi:2015dna}, with a redshifting of the line energy
consistent with cosmological origin, {\it i.e.}, not instrumental.

In another follow up, Anderson et al. \cite{Anderson:2014tza} stacked
approximately 80 galaxies from {\it XMM-Newton} and {\it Chandra}
data, and claimed a high-significance exclusion of the 3.5 keV line as
due to dark matter decay. The analysis relied on a model for the
continuum of the stacked spectrum that had large positive and negative
residuals, approximately the size of the 3.5 keV signal
(c.f.\ Anderson et al.'s Fig.\ 4). Despite this overwhelming
systematic uncertainty, this work used the statistical deviation from
the continuum to place their quoted limit. Another followup analysis
by Jeltema \& Profumo (JP) \cite{Jeltema:2014qfa} initially fit an
unconstrained potassium (K {\sc XVIII}) line to {\it XMM-Newton} data,
subtracted the fit from the data, and used the modified data to place
a constraint on the presence of dark matter decay. This is {\it prima
  facie} a highly problematic methodology. JP also used line ratios to
argue that the temperature of Perseus and stacked clusters could have
a large contribution from cool phase gas, and therefore the emissivity
of K {\sc XVIII} is not well known. Those line ratios were shown to be
inaccurate in Ref.~\cite{Bulbul:2014ala}. JP also used an energy
window in their M31 analysis that was too narrow to significantly
detect the line toward M31 \cite{Boyarsky:2014paa}.

In related work, Ref.~\cite{Carlson:2014lla} attempted a
characterization of the spatial origin of the 3.5 keV photons from the
Milky Way Galactic Center and Perseus Cluster. The Galactic Center
analysis in that work is compromised by the lack of incorporation of
X-ray absorption, which is very high in the Galactic Center direction,
and likely patchy and irregular, because of the irregular coverage by
molecular clouds \cite{Muno:2004wh}, 2 to 3 times higher than used in
Ref.~\cite{Carlson:2014lla}. The observed variation in hydrogen column
density gives a qualitative idea of the possible spatial variations of
the brightness of any extended signal, like dark matter, toward the
Galactic Center. Therefore, a proper dark matter template will not be
symmetric. The column density of hydrogen on the sky is likely
quadrupolar (c.f. Fig.~2 in Ref.~\cite{Carlson:2014lla}) since
molecular clouds tend to align with the Galactic plane. The spatial
analysis toward the Perseus Cluster is also problematic in that the
line in any spatial portion is of order 1\% of the continuum flux
(discussed in Appendix B of \cite{Franse:2016dln}). The spatial
continuum templates that Ref.~\cite{Carlson:2014lla} used to produce a
map of the 3.5 keV excess signal cannot be normalized to an accuracy
anywhere near that required, 1\% (given the broad range of continuum
models used by the authors). Since the continuum templates are
astrophysical, any residuals caused by their inaccurate normalization
would also follow the spatial distribution of that astrophysical
signal, regardless of the true distribution of the much weaker 3.5 keV
signal \cite{Franse:2016dln}.

The {\it Hitomi} Telescope did receive 275 ks of data toward the
Perseus Cluster before its failure which was analyzed for the presence
of the 3.5 keV line \cite{Aharonian:2016gzq}. The data was
unprecedented in its ability to do resolved X-ray spectrometry of the
Fe lines \cite{Aharonian:2016pyf}. Given that the protective closed
gate valve was in place, the exposure was equivalent to 70 ks of
normal operations, which was far short of what would be needed to be
highly sensitive. Ref.~\cite{Aharonian:2016gzq} analyzed {\it
  XMM-Newton} MOS data in the same field of view as the {\it Hitomi}
data, and found the MOS data to have a higher flux within that field.
{\it Hitomi} excluded the central value of the new MOS detection by
3$\sigma$. The prior detections were not appreciably constrained by
the {\it Hitomi} data, as shown in Fig.~\ref{fig:parspace}.

The NuSTAR telescope was found to be sufficiently sensitive to 3.5 keV
photons, with a wide field of view, $\sim$37 deg$^2$, from ``zero
bounce'' photons allowed into the detector because the design of
the telescope's optical bench allows for these photons in without
passing through the telescope's optics
\cite{Neronov:2016wdd,Perez:2016tcq}. This was used by Neronov et
al.~\cite{Neronov:2016wdd} to place constraints in the high 
mass range of sterile neutrino decay parameters space with {\it
  NuSTAR} data toward the COSMOS and CDFS empty sky fields. A few
unidentified lines that could be due to instrumental effects were also
detected. A line at $3.51\pm 0.02$ keV was detected at 11.1$\sigma$ in
that work, which is consistent with flux expected from previous
detections given the dark matter in the field of view. The response of
{\it NuSTAR} is very poorly known at the lower energies near 3-4 keV,
and it is thought that the line is likely instrumental since the line
is seen in Earth occulted data \cite{Perez:2016tcq}. Perez et
al.~\cite{Perez:2016tcq} placed constraints from observations of the
Galactic Center, which are shown in Fig.~\ref{fig:fullparams}.

The Deep Field exposures of the {\it Chandra} telescope were studied
to be potentially very sensitive to dark matter decays, and placed
limits on the parameter space \cite{abazajian:2006jc}. Recent work
by Cappelluti et al.~\cite{Cappelluti:2017ywp} used over three times
the exposure, with $\sim$10 Ms of Chandra observations towards the
COSMOS Legacy and CDFS survey fields. They see a line feature at
3$\sigma$ at $3.51 \pm 0.02\,\mathrm{keV}$ with flux (mixing angle)
consistent with prior detections. These parameters are shown in
Fig.~\ref{fig:parspace}. Significantly, Cappelluti et al. show that
the line feature cannot be due to instrumental effects given the
background analyses by Ref.~\cite{Bartalucci:2014mta}, and cannot be
due to K or Ar lines near 3.5 keV due to the lack of the presence of
partner lines of higher emissivity. Importantly, charge exchange lines
that could also be present near 3.5 keV \cite{Gu:2015gqm,Shah:2016efh}
were constrained by partner lines by Cappelluti et al., and more
analysis of the possibility of charge exchange emissivity
contributions are warranted.

\subsection{Future Observations}
There are a number of proposed and planned missions  that
are dedicated to the search of X-ray lines from dark matter or as
additional science. The XQC sounding rocket flight
\cite{mccammon:2002gb} is very sensitive to the presence of a line due
to its large field of view and microcalorimeter-based high energy
resolution \cite{abazajian:2006jc}. The XQC and Micro-X sounding
rocket X-ray experiments could be very sensitive to the presence of
the dark matter line when exposed toward or near the Galactic Center,
which is being proposed for summer of 2019 for a campaign in
Australia, where the Galactic Center is visible from the southern sky
\cite{Figueroa-Feliciano:2015gwa}. 

In 2016, the the {\it Hitomi} satellite was unfortunately lost. In
March 2017, Paul Hertz, director of NASA’s astrophysics division,
reported a formal start will happen in 2017 of the project known as
the {\it X-Ray Astronomy Recovery Mission (XARM)}, which will have a
soft X-ray spectrometer like that of {\it Hitomi}, prepared by NASA
and JAXA for launch in March 2021 \cite{XARM}. As shown in Bulbul et
al. \cite{bulbul:2014sua}, and discussed above, such a mission could
see the telltale sign of the velocity broadening of a dark matter
line. In addition, the high energy resolution velocity differential
between dark matter and gas in the Milky Way Galaxy could be used to
discriminate between gas and dark matter origins of the 3.5 keV line
\cite{Speckhard:2015eva,Powell:2016zbo}. The eROSITA is an all-sky
mission built by the Max Planck Institute for Extraterrestrial Physics
to follow up ROSAT, and is scheduled for launch in 2018
\cite{Merloni:2012uf}. eROSITA will have sensitivity to sterile
neutrino dark matter in cross-correlation analyses
\cite{Zandanel:2015xca}.

The future of X-ray astronomy after {\it Chandra} and {\it XMM-Newton}
is taking two paths. The {\it Athena X-ray Observatory} is planned for
launch in about 2028 as a European-Space-Agency-led mission that has
goals of microcalorimeter spectroscopy with resolving power of
$R\equiv \lambda/\delta\lambda \approx 1000$, similar to that of {\it
  XARM}, and a design for wide, medium-sensitivity surveys
\cite{Nandra:2013shg}. For these goals, area is built up at the
expense of angular resolution and sensitivity. The sensitivity of {\it
  Athena} to sterile neutrino dark matter was forecast by Neronov \&
Malyshev \cite{Neronov:2015kca} and is shown in
Fig.~\ref{fig:fullparams}. 

In a time frame beyond 2030, the second path after {\it Chandra} is
the {\it Lynx
  Telescope}\footnote{\texttt{https://wwwastro.msfc.nasa.gov/lynx}},
which is planned with 50 to 100 times the sensitivity over {\it
  Chandra} and {\it Athena}. {\it Lynx} has the goals of achieving
$R\approx 1000$ spectroscopy on 1'' scales, adding a true third
dimension to high-angular resolution X-ray data, as well as $R\approx
5000$ spectroscopy for point sources. For {\it Lynx}, area is built up
while preserving {\it Chandra} angular resolution of 0.5'', with 10
times the field of view. Since the technology and science goals are
still being developed, a detailed study of {\it Lynx} sensitivity to
dark matter has yet to be performed. It is certain, however, that if
there is a dark matter line signature well within the sensitivity of
{\it Lynx} like the 3.5 keV feature, the simultaneous high energy and
angular resolution of a telescope like {\it Lynx} will open up an era
of {\it dark matter astronomy} where the density and velocity space
mapping of dark matter could take place.

\section{Conclusions}

Sterile neutrinos are the leading natural extension of the Standard
Model of particle physics, accommodating neutrino mass.  Their
presence can have significant effects on the early Universe and
structure formation, at the eV scale, and could even be the dark
matter, at the keV scale. The light, eV-scale sterile neutrinos can
thermalize in the early Universe, affecting primordial nucleosynthesis
and structure formation.

Recent indication of tensions between large scale and local measures
of the amplitude of fluctuations on small scales ($\sigma_8$) and the
Hubble expansion rate ($H_0$) may be indicative of new physics related
to new neutrino physics, including eV-scale sterile neutrinos.  The
light sterile neutrinos can also be responsible for chaotic lepton
number generation or lepton number depletion. Detection of sterile
neutrinos in the laboratory would have significant implications for
our understanding of several aspects of cosmology.

Heavier, keV-scale sterile neutrinos can be produced to be the dark
matter via a number of mechanisms, through couplings to other
particles, or resonant and non-resonant oscillations. The evidence for
the candidate line signal near 3.5 keV is significant and building,
including detections in clusters of galaxies, field galaxies and
dwarfs, but the signal also has significant challenges, primarily {\it
  Chandra} observations of the Andromeda Galaxy and {\it XMM-Newton}
observations of the Draco Dwarf Galaxy. If sterile neutrinos are
definitively detected astrophysically or in the laboratory as all---or
as a portion---of the dark matter, a significant piece of our standard
cosmological and fundamental physics paradigm would be found.

\noindent{\bf Acknowledgments}

I would like to thank my fruitful collaborations and discussions with
Prateek Agrawal, Mike Boylan-Kolchin, Esra Bulbul, James Bullock,
Mu-Chun Chen, J.\ J.\ Cherry, Francis-Yan Cyr-Racine, Andr\'e de
Gouv\^ea, Bhupal Dev, Pasquale Di Bari, Scott Dodelson, Marco Drewes,
Enectali Figueroa-Feliciano, George M.\ Fuller, Chris Hirata, Shunsaku
Horiuchi, Patrick Huber, Manoj Kaplinghat, Alex Kusenko, Ranjan Laha,
Jon Link, Maxim Markevitch, Alexander Merle, Amol Patwardhan, Oleg
Ruchayskiy, Aaron Tohuvavohu, Teja Venumadhav, and Alexey Vikhlinin. I
would also like to thank the Mainz Institute for Theoretical Physics
program on ``The Energy Scales of the Universe'' for hospitality,
where many of these discussions took place and part of this review was
written. This work was supported in part by NSF Grants PHY-1316792 and
PHY-1620638.

\bibliographystyle{elsarticle-num}
\bibliography{/Users/aba/Dropbox/master}

\begin{thebibliography}{100}
\expandafter\ifx\csname url\endcsname\relax
  \def\url#1{\texttt{#1}}\fi
\expandafter\ifx\csname urlprefix\endcsname\relax\def\urlprefix{URL }\fi
\expandafter\ifx\csname href\endcsname\relax
  \def\href#1#2{#2} \def\path#1{#1}\fi

\bibitem{Olive:2016xmw}
C.~Patrignani, et~al., {Review of Particle Physics}, Chin. Phys. C40~(10)
  (2016) 100001.
\newblock \href {http://dx.doi.org/10.1088/1674-1137/40/10/100001}
  {\path{doi:10.1088/1674-1137/40/10/100001}}.

\bibitem{GellMann:1980vs}
M.~Gell-Mann, P.~Ramond, R.~Slansky, {Complex Spinors and Unified Theories},
  Conf. Proc. C790927 (1979) 315--321.
\newblock \href {http://arxiv.org/abs/1306.4669} {\path{arXiv:1306.4669}}.

\bibitem{Yanagida:1979as}
T.~Yanagida, {HORIZONTAL SYMMETRY AND MASSES OF NEUTRINOS}, Conf. Proc.
  C7902131 (1979) 95--99.

\bibitem{Mohapatra:1979ia}
R.~N. Mohapatra, G.~Senjanovic, {Neutrino Mass and Spontaneous Parity
  Violation}, Phys. Rev. Lett. 44 (1980) 912.
\newblock \href {http://dx.doi.org/10.1103/PhysRevLett.44.912}
  {\path{doi:10.1103/PhysRevLett.44.912}}.

\bibitem{Fukuda:1998mi}
Y.~Fukuda, et~al., {Evidence for oscillation of atmospheric neutrinos}, Phys.
  Rev. Lett. 81 (1998) 1562--1567.
\newblock \href {http://arxiv.org/abs/hep-ex/9807003}
  {\path{arXiv:hep-ex/9807003}}, \href
  {http://dx.doi.org/10.1103/PhysRevLett.81.1562}
  {\path{doi:10.1103/PhysRevLett.81.1562}}.

\bibitem{Cleveland:1998nv}
B.~T. Cleveland, T.~Daily, R.~Davis, Jr., J.~R. Distel, K.~Lande, C.~K. Lee,
  P.~S. Wildenhain, J.~Ullman, {Measurement of the solar electron neutrino flux
  with the Homestake chlorine detector}, Astrophys. J. 496 (1998) 505--526.
\newblock \href {http://dx.doi.org/10.1086/305343} {\path{doi:10.1086/305343}}.

\bibitem{Ahmad:2002jz}
Q.~R. Ahmad, et~al., {Direct evidence for neutrino flavor transformation from
  neutral current interactions in the Sudbury Neutrino Observatory}, Phys. Rev.
  Lett. 89 (2002) 011301.
\newblock \href {http://arxiv.org/abs/nucl-ex/0204008}
  {\path{arXiv:nucl-ex/0204008}}, \href
  {http://dx.doi.org/10.1103/PhysRevLett.89.011301}
  {\path{doi:10.1103/PhysRevLett.89.011301}}.

\bibitem{Aguilar:2001ty}
A.~Aguilar-Arevalo, et~al., {Evidence for neutrino oscillations from the
  observation of anti-neutrino(electron) appearance in a anti-neutrino(muon)
  beam}, Phys. Rev. D64 (2001) 112007.
\newblock \href {http://arxiv.org/abs/hep-ex/0104049}
  {\path{arXiv:hep-ex/0104049}}, \href
  {http://dx.doi.org/10.1103/PhysRevD.64.112007}
  {\path{doi:10.1103/PhysRevD.64.112007}}.

\bibitem{AguilarArevalo:2007it}
A.~A. Aguilar-Arevalo, et~al., A search for electron neutrino appearance at the
  delta m**2 ~ 1 ev**2 scale, Phys. Rev. Lett. 98 (2007) 231801.
\newblock \href {http://arxiv.org/abs/arXiv:0704.1500 [hep-ex]}
  {\path{arXiv:arXiv:0704.1500 [hep-ex]}}.

\bibitem{Huber:2011wv}
P.~Huber, {On the determination of anti-neutrino spectra from nuclear
  reactors}, Phys. Rev. C84 (2011) 024617, [Erratum: Phys.
  Rev.C85,029901(2012)].
\newblock \href {http://arxiv.org/abs/1106.0687} {\path{arXiv:1106.0687}},
  \href {http://dx.doi.org/10.1103/PhysRevC.85.029901,
  10.1103/PhysRevC.84.024617} {\path{doi:10.1103/PhysRevC.85.029901,
  10.1103/PhysRevC.84.024617}}.

\bibitem{Kopp:2013vaa}
J.~Kopp, P.~A.~N. Machado, M.~Maltoni, T.~Schwetz, {Sterile Neutrino
  Oscillations: The Global Picture}, JHEP 05 (2013) 050.
\newblock \href {http://arxiv.org/abs/1303.3011} {\path{arXiv:1303.3011}},
  \href {http://dx.doi.org/10.1007/JHEP05(2013)050}
  {\path{doi:10.1007/JHEP05(2013)050}}.

\bibitem{Abazajian:2012ys}
K.~N. Abazajian, et~al., {Light Sterile Neutrinos: A White Paper}\href
  {http://arxiv.org/abs/1204.5379} {\path{arXiv:1204.5379}}.

\bibitem{Lasserre:2014ita}
T.~Lasserre, {Light Sterile Neutrinos in Particle Physics: Experimental
  Status}, Phys. Dark Univ. 4 (2014) 81--85.
\newblock \href {http://arxiv.org/abs/1404.7352} {\path{arXiv:1404.7352}},
  \href {http://dx.doi.org/10.1016/j.dark.2014.10.001}
  {\path{doi:10.1016/j.dark.2014.10.001}}.

\bibitem{Hu:1997mj}
W.~Hu, D.~J. Eisenstein, M.~Tegmark, {Weighing neutrinos with galaxy surveys},
  Phys. Rev. Lett. 80 (1998) 5255--5258.
\newblock \href {http://arxiv.org/abs/astro-ph/9712057}
  {\path{arXiv:astro-ph/9712057}}, \href
  {http://dx.doi.org/10.1103/PhysRevLett.80.5255}
  {\path{doi:10.1103/PhysRevLett.80.5255}}.

\bibitem{Barger:1997yd}
V.~D. Barger, T.~J. Weiler, K.~Whisnant, {Four way neutrino oscillations},
  Phys. Lett. B427 (1998) 97--104.
\newblock \href {http://arxiv.org/abs/hep-ph/9712495}
  {\path{arXiv:hep-ph/9712495}}, \href
  {http://dx.doi.org/10.1016/S0370-2693(98)00325-6}
  {\path{doi:10.1016/S0370-2693(98)00325-6}}.

\bibitem{Langacker:1989sv}
P.~Langacker, {ON THE COSMOLOGICAL PRODUCTION OF LIGHT
  STERILE-NEUTRINOS}Preprint.

\bibitem{foot:1995bm}
R.~Foot, R.~R. Volkas, Reconciling sterile neutrinos with big bang
  nucleosynthesis, Phys. Rev. Lett. 75 (1995) 4350.
\newblock \href {http://arxiv.org/abs/hep-ph/9508275}
  {\path{arXiv:hep-ph/9508275}}.

\bibitem{abazajian:2004aj}
K.~Abazajian, N.~F. Bell, G.~M. Fuller, Y.~Y.~Y. Wong, Cosmological lepton
  asymmetry, primordial nucleosynthesis, and sterile neutrinos, Phys. Rev. D72
  (2005) 063004.
\newblock \href {http://arxiv.org/abs/astro-ph/0410175}
  {\path{arXiv:astro-ph/0410175}}.

\bibitem{Abazajian:2016hbv}
K.~N. Abazajian, M.~Kaplinghat, {Neutrino Physics from the Cosmic Microwave
  Background and Large-Scale Structure}, Ann. Rev. Nucl. Part. Sci. 66~(1)
  (2016) 401--420.
\newblock \href {http://dx.doi.org/10.1146/annurev-nucl-102014-021908}
  {\path{doi:10.1146/annurev-nucl-102014-021908}}.

\bibitem{dodelson:1993je}
S.~Dodelson, L.~M. Widrow, Sterile-neutrinos as dark matter, Phys. Rev. Lett.
  72 (1994) 17--20.
\newblock \href {http://arxiv.org/abs/hep-ph/9303287}
  {\path{arXiv:hep-ph/9303287}}.

\bibitem{Kusenko:2006rh}
A.~Kusenko, {Sterile neutrinos, dark matter, and the pulsar velocities in
  models with a Higgs singlet}, Phys. Rev. Lett. 97 (2006) 241301.
\newblock \href {http://arxiv.org/abs/hep-ph/0609081}
  {\path{arXiv:hep-ph/0609081}}, \href
  {http://dx.doi.org/10.1103/PhysRevLett.97.241301}
  {\path{doi:10.1103/PhysRevLett.97.241301}}.

\bibitem{Wolfenstein:1977ue}
L.~Wolfenstein, {Neutrino Oscillations in Matter}, Phys. Rev. D17 (1978)
  2369--2374.
\newblock \href {http://dx.doi.org/10.1103/PhysRevD.17.2369}
  {\path{doi:10.1103/PhysRevD.17.2369}}.

\bibitem{Mikheev:1986gs}
S.~P. Mikheev, A.~{\relax Yu}. Smirnov, {Resonance Amplification of
  Oscillations in Matter and Spectroscopy of Solar Neutrinos}, Sov. J. Nucl.
  Phys. 42 (1985) 913--917, [Yad. Fiz.42,1441(1985)].

\bibitem{blumenthal:1982mv}
G.~R. Blumenthal, H.~Pagels, J.~R. Primack, Galaxy formation by dissipationless
  particles heavier than neutrinos, Nature 299 (1982) 37--38.

\bibitem{colombi:1995ze}
S.~Colombi, S.~Dodelson, L.~M. Widrow, Large scale structure tests of warm dark
  matter, Astrophys. J. 458 (1996) 1.
\newblock \href {http://arxiv.org/abs/astro-ph/9505029}
  {\path{arXiv:astro-ph/9505029}}.

\bibitem{abazajian:2001nj}
K.~Abazajian, G.~M. Fuller, M.~Patel, Sterile neutrino hot, warm, and cold dark
  matter, Phys. Rev. D64 (2001) 023501.
\newblock \href {http://arxiv.org/abs/astro-ph/0101524}
  {\path{arXiv:astro-ph/0101524}}.

\bibitem{dolgov:2000ew}
A.~D. Dolgov, S.~H. Hansen, Massive sterile neutrinos as warm dark matter,
  Astropart. Phys. 16 (2002) 339--344.
\newblock \href {http://arxiv.org/abs/hep-ph/0009083}
  {\path{arXiv:hep-ph/0009083}}.

\bibitem{abazajian:2001vt}
K.~Abazajian, G.~M. Fuller, W.~H. Tucker, Direct detection of warm dark matter
  in the x-ray, Astrophys. J. 562 (2001) 593--604.
\newblock \href {http://arxiv.org/abs/astro-ph/0106002}
  {\path{arXiv:astro-ph/0106002}}.

\bibitem{bulbul:2014sua}
E.~Bulbul, M.~Markevitch, A.~Foster, R.~K. Smith, M.~Loewenstein, et~al.,
  {Detection of An Unidentified Emission Line in the Stacked X-ray spectrum of
  Galaxy Clusters}, Astrophys.J. 789 (2014) 13.
\newblock \href {http://arxiv.org/abs/1402.2301} {\path{arXiv:1402.2301}},
  \href {http://dx.doi.org/10.1088/0004-637X/789/1/13}
  {\path{doi:10.1088/0004-637X/789/1/13}}.

\bibitem{Boyarsky:2014jta}
A.~Boyarsky, O.~Ruchayskiy, D.~Iakubovskyi, J.~Franse, {Unidentified Line in
  X-Ray Spectra of the Andromeda Galaxy and Perseus Galaxy Cluster},
  Phys.Rev.Lett. 113 (2014) 251301.
\newblock \href {http://arxiv.org/abs/1402.4119} {\path{arXiv:1402.4119}},
  \href {http://dx.doi.org/10.1103/PhysRevLett.113.251301}
  {\path{doi:10.1103/PhysRevLett.113.251301}}.

\bibitem{Sakurai1995}
J.~J. Sakurai, Modern Quantum Mechanics, Addison-Wesley, Reading,
  Massachusetts, 1995.

\bibitem{Fuller1992}
G.~M. {Fuller}, R.~{Mayle}, B.~S. {Meyer}, J.~R. {Wilson}, Can a closure mass
  neutrino help solve the supernova shock reheating problem?, Astrophys. J. 389
  (1992) 517--526.

\bibitem{Savage:1990by}
M.~J. Savage, R.~A. Malaney, G.~M. Fuller, {Neutrino Oscillations and the
  Leptonic Charge of the Universe}, Astrophys. J. 368 (1991) 1--11.
\newblock \href {http://dx.doi.org/10.1086/169665} {\path{doi:10.1086/169665}}.

\bibitem{Foot:1996qk}
R.~Foot, M.~J. Thomson, R.~R. Volkas, Large neutrino asymmetries from neutrino
  oscillations, Phys. Rev. D53 (1996) 5349--5353.
\newblock \href {http://arxiv.org/abs/hep-ph/9509327}
  {\path{arXiv:hep-ph/9509327}}.

\bibitem{Shi:1996ic}
X.-D. Shi, {Chaotic amplification of neutrino chemical potentials by neutrino
  oscillations in big bang nucleosynthesis}, Phys. Rev. D54 (1996) 2753--2760.
\newblock \href {http://arxiv.org/abs/astro-ph/9602135}
  {\path{arXiv:astro-ph/9602135}}, \href
  {http://dx.doi.org/10.1103/PhysRevD.54.2753}
  {\path{doi:10.1103/PhysRevD.54.2753}}.

\bibitem{Davis:1994jw}
R.~Davis, {A review of the Homestake solar neutrino experiment}, Prog. Part.
  Nucl. Phys. 32 (1994) 13--32.
\newblock \href {http://dx.doi.org/10.1016/0146-6410(94)90004-3}
  {\path{doi:10.1016/0146-6410(94)90004-3}}.

\bibitem{Bahcall:1989ks}
J.~N. Bahcall, {Neutrino Astrophysics}, Cambridge University Press, 1989.

\bibitem{Haxton:1998sd}
W.~C. Haxton, \href{http://alice.cern.ch/format/showfull?sysnb=0303111}{{Topics
  in neutrino astrophysics}}, in: {Neutrinos in physics and astrophysics from
  10**(-33) to 10**28 CM. Proceedings, Conference, TASI'98, Boulder, USA, June
  1-26, 1998}, 1998, pp. 432--487.
\newblock \href {http://arxiv.org/abs/nucl-th/9901076}
  {\path{arXiv:nucl-th/9901076}}.
\newline\urlprefix\url{http://alice.cern.ch/format/showfull?sysnb=0303111}

\bibitem{Balantekin:1999re}
A.~B. Balantekin, W.~C. Haxton,
  \href{http://alice.cern.ch/format/showfull?sysnb=0307748}{{Solar, supernova,
  and atmospheric neutrinos}}, in: {Frontiers in nuclear physics. Proceedings,
  11th Physics Summer School, Canberra, Australia, January 12-23, 1998}, 1999,
  pp. 268--350.
\newblock \href {http://arxiv.org/abs/nucl-th/9903038}
  {\path{arXiv:nucl-th/9903038}}.
\newline\urlprefix\url{http://alice.cern.ch/format/showfull?sysnb=0307748}

\bibitem{Eguchi:2002dm}
K.~Eguchi, et~al., {First results from KamLAND: Evidence for reactor
  anti-neutrino disappearance}, Phys. Rev. Lett. 90 (2003) 021802.
\newblock \href {http://arxiv.org/abs/hep-ex/0212021}
  {\path{arXiv:hep-ex/0212021}}, \href
  {http://dx.doi.org/10.1103/PhysRevLett.90.021802}
  {\path{doi:10.1103/PhysRevLett.90.021802}}.

\bibitem{Ahn:2002up}
M.~H. Ahn, et~al., {Indications of neutrino oscillation in a 250 km long
  baseline experiment}, Phys. Rev. Lett. 90 (2003) 041801.
\newblock \href {http://arxiv.org/abs/hep-ex/0212007}
  {\path{arXiv:hep-ex/0212007}}, \href
  {http://dx.doi.org/10.1103/PhysRevLett.90.041801}
  {\path{doi:10.1103/PhysRevLett.90.041801}}.

\bibitem{Michael:2006rx}
D.~G. Michael, et~al., {Observation of muon neutrino disappearance with the
  MINOS detectors and the NuMI neutrino beam}, Phys. Rev. Lett. 97 (2006)
  191801.
\newblock \href {http://arxiv.org/abs/hep-ex/0607088}
  {\path{arXiv:hep-ex/0607088}}, \href
  {http://dx.doi.org/10.1103/PhysRevLett.97.191801}
  {\path{doi:10.1103/PhysRevLett.97.191801}}.

\bibitem{Capozzi:2016rtj}
F.~Capozzi, E.~Lisi, A.~Marrone, D.~Montanino, A.~Palazzo, {Neutrino masses and
  mixings: Status of known and unknown $3\nu$ parameters}, Nucl. Phys. B908
  (2016) 218--234.
\newblock \href {http://arxiv.org/abs/1601.07777} {\path{arXiv:1601.07777}},
  \href {http://dx.doi.org/10.1016/j.nuclphysb.2016.02.016}
  {\path{doi:10.1016/j.nuclphysb.2016.02.016}}.

\bibitem{Beck:2010zzb}
M.~Beck, {The KATRIN Experiment}, J. Phys. Conf. Ser. 203 (2010) 012097.
\newblock \href {http://arxiv.org/abs/0910.4862} {\path{arXiv:0910.4862}},
  \href {http://dx.doi.org/10.1088/1742-6596/203/1/012097}
  {\path{doi:10.1088/1742-6596/203/1/012097}}.

\bibitem{Avignone:2007fu}
F.~T. Avignone, III, S.~R. Elliott, J.~Engel, {Double Beta Decay, Majorana
  Neutrinos, and Neutrino Mass}, Rev. Mod. Phys. 80 (2008) 481--516.
\newblock \href {http://arxiv.org/abs/0708.1033} {\path{arXiv:0708.1033}},
  \href {http://dx.doi.org/10.1103/RevModPhys.80.481}
  {\path{doi:10.1103/RevModPhys.80.481}}.

\bibitem{Mohapatra:2006gs}
R.~N. Mohapatra, A.~Y. Smirnov, {Neutrino Mass and New Physics}, Ann. Rev.
  Nucl. Part. Sci. 56 (2006) 569--628.
\newblock \href {http://arxiv.org/abs/hep-ph/0603118}
  {\path{arXiv:hep-ph/0603118}}, \href
  {http://dx.doi.org/10.1146/annurev.nucl.56.080805.140534}
  {\path{doi:10.1146/annurev.nucl.56.080805.140534}}.

\bibitem{Athanassopoulos:1997pv}
C.~Athanassopoulos, et~al., Evidence for nu/mu --> nu/e neutrino oscillations
  from lsnd, Phys. Rev. Lett. 81 (1998) 1774--1777.
\newblock \href {http://arxiv.org/abs/nucl-ex/9709006}
  {\path{arXiv:nucl-ex/9709006}}.

\bibitem{Aguilar-Arevalo:2013pmq}
A.~A. Aguilar-Arevalo, et~al., {Improved Search for $\bar \nu_\mu \rightarrow
  \bar \nu_e$ Oscillations in the MiniBooNE Experiment}, Phys. Rev. Lett. 110
  (2013) 161801.
\newblock \href {http://arxiv.org/abs/1207.4809} {\path{arXiv:1207.4809}},
  \href {http://dx.doi.org/10.1103/PhysRevLett.110.161801}
  {\path{doi:10.1103/PhysRevLett.110.161801}}.

\bibitem{Gariazzo:2017fdh}
S.~Gariazzo, C.~Giunti, M.~Laveder, Y.~F. Li, {Updated Global 3+1 Analysis of
  Short-BaseLine Neutrino Oscillations}\href {http://arxiv.org/abs/1703.00860}
  {\path{arXiv:1703.00860}}.

\bibitem{Langacker:1998fq}
P.~Langacker, {Overview of neutrino physics and astrophysics}, in: {Neutrinos
  in physics and astrophysics from 10**(-33) to 10**28 CM. Proceedings,
  Conference, TASI'98, Boulder, USA, June 1-26, 1998}, 1998, pp. 1--26.

\bibitem{Pontecorvo:1957cp}
B.~Pontecorvo, {Mesonium and anti-mesonium}, Sov. Phys. JETP 6 (1957) 429, [Zh.
  Eksp. Teor. Fiz.33,549(1957)].

\bibitem{Pontecorvo:1957qd}
B.~Pontecorvo, {Inverse beta processes and nonconservation of lepton charge},
  Sov. Phys. JETP 7 (1958) 172--173, [Zh. Eksp. Teor. Fiz.34,247(1957)].

\bibitem{Maki:1962mu}
Z.~Maki, M.~Nakagawa, S.~Sakata, {Remarks on the unified model of elementary
  particles}, Prog. Theor. Phys. 28 (1962) 870--880.
\newblock \href {http://dx.doi.org/10.1143/PTP.28.870}
  {\path{doi:10.1143/PTP.28.870}}.

\bibitem{Kraus:2004zw}
C.~Kraus, et~al., {Final results from phase II of the Mainz neutrino mass
  search in tritium beta decay}, Eur. Phys. J. C40 (2005) 447--468.
\newblock \href {http://arxiv.org/abs/hep-ex/0412056}
  {\path{arXiv:hep-ex/0412056}}, \href
  {http://dx.doi.org/10.1140/epjc/s2005-02139-7}
  {\path{doi:10.1140/epjc/s2005-02139-7}}.

\bibitem{Aseev:2011dq}
V.~N. Aseev, et~al., {An upper limit on electron antineutrino mass from Troitsk
  experiment}, Phys. Rev. D84 (2011) 112003.
\newblock \href {http://arxiv.org/abs/1108.5034} {\path{arXiv:1108.5034}},
  \href {http://dx.doi.org/10.1103/PhysRevD.84.112003}
  {\path{doi:10.1103/PhysRevD.84.112003}}.

\bibitem{deGouvea:2005er}
A.~de~Gouvea, {See-saw energy scale and the LSND anomaly}, Phys. Rev. D72
  (2005) 033005.
\newblock \href {http://arxiv.org/abs/hep-ph/0501039}
  {\path{arXiv:hep-ph/0501039}}, \href
  {http://dx.doi.org/10.1103/PhysRevD.72.033005}
  {\path{doi:10.1103/PhysRevD.72.033005}}.

\bibitem{asaka:2005an}
T.~Asaka, S.~Blanchet, M.~Shaposhnikov, The numsm, dark matter and neutrino
  masses, Phys. Lett. B631 (2005) 151--156.
\newblock \href {http://arxiv.org/abs/hep-ph/0503065}
  {\path{arXiv:hep-ph/0503065}}.

\bibitem{deGouvea:2007hks}
A.~de~Gouvea, {GeV seesaw, accidentally small neutrino masses, and Higgs decays
  to neutrinos}\href {http://arxiv.org/abs/0706.1732} {\path{arXiv:0706.1732}}.

\bibitem{Merle:2013gea}
A.~Merle, {keV Neutrino Model Building}, Int. J. Mod. Phys. D22 (2013) 1330020.
\newblock \href {http://arxiv.org/abs/1302.2625} {\path{arXiv:1302.2625}},
  \href {http://dx.doi.org/10.1142/S0218271813300206}
  {\path{doi:10.1142/S0218271813300206}}.

\bibitem{Adhikari:2016bei}
R.~Adhikari, et~al., {A White Paper on keV Sterile Neutrino Dark Matter}, JCAP
  1701~(01) (2017) 025.
\newblock \href {http://arxiv.org/abs/1602.04816} {\path{arXiv:1602.04816}},
  \href {http://dx.doi.org/10.1088/1475-7516/2017/01/025}
  {\path{doi:10.1088/1475-7516/2017/01/025}}.

\bibitem{Lazarides:1980nt}
G.~Lazarides, Q.~Shafi, C.~Wetterich, {Proton Lifetime and Fermion Masses in an
  SO(10) Model}, Nucl. Phys. B181 (1981) 287--300.
\newblock \href {http://dx.doi.org/10.1016/0550-3213(81)90354-0}
  {\path{doi:10.1016/0550-3213(81)90354-0}}.

\bibitem{Magg:1980ut}
M.~Magg, C.~Wetterich, {Neutrino Mass Problem and Gauge Hierarchy}, Phys. Lett.
  B94 (1980) 61--64.
\newblock \href {http://dx.doi.org/10.1016/0370-2693(80)90825-4}
  {\path{doi:10.1016/0370-2693(80)90825-4}}.

\bibitem{Mohapatra:1980yp}
R.~N. Mohapatra, G.~Senjanovic, {Neutrino Masses and Mixings in Gauge Models
  with Spontaneous Parity Violation}, Phys. Rev. D23 (1981) 165.
\newblock \href {http://dx.doi.org/10.1103/PhysRevD.23.165}
  {\path{doi:10.1103/PhysRevD.23.165}}.

\bibitem{Schechter:1980gr}
J.~Schechter, J.~W.~F. Valle, {Neutrino Masses in SU(2) x U(1) Theories}, Phys.
  Rev. D22 (1980) 2227.
\newblock \href {http://dx.doi.org/10.1103/PhysRevD.22.2227}
  {\path{doi:10.1103/PhysRevD.22.2227}}.

\bibitem{Mohapatra:1986bd}
R.~N. Mohapatra, J.~W.~F. Valle, {Neutrino Mass and Baryon Number
  Nonconservation in Superstring Models}, Phys. Rev. D34 (1986) 1642.
\newblock \href {http://dx.doi.org/10.1103/PhysRevD.34.1642}
  {\path{doi:10.1103/PhysRevD.34.1642}}.

\bibitem{Shaposhnikov:2006nn}
M.~Shaposhnikov, {A Possible symmetry of the nuMSM}, Nucl. Phys. B763 (2007)
  49--59.
\newblock \href {http://arxiv.org/abs/hep-ph/0605047}
  {\path{arXiv:hep-ph/0605047}}, \href
  {http://dx.doi.org/10.1016/j.nuclphysb.2006.11.003}
  {\path{doi:10.1016/j.nuclphysb.2006.11.003}}.

\bibitem{Lindner:2010wr}
M.~Lindner, A.~Merle, V.~Niro, {Soft $L_e - L_\mu - L_\tau$ flavour symmetry
  breaking and sterile neutrino keV Dark Matter}, JCAP 1101 (2011) 034,
  [Erratum: JCAP1407,E01(2014)].
\newblock \href {http://arxiv.org/abs/1011.4950} {\path{arXiv:1011.4950}},
  \href {http://dx.doi.org/10.1088/1475-7516/2011/01/034,
  10.1088/1475-7516/2014/07/E01} {\path{doi:10.1088/1475-7516/2011/01/034,
  10.1088/1475-7516/2014/07/E01}}.

\bibitem{Grossman:1999ra}
Y.~Grossman, M.~Neubert, {Neutrino masses and mixings in nonfactorizable
  geometry}, Phys. Lett. B474 (2000) 361--371.
\newblock \href {http://arxiv.org/abs/hep-ph/9912408}
  {\path{arXiv:hep-ph/9912408}}, \href
  {http://dx.doi.org/10.1016/S0370-2693(00)00054-X}
  {\path{doi:10.1016/S0370-2693(00)00054-X}}.

\bibitem{Zee:1980ai}
A.~Zee, {A Theory of Lepton Number Violation, Neutrino Majorana Mass, and
  Oscillation}, Phys. Lett. B93 (1980) 389, [Erratum: Phys.
  Lett.B95,461(1980)].
\newblock \href {http://dx.doi.org/10.1016/0370-2693(80)90349-4,
  10.1016/0370-2693(80)90193-8} {\path{doi:10.1016/0370-2693(80)90349-4,
  10.1016/0370-2693(80)90193-8}}.

\bibitem{Kusenko:2010ik}
A.~Kusenko, F.~Takahashi, T.~T. Yanagida, {Dark Matter from Split Seesaw},
  Phys. Lett. B693 (2010) 144--148.
\newblock \href {http://arxiv.org/abs/1006.1731} {\path{arXiv:1006.1731}},
  \href {http://dx.doi.org/10.1016/j.physletb.2010.08.031}
  {\path{doi:10.1016/j.physletb.2010.08.031}}.

\bibitem{Adulpravitchai:2011rq}
A.~Adulpravitchai, R.~Takahashi, {A4 Flavor Models in Split Seesaw Mechanism},
  JHEP 09 (2011) 127.
\newblock \href {http://arxiv.org/abs/1107.3829} {\path{arXiv:1107.3829}},
  \href {http://dx.doi.org/10.1007/JHEP09(2011)127}
  {\path{doi:10.1007/JHEP09(2011)127}}.

\bibitem{Takahashi:2013eva}
R.~Takahashi, {Separate seesaw and its applications to dark matter and
  baryogenesis}, PTEP 2013~(6) (2013) 063B04.
\newblock \href {http://arxiv.org/abs/1303.0108} {\path{arXiv:1303.0108}},
  \href {http://dx.doi.org/10.1093/ptep/ptt042}
  {\path{doi:10.1093/ptep/ptt042}}.

\bibitem{Chen:2006hn}
M.-C. Chen, A.~de~Gouvea, B.~A. Dobrescu, {Gauge Trimming of Neutrino Masses},
  Phys. Rev. D75 (2007) 055009.
\newblock \href {http://arxiv.org/abs/hep-ph/0612017}
  {\path{arXiv:hep-ph/0612017}}, \href
  {http://dx.doi.org/10.1103/PhysRevD.75.055009}
  {\path{doi:10.1103/PhysRevD.75.055009}}.

\bibitem{Kamikado:2008jx}
H.~Kamikado, T.~Shindou, E.~Takasugi, {Froggatt-Nielsen hierarchy and the
  neutrino mass matrix}\href {http://arxiv.org/abs/0805.1338}
  {\path{arXiv:0805.1338}}.

\bibitem{Merle:2011yv}
A.~Merle, V.~Niro, {Deriving Models for keV sterile Neutrino Dark Matter with
  the Froggatt-Nielsen mechanism}, JCAP 1107 (2011) 023.
\newblock \href {http://arxiv.org/abs/1105.5136} {\path{arXiv:1105.5136}},
  \href {http://dx.doi.org/10.1088/1475-7516/2011/07/023}
  {\path{doi:10.1088/1475-7516/2011/07/023}}.

\bibitem{Chun:1995js}
E.~J. Chun, A.~S. Joshipura, A.~{\relax Yu}. Smirnov, {Models of light singlet
  fermion and neutrino phenomenology}, Phys. Lett. B357 (1995) 608--615.
\newblock \href {http://arxiv.org/abs/hep-ph/9505275}
  {\path{arXiv:hep-ph/9505275}}, \href
  {http://dx.doi.org/10.1016/0370-2693(95)00967-P}
  {\path{doi:10.1016/0370-2693(95)00967-P}}.

\bibitem{Barry:2011wb}
J.~Barry, W.~Rodejohann, H.~Zhang, {Light Sterile Neutrinos: Models and
  Phenomenology}, JHEP 07 (2011) 091.
\newblock \href {http://arxiv.org/abs/1105.3911} {\path{arXiv:1105.3911}},
  \href {http://dx.doi.org/10.1007/JHEP07(2011)091}
  {\path{doi:10.1007/JHEP07(2011)091}}.

\bibitem{Zhang:2011vh}
H.~Zhang, {Light Sterile Neutrino in the Minimal Extended Seesaw}, Phys. Lett.
  B714 (2012) 262--266.
\newblock \href {http://arxiv.org/abs/1110.6838} {\path{arXiv:1110.6838}},
  \href {http://dx.doi.org/10.1016/j.physletb.2012.06.074}
  {\path{doi:10.1016/j.physletb.2012.06.074}}.

\bibitem{Dev:2012bd}
P.~S. Bhupal~Dev, A.~Pilaftsis, {Light and Superlight Sterile Neutrinos in the
  Minimal Radiative Inverse Seesaw Model}, Phys. Rev. D87~(5) (2013) 053007.
\newblock \href {http://arxiv.org/abs/1212.3808} {\path{arXiv:1212.3808}},
  \href {http://dx.doi.org/10.1103/PhysRevD.87.053007}
  {\path{doi:10.1103/PhysRevD.87.053007}}.

\bibitem{chun:1999cq}
E.~J. Chun, H.~B. Kim, Nonthermal axino as cool dark matter in supersymmetric
  standard model without r-parity, Phys. Rev. D60 (1999) 095006.
\newblock \href {http://arxiv.org/abs/hep-ph/9906392}
  {\path{arXiv:hep-ph/9906392}}.

\bibitem{Foot:1995pa}
R.~Foot, R.~R. Volkas, {Neutrino physics and the mirror world: How exact parity
  symmetry explains the solar neutrino deficit, the atmospheric neutrino
  anomaly and the LSND experiment}, Phys. Rev. D52 (1995) 6595--6606.
\newblock \href {http://arxiv.org/abs/hep-ph/9505359}
  {\path{arXiv:hep-ph/9505359}}, \href
  {http://dx.doi.org/10.1103/PhysRevD.52.6595}
  {\path{doi:10.1103/PhysRevD.52.6595}}.

\bibitem{Berezhiani:1995yi}
Z.~G. Berezhiani, R.~N. Mohapatra, {Reconciling present neutrino puzzles:
  Sterile neutrinos as mirror neutrinos}, Phys. Rev. D52 (1995) 6607--6611.
\newblock \href {http://arxiv.org/abs/hep-ph/9505385}
  {\path{arXiv:hep-ph/9505385}}, \href
  {http://dx.doi.org/10.1103/PhysRevD.52.6607}
  {\path{doi:10.1103/PhysRevD.52.6607}}.

\bibitem{Bezrukov:2009th}
F.~Bezrukov, H.~Hettmansperger, M.~Lindner, {keV sterile neutrino Dark Matter
  in gauge extensions of the Standard Model}, Phys. Rev. D81 (2010) 085032.
\newblock \href {http://arxiv.org/abs/0912.4415} {\path{arXiv:0912.4415}},
  \href {http://dx.doi.org/10.1103/PhysRevD.81.085032}
  {\path{doi:10.1103/PhysRevD.81.085032}}.

\bibitem{Nemevsek:2012cd}
M.~Nemevsek, G.~Senjanovic, Y.~Zhang, {Warm Dark Matter in Low Scale Left-Right
  Theory}, JCAP 1207 (2012) 006.
\newblock \href {http://arxiv.org/abs/1205.0844} {\path{arXiv:1205.0844}},
  \href {http://dx.doi.org/10.1088/1475-7516/2012/07/006}
  {\path{doi:10.1088/1475-7516/2012/07/006}}.

\bibitem{Borah:2016lrl}
D.~Borah, {Light sterile neutrino and dark matter in left-right symmetric
  models without a Higgs bidoublet}, Phys. Rev. D94~(7) (2016) 075024.
\newblock \href {http://arxiv.org/abs/1607.00244} {\path{arXiv:1607.00244}},
  \href {http://dx.doi.org/10.1103/PhysRevD.94.075024}
  {\path{doi:10.1103/PhysRevD.94.075024}}.

\bibitem{Chen:2012jg}
M.-C. Chen, M.~Ratz, C.~Staudt, P.~K.~S. Vaudrevange, {The mu Term and Neutrino
  Masses}, Nucl. Phys. B866 (2013) 157--176.
\newblock \href {http://arxiv.org/abs/1206.5375} {\path{arXiv:1206.5375}},
  \href {http://dx.doi.org/10.1016/j.nuclphysb.2012.08.018}
  {\path{doi:10.1016/j.nuclphysb.2012.08.018}}.

\bibitem{Hu:2001bc}
W.~Hu, S.~Dodelson, {Cosmic microwave background anisotropies}, Ann. Rev.
  Astron. Astrophys. 40 (2002) 171--216.
\newblock \href {http://arxiv.org/abs/astro-ph/0110414}
  {\path{arXiv:astro-ph/0110414}}, \href
  {http://dx.doi.org/10.1146/annurev.astro.40.060401.093926}
  {\path{doi:10.1146/annurev.astro.40.060401.093926}}.

\bibitem{Fixsen:2009ug}
D.~J. Fixsen, {The Temperature of the Cosmic Microwave Background}, Astrophys.
  J. 707 (2009) 916--920.
\newblock \href {http://arxiv.org/abs/0911.1955} {\path{arXiv:0911.1955}},
  \href {http://dx.doi.org/10.1088/0004-637X/707/2/916}
  {\path{doi:10.1088/0004-637X/707/2/916}}.

\bibitem{KolbTurner1990}
E.~W. {Kolb}, M.~S. {Turner}, {The early universe.}, Addison Wesley, 1990.

\bibitem{dolgov:2002ab}
A.~D. Dolgov, et~al., Cosmological bounds on neutrino degeneracy improved by
  flavor oscillations, Nucl. Phys. B632 (2002) 363--382.
\newblock \href {http://arxiv.org/abs/hep-ph/0201287}
  {\path{arXiv:hep-ph/0201287}}.

\bibitem{wong:2002fa}
Y.~Y.~Y. Wong, Analytical treatment of neutrino asymmetry equilibration from
  flavour oscillations in the early universe, Phys. Rev. D66 (2002) 025015.
\newblock \href {http://arxiv.org/abs/hep-ph/0203180}
  {\path{arXiv:hep-ph/0203180}}.

\bibitem{abazajian:2002qx}
K.~N. Abazajian, J.~F. Beacom, N.~F. Bell, Stringent constraints on
  cosmological neutrino antineutrino asymmetries from synchronized flavor
  transformation, Phys. Rev. D66 (2002) 013008.
\newblock \href {http://arxiv.org/abs/astro-ph/0203442}
  {\path{arXiv:astro-ph/0203442}}.

\bibitem{Dicus:1982bz}
D.~A. Dicus, E.~W. Kolb, A.~M. Gleeson, E.~C.~G. Sudarshan, V.~L. Teplitz,
  M.~S. Turner, {Primordial Nucleosynthesis Including Radiative, Coulomb, and
  Finite Temperature Corrections to Weak Rates}, Phys. Rev. D26 (1982) 2694.
\newblock \href {http://dx.doi.org/10.1103/PhysRevD.26.2694}
  {\path{doi:10.1103/PhysRevD.26.2694}}.

\bibitem{Mangano:2005cc}
G.~Mangano, G.~Miele, S.~Pastor, T.~Pinto, O.~Pisanti, P.~D. Serpico, {Relic
  neutrino decoupling including flavor oscillations}, Nucl. Phys. B729 (2005)
  221--234.
\newblock \href {http://arxiv.org/abs/hep-ph/0506164}
  {\path{arXiv:hep-ph/0506164}}, \href
  {http://dx.doi.org/10.1016/j.nuclphysb.2005.09.041}
  {\path{doi:10.1016/j.nuclphysb.2005.09.041}}.

\bibitem{Landau:qm}
L.~D. Landau, E.~M. Lifshitz, Quantum Mechanics: Non-relativistic Theory,
  Pergamon Press, New York, 1977.

\bibitem{Dolgov:1980cq}
A.~D. Dolgov, {Neutrinos in the Early Universe}, Sov. J. Nucl. Phys. 33 (1981)
  700--706, [Yad. Fiz.33,1309(1981)].

\bibitem{McKellar:1992ja}
B.~H.~J. McKellar, M.~J. Thomson, {Oscillating doublet neutrinos in the early
  universe}, Phys. Rev. D49 (1994) 2710--2728.
\newblock \href {http://dx.doi.org/10.1103/PhysRevD.49.2710}
  {\path{doi:10.1103/PhysRevD.49.2710}}.

\bibitem{Sigl:1992fn}
G.~Sigl, G.~Raffelt, {General kinetic description of relativistic mixed
  neutrinos}, Nucl. Phys. B406 (1993) 423--451.
\newblock \href {http://dx.doi.org/10.1016/0550-3213(93)90175-O}
  {\path{doi:10.1016/0550-3213(93)90175-O}}.

\bibitem{Bell:1998ds}
N.~F. Bell, R.~R. Volkas, Y.~Y.~Y. Wong, Relic neutrino asymmetry evolution
  from first principles, Phys. Rev. D59 (1999) 113001.
\newblock \href {http://arxiv.org/abs/hep-ph/9809363}
  {\path{arXiv:hep-ph/9809363}}.

\bibitem{Asaka:2006rw}
T.~Asaka, M.~Laine, M.~Shaposhnikov, {On the hadronic contribution to sterile
  neutrino production}, JHEP 06 (2006) 053.
\newblock \href {http://arxiv.org/abs/hep-ph/0605209}
  {\path{arXiv:hep-ph/0605209}}, \href
  {http://dx.doi.org/10.1088/1126-6708/2006/06/053}
  {\path{doi:10.1088/1126-6708/2006/06/053}}.

\bibitem{Asaka:2006nq}
T.~Asaka, M.~Laine, M.~Shaposhnikov, {Lightest sterile neutrino abundance
  within the nuMSM}, JHEP 01 (2007) 091, [Erratum: JHEP02,028(2015)].
\newblock \href {http://arxiv.org/abs/hep-ph/0612182}
  {\path{arXiv:hep-ph/0612182}}, \href
  {http://dx.doi.org/10.1088/1126-6708/2007/01/091, 10.1007/JHEP02(2015)028}
  {\path{doi:10.1088/1126-6708/2007/01/091, 10.1007/JHEP02(2015)028}}.

\bibitem{Ghiglieri:2015jua}
J.~Ghiglieri, M.~Laine, {Improved determination of sterile neutrino dark matter
  spectrum}, JHEP 11 (2015) 171.
\newblock \href {http://arxiv.org/abs/1506.06752} {\path{arXiv:1506.06752}},
  \href {http://dx.doi.org/10.1007/JHEP11(2015)171}
  {\path{doi:10.1007/JHEP11(2015)171}}.

\bibitem{notzold:1987ik}
D.~Notzold, G.~Raffelt, Neutrino dispersion at finite temperature and density,
  Nucl. Phys. B307 (1988) 924.

\bibitem{Venumadhav:2015pla}
T.~Venumadhav, F.-Y. Cyr-Racine, K.~N. Abazajian, C.~M. Hirata, {Sterile
  neutrino dark matter: Weak interactions in the strong coupling epoch}, Phys.
  Rev. D94~(4) (2016) 043515.
\newblock \href {http://arxiv.org/abs/1507.06655} {\path{arXiv:1507.06655}},
  \href {http://dx.doi.org/10.1103/PhysRevD.94.043515}
  {\path{doi:10.1103/PhysRevD.94.043515}}.

\bibitem{Barbieri:1989ti}
R.~Barbieri, A.~Dolgov, {Bounds on Sterile-neutrinos from Nucleosynthesis},
  Phys. Lett. B237 (1990) 440--445.
\newblock \href {http://dx.doi.org/10.1016/0370-2693(90)91203-N}
  {\path{doi:10.1016/0370-2693(90)91203-N}}.

\bibitem{Barbieri:1990vx}
R.~Barbieri, A.~Dolgov, {Neutrino oscillations in the early universe}, Nucl.
  Phys. B349 (1991) 743--753.
\newblock \href {http://dx.doi.org/10.1016/0550-3213(91)90396-F}
  {\path{doi:10.1016/0550-3213(91)90396-F}}.

\bibitem{Enqvist:1990dq}
K.~Enqvist, K.~Kainulainen, J.~Maalampi, {Neutrino Asymmetry and Oscillations
  in the Early Universe}, Phys. Lett. B244 (1990) 186--190.
\newblock \href {http://dx.doi.org/10.1016/0370-2693(90)90053-9}
  {\path{doi:10.1016/0370-2693(90)90053-9}}.

\bibitem{Enqvist:1990ek}
K.~Enqvist, K.~Kainulainen, J.~Maalampi, {Resonant neutrino transitions and
  nucleosynthesis}, Phys. Lett. B249 (1990) 531--534.
\newblock \href {http://dx.doi.org/10.1016/0370-2693(90)91030-F}
  {\path{doi:10.1016/0370-2693(90)91030-F}}.

\bibitem{Enqvist:1991qj}
K.~Enqvist, K.~Kainulainen, M.~J. Thomson, {Stringent cosmological bounds on
  inert neutrino mixing}, Nucl. Phys. B373 (1992) 498--528.
\newblock \href {http://dx.doi.org/10.1016/0550-3213(92)90442-E}
  {\path{doi:10.1016/0550-3213(92)90442-E}}.

\bibitem{Cline:1991zb}
J.~M. Cline, {Constraints on almost Dirac neutrinos from neutrino -
  anti-neutrino oscillations}, Phys. Rev. Lett. 68 (1992) 3137--3140.
\newblock \href {http://dx.doi.org/10.1103/PhysRevLett.68.3137}
  {\path{doi:10.1103/PhysRevLett.68.3137}}.

\bibitem{Stodolsky:1986dx}
L.~Stodolsky, {On the Treatment of Neutrino Oscillations in a Thermal
  Environment}, Phys. Rev. D36 (1987) 2273.
\newblock \href {http://dx.doi.org/10.1103/PhysRevD.36.2273}
  {\path{doi:10.1103/PhysRevD.36.2273}}.

\bibitem{Harris:1980zi}
R.~A. Harris, L.~Stodolsky, {Two State Systems in Media and `Turing's
  Paradox'}, Phys. Lett. B116 (1982) 464--468.
\newblock \href {http://dx.doi.org/10.1016/0370-2693(82)90169-1}
  {\path{doi:10.1016/0370-2693(82)90169-1}}.

\bibitem{Kainulainen:2001cb}
K.~Kainulainen, A.~Sorri, {Oscillation induced neutrino asymmetry growth in the
  early universe}, JHEP 02 (2002) 020.
\newblock \href {http://arxiv.org/abs/hep-ph/0112158}
  {\path{arXiv:hep-ph/0112158}}, \href
  {http://dx.doi.org/10.1088/1126-6708/2002/02/020}
  {\path{doi:10.1088/1126-6708/2002/02/020}}.

\bibitem{Abazajian:2008dz}
K.~N. Abazajian, P.~Agrawal, {Chaos, Determinacy and Fractals in Active-Sterile
  Neutrino Oscillations in the Early Universe}, JCAP 0810 (2008) 006.
\newblock \href {http://arxiv.org/abs/0807.0456} {\path{arXiv:0807.0456}},
  \href {http://dx.doi.org/10.1088/1475-7516/2008/10/006}
  {\path{doi:10.1088/1475-7516/2008/10/006}}.

\bibitem{Bell:1998sr}
N.~F. Bell, R.~Foot, R.~R. Volkas, {Relic neutrino asymmetries and big bang
  nucleosynthesis in a four neutrino model}, Phys. Rev. D58 (1998) 105010.
\newblock \href {http://arxiv.org/abs/hep-ph/9805259}
  {\path{arXiv:hep-ph/9805259}}, \href
  {http://dx.doi.org/10.1103/PhysRevD.58.105010}
  {\path{doi:10.1103/PhysRevD.58.105010}}.

\bibitem{Abazajian:1999tc}
K.~Abazajian, G.~M. Fuller, X.~Shi, {Increase in the primordial He-4 yield in
  the two-doublet four neutrino mixing scheme}, Phys. Rev. D62 (2000) 093003.
\newblock \href {http://arxiv.org/abs/astro-ph/9908081}
  {\path{arXiv:astro-ph/9908081}}, \href
  {http://dx.doi.org/10.1103/PhysRevD.62.093003}
  {\path{doi:10.1103/PhysRevD.62.093003}}.

\bibitem{dibari:2001ua}
P.~Di~Bari, Update on neutrino mixing in the early universe, Phys. Rev. D65
  (2002) 043509.
\newblock \href {http://arxiv.org/abs/hep-ph/0108182}
  {\path{arXiv:hep-ph/0108182}}.

\bibitem{abazajian:2002bj}
K.~N. Abazajian, Telling three from four neutrinos with cosmology, Astropart.
  Phys. 19 (2003) 303--312.
\newblock \href {http://arxiv.org/abs/astro-ph/0205238}
  {\path{arXiv:astro-ph/0205238}}.

\bibitem{Smith:2006uw}
C.~J. Smith, G.~M. Fuller, C.~T. Kishimoto, K.~N. Abazajian, {Light Element
  Signatures of Sterile Neutrinos and Cosmological Lepton Numbers}, Phys. Rev.
  D74 (2006) 085008.
\newblock \href {http://arxiv.org/abs/astro-ph/0608377}
  {\path{arXiv:astro-ph/0608377}}, \href
  {http://dx.doi.org/10.1103/PhysRevD.74.085008}
  {\path{doi:10.1103/PhysRevD.74.085008}}.

\bibitem{abazajian:2005gj}
K.~Abazajian, {Production and evolution of perturbations of sterile neutrino
  dark matter}, Phys.Rev. D73 (2006) 063506.
\newblock \href {http://arxiv.org/abs/astro-ph/0511630}
  {\path{arXiv:astro-ph/0511630}}, \href
  {http://dx.doi.org/10.1103/PhysRevD.73.063506}
  {\path{doi:10.1103/PhysRevD.73.063506}}.

\bibitem{shi:1998km}
X.-d. Shi, G.~M. Fuller, A new dark matter candidate: Non-thermal sterile
  neutrinos, Phys. Rev. Lett. 82 (1999) 2832--2835.
\newblock \href {http://arxiv.org/abs/astro-ph/9810076}
  {\path{arXiv:astro-ph/9810076}}.

\bibitem{Volkas:2000ei}
R.~R. Volkas, Y.~Y.~Y. Wong, {Further studies on relic neutrino asymmetry
  generation. 1. The adiabatic Boltzmann limit, nonadiabatic evolution, and the
  classical harmonic oscillator analog of the quantum kinetic equations}, Phys.
  Rev. D62 (2000) 093024.
\newblock \href {http://arxiv.org/abs/hep-ph/0007185}
  {\path{arXiv:hep-ph/0007185}}, \href
  {http://dx.doi.org/10.1103/PhysRevD.62.093024}
  {\path{doi:10.1103/PhysRevD.62.093024}}.

\bibitem{Lee:2000ej}
K.~S.~M. Lee, R.~R. Volkas, Y.~Y.~Y. Wong, {Further studies on relic neutrino
  asymmetry generation. 2. A Rigorous treatment of repopulation in the
  adiabatic limit}, Phys. Rev. D62 (2000) 093025.
\newblock \href {http://arxiv.org/abs/hep-ph/0007186}
  {\path{arXiv:hep-ph/0007186}}, \href
  {http://dx.doi.org/10.1103/PhysRevD.62.093025}
  {\path{doi:10.1103/PhysRevD.62.093025}}.

\bibitem{Raffelt:1991ck}
G.~Raffelt, G.~Sigl, L.~Stodolsky, {Quantum statistics in particle mixing
  phenomena}, Phys. Rev. D45 (1992) 1782--1788.
\newblock \href {http://dx.doi.org/10.1103/PhysRevD.45.1782}
  {\path{doi:10.1103/PhysRevD.45.1782}}.

\bibitem{boyanovsky:2006it}
D.~Boyanovsky, C.~Ho, {Sterile neutrino production via active-sterile
  oscillations: The Quantum Zeno effect}, JHEP 0707 (2007) 030.
\newblock \href {http://arxiv.org/abs/hep-ph/0612092}
  {\path{arXiv:hep-ph/0612092}}, \href
  {http://dx.doi.org/10.1088/1126-6708/2007/07/030}
  {\path{doi:10.1088/1126-6708/2007/07/030}}.

\bibitem{kishimoto:2008ic}
C.~T. Kishimoto, G.~M. Fuller, {Lepton Number-Driven Sterile Neutrino
  Production in the Early Universe}, Phys.Rev. D78 (2008) 023524.
\newblock \href {http://arxiv.org/abs/0802.3377} {\path{arXiv:0802.3377}},
  \href {http://dx.doi.org/10.1103/PhysRevD.78.023524}
  {\path{doi:10.1103/PhysRevD.78.023524}}.

\bibitem{Gelmini:2004ah}
G.~Gelmini, S.~Palomares-Ruiz, S.~Pascoli, {Low reheating temperature and the
  visible sterile neutrino}, Phys. Rev. Lett. 93 (2004) 081302.
\newblock \href {http://arxiv.org/abs/astro-ph/0403323}
  {\path{arXiv:astro-ph/0403323}}, \href
  {http://dx.doi.org/10.1103/PhysRevLett.93.081302}
  {\path{doi:10.1103/PhysRevLett.93.081302}}.

\bibitem{Ade:2015xua}
P.~A.~R. Ade, et~al., {Planck 2015 results. XIII. Cosmological parameters}\href
  {http://arxiv.org/abs/1502.01589} {\path{arXiv:1502.01589}}.

\bibitem{Abazajian:2016yjj}
K.~N. Abazajian, et~al., {CMB-S4 Science Book, First Edition}\href
  {http://arxiv.org/abs/1610.02743} {\path{arXiv:1610.02743}}.

\bibitem{affleck:1984fy}
I.~Affleck, M.~Dine, {A New Mechanism for Baryogenesis}, Nucl.Phys. B249 (1985)
  361.
\newblock \href {http://dx.doi.org/10.1016/0550-3213(85)90021-5}
  {\path{doi:10.1016/0550-3213(85)90021-5}}.

\bibitem{KamLAND-Zen:2016pfg}
A.~Gando, et~al., {Search for Majorana Neutrinos near the Inverted Mass
  Hierarchy Region with KamLAND-Zen}, Phys. Rev. Lett. 117~(8) (2016) 082503,
  [Addendum: Phys. Rev. Lett.117,no.10,109903(2016)].
\newblock \href {http://arxiv.org/abs/1605.02889} {\path{arXiv:1605.02889}},
  \href {http://dx.doi.org/10.1103/PhysRevLett.117.109903,
  10.1103/PhysRevLett.117.082503} {\path{doi:10.1103/PhysRevLett.117.109903,
  10.1103/PhysRevLett.117.082503}}.

\bibitem{Steinbrink:2017ung}
N.~M.~N. Steinbrink, J.~D. Behrens, S.~Mertens, P.~C.~O. Ranitzsch,
  C.~Weinheimer, {keV-Scale Sterile Neutrino Sensitivity Estimation with
  Time-Of-Flight Spectroscopy in KATRIN using Self Consistent Approximate Monte
  Carlo}\href {http://arxiv.org/abs/1710.04939} {\path{arXiv:1710.04939}}.

\bibitem{anderhalden:2012jc}
D.~Anderhalden, A.~Schneider, A.~V. Maccio, J.~Diemand, G.~Bertone, {Hints on
  the Nature of Dark Matter from the Properties of Milky Way Satellites}, JCAP
  1303 (2013) 014.
\newblock \href {http://arxiv.org/abs/1212.2967} {\path{arXiv:1212.2967}},
  \href {http://dx.doi.org/10.1088/1475-7516/2013/03/014}
  {\path{doi:10.1088/1475-7516/2013/03/014}}.

\bibitem{horiuchi:2013noa}
S.~Horiuchi, P.~J. Humphrey, J.~Onorbe, K.~N. Abazajian, M.~Kaplinghat, et~al.,
  {Sterile neutrino dark matter bounds from galaxies of the Local Group},
  Phys.Rev. D89 (2014) 025017.
\newblock \href {http://arxiv.org/abs/1311.0282} {\path{arXiv:1311.0282}},
  \href {http://dx.doi.org/10.1103/PhysRevD.89.025017}
  {\path{doi:10.1103/PhysRevD.89.025017}}.

\bibitem{Hidaka:2006sg}
J.~Hidaka, G.~M. Fuller, {Dark matter sterile neutrinos in stellar collapse:
  Alteration of energy/lepton number transport and a mechanism for supernova
  explosion enhancement}, Phys. Rev. D74 (2006) 125015.
\newblock \href {http://arxiv.org/abs/astro-ph/0609425}
  {\path{arXiv:astro-ph/0609425}}, \href
  {http://dx.doi.org/10.1103/PhysRevD.74.125015}
  {\path{doi:10.1103/PhysRevD.74.125015}}.

\bibitem{Arguelles:2016uwb}
C.~A. Argüelles, V.~Brdar, J.~Kopp, {Production of keV Sterile Neutrinos in
  Supernovae: New Constraints and Gamma Ray Observables}\href
  {http://arxiv.org/abs/1605.00654} {\path{arXiv:1605.00654}}.

\bibitem{Cherry:2017dwu}
J.~F. Cherry, S.~Horiuchi, {Closing in on Resonantly Produced Sterile Neutrino
  Dark Matter}\href {http://arxiv.org/abs/1701.07874}
  {\path{arXiv:1701.07874}}.

\bibitem{Kainulainen:1990bn}
K.~Kainulainen, J.~Maalampi, J.~T. Peltoniemi, {Inert neutrinos in supernovae},
  Nucl. Phys. B358 (1991) 435--446.
\newblock \href {http://dx.doi.org/10.1016/0550-3213(91)90354-Z}
  {\path{doi:10.1016/0550-3213(91)90354-Z}}.

\bibitem{Mertens:2014nha}
S.~Mertens, T.~Lasserre, S.~Groh, G.~Drexlin, F.~Glueck, A.~Huber, A.~W.~P.
  Poon, M.~Steidl, N.~Steinbrink, C.~Weinheimer, {Sensitivity of
  Next-Generation Tritium Beta-Decay Experiments for keV-Scale Sterile
  Neutrinos}, JCAP 1502~(02) (2015) 020.
\newblock \href {http://arxiv.org/abs/1409.0920} {\path{arXiv:1409.0920}},
  \href {http://dx.doi.org/10.1088/1475-7516/2015/02/020}
  {\path{doi:10.1088/1475-7516/2015/02/020}}.

\bibitem{Mertens:2014osa}
S.~Mertens, K.~Dolde, M.~Korzeczek, F.~Glueck, S.~Groh, R.~D. Martin, A.~W.~P.
  Poon, M.~Steidl, {Wavelet approach to search for sterile neutrinos in tritium
  $\beta$-decay spectra}, Phys. Rev. D91 (2015) 042005.
\newblock \href {http://arxiv.org/abs/1410.7684} {\path{arXiv:1410.7684}},
  \href {http://dx.doi.org/10.1103/PhysRevD.91.042005}
  {\path{doi:10.1103/PhysRevD.91.042005}}.

\bibitem{Finocchiaro:1992hy}
G.~Finocchiaro, R.~E. Shrock, {An Experiment to search for a massive admixed
  neutrino in nuclear beta decay by complete kinematic reconstruction of the
  final state}, Phys. Rev. D46 (1992) R888--R891.
\newblock \href {http://dx.doi.org/10.1103/PhysRevD.46.R888}
  {\path{doi:10.1103/PhysRevD.46.R888}}.

\bibitem{Hindi:1998ym}
M.~M. Hindi, R.~L. Kozub, P.~Miocinovic, R.~Acvi, L.~Zhu, A.~H. Hussein,
  {Search for the admixture of heavy neutrinos in the recoil spectra of Ar-37
  decay}, Phys. Rev. C58 (1998) 2512--2525.
\newblock \href {http://dx.doi.org/10.1103/PhysRevC.58.2512}
  {\path{doi:10.1103/PhysRevC.58.2512}}.

\bibitem{Smith:2016vku}
P.~F. Smith, {Proposed experiments to detect keV range sterile neutrinos using
  energy-momentum reconstruction of beta decay or K-capture events}\href
  {http://arxiv.org/abs/1607.06876} {\path{arXiv:1607.06876}}.

\bibitem{shaposhnikov:2006xi}
M.~Shaposhnikov, I.~Tkachev, The numsm, inflation, and dark matter, Phys. Lett.
  B639 (2006) 414--417.
\newblock \href {http://arxiv.org/abs/hep-ph/0604236}
  {\path{arXiv:hep-ph/0604236}}.

\bibitem{Petraki:2007gq}
K.~Petraki, A.~Kusenko, {Dark-matter sterile neutrinos in models with a gauge
  singlet in the Higgs sector}, Phys. Rev. D77 (2008) 065014.
\newblock \href {http://arxiv.org/abs/0711.4646} {\path{arXiv:0711.4646}},
  \href {http://dx.doi.org/10.1103/PhysRevD.77.065014}
  {\path{doi:10.1103/PhysRevD.77.065014}}.

\bibitem{Bezrukov:2009yw}
F.~Bezrukov, D.~Gorbunov, {Light inflaton Hunter's Guide}, JHEP 05 (2010) 010.
\newblock \href {http://arxiv.org/abs/0912.0390} {\path{arXiv:0912.0390}},
  \href {http://dx.doi.org/10.1007/JHEP05(2010)010}
  {\path{doi:10.1007/JHEP05(2010)010}}.

\bibitem{Kusenko:2012ch}
A.~Kusenko, M.~Loewenstein, T.~T. Yanagida, {Moduli dark matter and the search
  for its decay line using Suzaku X-ray telescope}, Phys. Rev. D87~(4) (2013)
  043508.
\newblock \href {http://arxiv.org/abs/1209.6403} {\path{arXiv:1209.6403}},
  \href {http://dx.doi.org/10.1103/PhysRevD.87.043508}
  {\path{doi:10.1103/PhysRevD.87.043508}}.

\bibitem{Merle:2015oja}
A.~Merle, M.~Totzauer, {keV Sterile Neutrino Dark Matter from Singlet Scalar
  Decays: Basic Concepts and Subtle Features}, JCAP 1506 (2015) 011.
\newblock \href {http://arxiv.org/abs/1502.01011} {\path{arXiv:1502.01011}},
  \href {http://dx.doi.org/10.1088/1475-7516/2015/06/011}
  {\path{doi:10.1088/1475-7516/2015/06/011}}.

\bibitem{Shuve:2014doa}
B.~Shuve, I.~Yavin, {Dark matter progenitor: Light vector boson decay into
  sterile neutrinos}, Phys. Rev. D89~(11) (2014) 113004.
\newblock \href {http://arxiv.org/abs/1403.2727} {\path{arXiv:1403.2727}},
  \href {http://dx.doi.org/10.1103/PhysRevD.89.113004}
  {\path{doi:10.1103/PhysRevD.89.113004}}.

\bibitem{Abada:2014zra}
A.~Abada, G.~Arcadi, M.~Lucente, {Dark Matter in the minimal Inverse Seesaw
  mechanism}, JCAP 1410 (2014) 001.
\newblock \href {http://arxiv.org/abs/1406.6556} {\path{arXiv:1406.6556}},
  \href {http://dx.doi.org/10.1088/1475-7516/2014/10/001}
  {\path{doi:10.1088/1475-7516/2014/10/001}}.

\bibitem{Asaka:2006ek}
T.~Asaka, M.~Shaposhnikov, A.~Kusenko, {Opening a new window for warm dark
  matter}, Phys. Lett. B638 (2006) 401--406.
\newblock \href {http://arxiv.org/abs/hep-ph/0602150}
  {\path{arXiv:hep-ph/0602150}}, \href
  {http://dx.doi.org/10.1016/j.physletb.2006.05.067}
  {\path{doi:10.1016/j.physletb.2006.05.067}}.

\bibitem{Petraki:2008ef}
K.~Petraki, {Small-scale structure formation properties of chilled sterile
  neutrinos as dark matter}, Phys. Rev. D77 (2008) 105004.
\newblock \href {http://arxiv.org/abs/0801.3470} {\path{arXiv:0801.3470}},
  \href {http://dx.doi.org/10.1103/PhysRevD.77.105004}
  {\path{doi:10.1103/PhysRevD.77.105004}}.

\bibitem{Boyanovsky:2008nc}
D.~Boyanovsky, {Clustering properties of a sterile neutrino dark matter
  candidate}, Phys. Rev. D78 (2008) 103505.
\newblock \href {http://arxiv.org/abs/0807.0646} {\path{arXiv:0807.0646}},
  \href {http://dx.doi.org/10.1103/PhysRevD.78.103505}
  {\path{doi:10.1103/PhysRevD.78.103505}}.

\bibitem{Patwardhan:2015kga}
A.~V. Patwardhan, G.~M. Fuller, C.~T. Kishimoto, A.~Kusenko, {Diluted
  equilibrium sterile neutrino dark matter}, Phys. Rev. D92~(10) (2015) 103509.
\newblock \href {http://arxiv.org/abs/1507.01977} {\path{arXiv:1507.01977}},
  \href {http://dx.doi.org/10.1103/PhysRevD.92.103509}
  {\path{doi:10.1103/PhysRevD.92.103509}}.

\bibitem{abazajian:2006yn}
K.~Abazajian, S.~M. Koushiappas, {Constraints on sterile neutrino dark matter},
  Phys. Rev. D74 (2006) 023527.
\newblock \href {http://arxiv.org/abs/astro-ph/0605271}
  {\path{arXiv:astro-ph/0605271}}, \href
  {http://dx.doi.org/10.1103/PhysRevD.74.023527}
  {\path{doi:10.1103/PhysRevD.74.023527}}.

\bibitem{Menci:2017nsr}
N.~Menci, A.~Merle, M.~Totzauer, A.~Schneider, A.~Grazian, M.~Castellano, N.~G.
  Sanchez, {Fundamental physics with the Hubble Frontier Fields: constraining
  Dark Matter models with the abundance of extremely faint and distant
  galaxies}, Astrophys. J. 836~(1) (2017) 61.
\newblock \href {http://arxiv.org/abs/1701.01339} {\path{arXiv:1701.01339}},
  \href {http://dx.doi.org/10.3847/1538-4357/836/1/61}
  {\path{doi:10.3847/1538-4357/836/1/61}}.

\bibitem{lewis02}
A.~Lewis, S.~Bridle, Cosmological parameters from cmb and other data: a monte-
  carlo approach, Phys. Rev. D66 (2002) 103511.
\newblock \href {http://arxiv.org/abs/astro-ph/0205436}
  {\path{arXiv:astro-ph/0205436}}.

\bibitem{abazajian:2014gza}
K.~N. Abazajian, {Resonantly-Produced 7 keV Sterile Neutrino Dark Matter Models
  and the Properties of Milky Way Satellites}, Phys.Rev.Lett. 112 (2014)
  161303.
\newblock \href {http://arxiv.org/abs/1403.0954} {\path{arXiv:1403.0954}},
  \href {http://dx.doi.org/10.1103/PhysRevLett.112.161303}
  {\path{doi:10.1103/PhysRevLett.112.161303}}.

\bibitem{Steigman:1977kc}
G.~Steigman, D.~N. Schramm, J.~E. Gunn, {Cosmological Limits to the Number of
  Massive Leptons}, Phys. Lett. B66 (1977) 202--204.
\newblock \href {http://dx.doi.org/10.1016/0370-2693(77)90176-9}
  {\path{doi:10.1016/0370-2693(77)90176-9}}.

\bibitem{Cyburt:2015mya}
R.~H. Cyburt, B.~D. Fields, K.~A. Olive, T.-H. Yeh, {Big Bang Nucleosynthesis:
  2015}, Rev. Mod. Phys. 88 (2016) 015004.
\newblock \href {http://arxiv.org/abs/1505.01076} {\path{arXiv:1505.01076}},
  \href {http://dx.doi.org/10.1103/RevModPhys.88.015004}
  {\path{doi:10.1103/RevModPhys.88.015004}}.

\bibitem{2003moco.book.....D}
S.~{Dodelson}, {Modern cosmology}, 2003.

\bibitem{Battye:2013xqa}
R.~A. Battye, A.~Moss, {Evidence for Massive Neutrinos from Cosmic Microwave
  Background and Lensing Observations}, Phys. Rev. Lett. 112~(5) (2014) 051303.
\newblock \href {http://arxiv.org/abs/1308.5870} {\path{arXiv:1308.5870}},
  \href {http://dx.doi.org/10.1103/PhysRevLett.112.051303}
  {\path{doi:10.1103/PhysRevLett.112.051303}}.

\bibitem{Wyman:2013lza}
M.~Wyman, D.~H. Rudd, R.~A. Vanderveld, W.~Hu, {Neutrinos Help Reconcile Planck
  Measurements with the Local Universe}, Phys. Rev. Lett. 112~(5) (2014)
  051302.
\newblock \href {http://arxiv.org/abs/1307.7715} {\path{arXiv:1307.7715}},
  \href {http://dx.doi.org/10.1103/PhysRevLett.112.051302}
  {\path{doi:10.1103/PhysRevLett.112.051302}}.

\bibitem{Dvorkin:2014lea}
C.~Dvorkin, M.~Wyman, D.~H. Rudd, W.~Hu, {Neutrinos help reconcile Planck
  measurements with both the early and local Universe}, Phys. Rev. D90~(8)
  (2014) 083503.
\newblock \href {http://arxiv.org/abs/1403.8049} {\path{arXiv:1403.8049}},
  \href {http://dx.doi.org/10.1103/PhysRevD.90.083503}
  {\path{doi:10.1103/PhysRevD.90.083503}}.

\bibitem{Beutler:2014yhv}
F.~Beutler, et~al., {The clustering of galaxies in the SDSS-III Baryon
  Oscillation Spectroscopic Survey: Signs of neutrino mass in current
  cosmological datasets}, Mon. Not. Roy. Astron. Soc. 444 (2014) 3501.
\newblock \href {http://arxiv.org/abs/1403.4599} {\path{arXiv:1403.4599}},
  \href {http://dx.doi.org/10.1093/mnras/stu1702}
  {\path{doi:10.1093/mnras/stu1702}}.

\bibitem{Giusarma:2014zza}
E.~Giusarma, E.~Di~Valentino, M.~Lattanzi, A.~Melchiorri, O.~Mena, {Relic
  Neutrinos, thermal axions and cosmology in early 2014}, Phys. Rev. D90~(4)
  (2014) 043507.
\newblock \href {http://arxiv.org/abs/1403.4852} {\path{arXiv:1403.4852}},
  \href {http://dx.doi.org/10.1103/PhysRevD.90.043507}
  {\path{doi:10.1103/PhysRevD.90.043507}}.

\bibitem{Jacques:2013xr}
T.~D. Jacques, L.~M. Krauss, C.~Lunardini, {Additional Light Sterile Neutrinos
  and Cosmology}, Phys. Rev. D87~(8) (2013) 083515, [Erratum: Phys.
  Rev.D88,no.10,109901(2013)].
\newblock \href {http://arxiv.org/abs/1301.3119} {\path{arXiv:1301.3119}},
  \href {http://dx.doi.org/10.1103/PhysRevD.87.083515,
  10.1103/PhysRevD.88.109901} {\path{doi:10.1103/PhysRevD.87.083515,
  10.1103/PhysRevD.88.109901}}.

\bibitem{Canac:2016smv}
N.~Canac, G.~Aslanyan, K.~N. Abazajian, R.~Easther, L.~C. Price, {Testing for
  New Physics: Neutrinos and the Primordial Power Spectrum}, JCAP 1609~(09)
  (2016) 022.
\newblock \href {http://arxiv.org/abs/1606.03057} {\path{arXiv:1606.03057}},
  \href {http://dx.doi.org/10.1088/1475-7516/2016/09/022}
  {\path{doi:10.1088/1475-7516/2016/09/022}}.

\bibitem{DiValentino:2016hlg}
E.~Di~Valentino, A.~Melchiorri, J.~Silk, {Reconciling Planck with the local
  value of $H_0$ in extended parameter space}, Phys. Lett. B761 (2016)
  242--246.
\newblock \href {http://arxiv.org/abs/1606.00634} {\path{arXiv:1606.00634}},
  \href {http://dx.doi.org/10.1016/j.physletb.2016.08.043}
  {\path{doi:10.1016/j.physletb.2016.08.043}}.

\bibitem{viel:2005qj}
M.~Viel, J.~Lesgourgues, M.~G. Haehnelt, S.~Matarrese, A.~Riotto, Constraining
  warm dark matter candidates including sterile neutrinos and light gravitinos
  with wmap and the lyman- alpha forest, Phys. Rev. D71 (2005) 063534.
\newblock \href {http://arxiv.org/abs/astro-ph/0501562}
  {\path{arXiv:astro-ph/0501562}}.

\bibitem{abazajian:2002yz}
K.~N. Abazajian, G.~M. Fuller, Bulk qcd thermodynamics and sterile neutrino
  dark matter, Phys. Rev. D66 (2002) 023526.
\newblock \href {http://arxiv.org/abs/astro-ph/0204293}
  {\path{arXiv:astro-ph/0204293}}.

\bibitem{abazajian:2005xn}
K.~Abazajian, Linear cosmological structure limits on warm dark matter, Phys.
  Rev. D73 (2006) 063513.
\newblock \href {http://arxiv.org/abs/astro-ph/0512631}
  {\path{arXiv:astro-ph/0512631}}.

\bibitem{Schneider:2016uqi}
A.~Schneider, {Astrophysical constraints on resonantly produced sterile
  neutrino dark matter}, JCAP 1604~(04) (2016) 059.
\newblock \href {http://arxiv.org/abs/1601.07553} {\path{arXiv:1601.07553}},
  \href {http://dx.doi.org/10.1088/1475-7516/2016/04/059}
  {\path{doi:10.1088/1475-7516/2016/04/059}}.

\bibitem{croft:1999mm}
R.~A.~C. Croft, W.~Hu, R.~Dave, {Cosmological Limits on the Neutrino Mass from
  the Lya Forest}, Phys. Rev. Lett. 83 (1999) 1092--1095.
\newblock \href {http://arxiv.org/abs/astro-ph/9903335}
  {\path{arXiv:astro-ph/9903335}}, \href
  {http://dx.doi.org/10.1103/PhysRevLett.83.1092}
  {\path{doi:10.1103/PhysRevLett.83.1092}}.

\bibitem{mcdonald:2004xn}
P.~{McDonald}, U.~{Seljak}, R.~{Cen}, D.~{Shih}, D.~H. {Weinberg}, S.~{Burles},
  D.~P. {Schneider}, D.~J. {Schlegel}, N.~A. {Bahcall}, J.~W. {Briggs},
  J.~{Brinkmann}, M.~{Fukugita}, {\v Z}.~{Ivezi{\'c}}, S.~{Kent}, D.~E. {Vanden
  Berk}, {The Linear Theory Power Spectrum from the Ly{$\alpha$} Forest in the
  Sloan Digital Sky Survey}, \apj 635 (2005) 761--783.

\bibitem{seljak:2006qw}
U.~Seljak, A.~Makarov, P.~McDonald, H.~Trac, {Can sterile neutrinos be the dark
  matter?}, Phys. Rev. Lett. 97 (2006) 191303.
\newblock \href {http://arxiv.org/abs/astro-ph/0602430}
  {\path{arXiv:astro-ph/0602430}}, \href
  {http://dx.doi.org/10.1103/PhysRevLett.97.191303}
  {\path{doi:10.1103/PhysRevLett.97.191303}}.

\bibitem{viel:2006kd}
M.~Viel, J.~Lesgourgues, M.~G. Haehnelt, S.~Matarrese, A.~Riotto, Can sterile
  neutrinos be ruled out as warm dark matter candidates?, Phys. Rev. Lett. 97
  (2006) 071301.
\newblock \href {http://arxiv.org/abs/astro-ph/0605706}
  {\path{arXiv:astro-ph/0605706}}.

\bibitem{viel:2007mv}
M.~Viel, et~al., {How cold is cold dark matter? Small scales constraints from
  the flux power spectrum of the high-redshift Lyman- alpha forest}, Phys. Rev.
  Lett. 100 (2008) 041304.
\newblock \href {http://arxiv.org/abs/0709.0131} {\path{arXiv:0709.0131}},
  \href {http://dx.doi.org/10.1103/PhysRevLett.100.041304}
  {\path{doi:10.1103/PhysRevLett.100.041304}}.

\bibitem{viel:2013fqw}
M.~Viel, G.~D. Becker, J.~S. Bolton, M.~G. Haehnelt, {Warm Dark Matter as a
  solution to the small scale crisis: new constraints from high redshift
  Lyman-alpha forest data}, Physical Review D88~(4) (2013) 043502.
\newblock \href {http://arxiv.org/abs/1306.2314} {\path{arXiv:1306.2314}}.

\bibitem{Baur:2015jsy}
J.~Baur, N.~Palanque-Delabrouille, C.~Yèche, C.~Magneville, M.~Viel,
  {Lyman-alpha Forests cool Warm Dark Matter}, JCAP 1608~(08) (2016) 012.
\newblock \href {http://arxiv.org/abs/1512.01981} {\path{arXiv:1512.01981}},
  \href {http://dx.doi.org/10.1088/1475-7516/2016/08/012}
  {\path{doi:10.1088/1475-7516/2016/08/012}}.

\bibitem{Irsic:2017ixq}
V.~Iršič, et~al., {New Constraints on the free-streaming of warm dark matter
  from intermediate and small scale Lyman-$\alpha$ forest data}\href
  {http://arxiv.org/abs/1702.01764} {\path{arXiv:1702.01764}}.

\bibitem{Boyarsky:2008xj}
A.~Boyarsky, J.~Lesgourgues, O.~Ruchayskiy, M.~Viel, {Lyman-alpha constraints
  on warm and on warm-plus-cold dark matter models}, JCAP 0905 (2009) 012.
\newblock \href {http://arxiv.org/abs/0812.0010} {\path{arXiv:0812.0010}},
  \href {http://dx.doi.org/10.1088/1475-7516/2009/05/012}
  {\path{doi:10.1088/1475-7516/2009/05/012}}.

\bibitem{gnedin:2001wg}
N.~Y. Gnedin, A.~J.~S. Hamilton, Matter power spectrum from the lyman-alpha
  forest: Myth or reality?, Mon. Not. Roy. Astron. Soc. 334 (2002) 107--116.
\newblock \href {http://arxiv.org/abs/astro-ph/0111194}
  {\path{arXiv:astro-ph/0111194}}.

\bibitem{Kulkarni:2015fga}
G.~Kulkarni, J.~F. Hennawi, J.~Oñorbe, A.~Rorai, V.~Springel, {Characterizing
  the Pressure Smoothing Scale of the Intergalactic Medium}, Astrophys. J. 812
  (2015) 30.
\newblock \href {http://arxiv.org/abs/1504.00366} {\path{arXiv:1504.00366}},
  \href {http://dx.doi.org/10.1088/0004-637X/812/1/30}
  {\path{doi:10.1088/0004-637X/812/1/30}}.

\bibitem{Onorbe:2017ftn}
J.~Oñorbe, J.~F. Hennawi, Z.~Lukić, M.~Walther, {Constraining Reionization
  with the $z \sim 5-6$ Lyman-$\alpha$ Forest Power Spectrum: the Outlook after
  Planck}\href {http://arxiv.org/abs/1703.08633} {\path{arXiv:1703.08633}}.

\bibitem{schultz:2014eia}
C.~Schultz, J.~Onorbe, K.~N. Abazajian, J.~S. Bullock, {The High-$z$ Universe
  Confronts Warm Dark Matter: Galaxy Counts, Reionization and the Nature of
  Dark Matter}, Mon.Not.Roy.Astron.Soc. 442 (2014) 1597--1609.
\newblock \href {http://arxiv.org/abs/1401.3769} {\path{arXiv:1401.3769}},
  \href {http://dx.doi.org/10.1093/mnras/stu976}
  {\path{doi:10.1093/mnras/stu976}}.

\bibitem{Barkana:2001gra}
R.~Barkana, Z.~Haiman, J.~P. Ostriker, {Constraints on warm dark matter from
  cosmological reionization}, Astrophys. J. 558 (2001) 482.
\newblock \href {http://arxiv.org/abs/astro-ph/0102304}
  {\path{arXiv:astro-ph/0102304}}, \href {http://dx.doi.org/10.1086/322393}
  {\path{doi:10.1086/322393}}.

\bibitem{Bozek:2015bdo}
B.~Bozek, M.~Boylan-Kolchin, S.~Horiuchi, S.~Garrison-Kimmel, K.~Abazajian,
  J.~S. Bullock, {Resonant Sterile Neutrino Dark Matter in the Local and High-z
  Universe}, Mon. Not. Roy. Astron. Soc. 459 (2016) 1489.
\newblock \href {http://arxiv.org/abs/1512.04544} {\path{arXiv:1512.04544}},
  \href {http://dx.doi.org/10.1093/mnras/stw688}
  {\path{doi:10.1093/mnras/stw688}}.

\bibitem{BullockARAA}
J.~Bullock, M.~Boylan-Kolchin, {Small-Scale Challenges to the $\Lambda$CDM
  Paradigm}, Ann. Rev. Astron. Astrophys. 55, in press.

\bibitem{lovell:2011rd}
M.~R. Lovell, V.~Eke, C.~S. Frenk, L.~Gao, A.~Jenkins, et~al., {The Haloes of
  Bright Satellite Galaxies in a Warm Dark Matter Universe},
  Mon.Not.Roy.Astron.Soc. 420 (2012) 2318--2324.
\newblock \href {http://arxiv.org/abs/1104.2929} {\path{arXiv:1104.2929}},
  \href {http://dx.doi.org/10.1111/j.1365-2966.2011.20200.x}
  {\path{doi:10.1111/j.1365-2966.2011.20200.x}}.

\bibitem{Horiuchi:2015qri}
S.~Horiuchi, B.~Bozek, K.~N. Abazajian, M.~Boylan-Kolchin, J.~S. Bullock,
  S.~Garrison-Kimmel, J.~Onorbe, {Properties of resonantly produced sterile
  neutrino dark matter subhaloes}, Mon. Not. Roy. Astron. Soc. 456~(4) (2016)
  4346--4353.
\newblock \href {http://arxiv.org/abs/1512.04548} {\path{arXiv:1512.04548}},
  \href {http://dx.doi.org/10.1093/mnras/stv2922}
  {\path{doi:10.1093/mnras/stv2922}}.

\bibitem{bode:2000gq}
P.~Bode, J.~P. Ostriker, N.~Turok, Halo formation in warm dark matter models,
  Astrophys. J. 556 (2001) 93--107.
\newblock \href {http://arxiv.org/abs/astro-ph/0010389}
  {\path{arXiv:astro-ph/0010389}}.

\bibitem{polisensky:2010rw}
E.~Polisensky, M.~Ricotti, {Constraints on the Dark Matter Particle Mass from
  the Number of Milky Way Satellites}, Phys.Rev. D83 (2011) 043506.
\newblock \href {http://arxiv.org/abs/1004.1459} {\path{arXiv:1004.1459}},
  \href {http://dx.doi.org/10.1103/PhysRevD.83.043506}
  {\path{doi:10.1103/PhysRevD.83.043506}}.

\bibitem{Shrock:1974nd}
R.~Shrock, {Decay l0 ---> nu(lepton) gamma in gauge theories of weak and
  electromagnetic interactions}, Phys. Rev. D9 (1974) 743--748.
\newblock \href {http://dx.doi.org/10.1103/PhysRevD.9.743}
  {\path{doi:10.1103/PhysRevD.9.743}}.

\bibitem{Pal:1981rm}
P.~B. Pal, L.~Wolfenstein, Radiative decays of massive neutrinos, Phys. Rev.
  D25 (1982) 766.

\bibitem{Drees:2000qr}
M.~Drees, {Comment on `A New dark matter candidate: Nonthermal sterile
  neutrinos'}\href {http://arxiv.org/abs/hep-ph/0003127}
  {\path{arXiv:hep-ph/0003127}}.

\bibitem{Boyarsky:2005us}
A.~Boyarsky, A.~Neronov, O.~Ruchayskiy, M.~Shaposhnikov, {Constraints on
  sterile neutrino as a dark matter candidate from the diffuse x-ray
  background}, Mon.Not.Roy.Astron.Soc. 370 (2006) 213--218.
\newblock \href {http://arxiv.org/abs/astro-ph/0512509}
  {\path{arXiv:astro-ph/0512509}}, \href
  {http://dx.doi.org/10.1111/j.1365-2966.2006.10458.x}
  {\path{doi:10.1111/j.1365-2966.2006.10458.x}}.

\bibitem{Watson:2006qb}
C.~R. Watson, J.~F. Beacom, H.~Yuksel, T.~P. Walker, Direct x-ray constraints
  on sterile neutrino warm dark matter, Phys. Rev. D74 (2006) 033009.
\newblock \href {http://arxiv.org/abs/astro-ph/0605424}
  {\path{arXiv:astro-ph/0605424}}.

\bibitem{Boyarsky:2006fg}
A.~Boyarsky, A.~Neronov, O.~Ruchayskiy, M.~Shaposhnikov, I.~Tkachev, {Where to
  find a dark matter sterile neutrino?}, Phys.Rev.Lett. 97 (2006) 261302.
\newblock \href {http://arxiv.org/abs/astro-ph/0603660}
  {\path{arXiv:astro-ph/0603660}}, \href
  {http://dx.doi.org/10.1103/PhysRevLett.97.261302}
  {\path{doi:10.1103/PhysRevLett.97.261302}}.

\bibitem{loewenstein:2008yi}
M.~Loewenstein, A.~Kusenko, P.~L. Biermann, {New Limits on Sterile Neutrinos
  from Suzaku Observations of the Ursa Minor Dwarf Spheroidal Galaxy},
  Astrophys.J. 700 (2009) 426--435.
\newblock \href {http://arxiv.org/abs/0812.2710} {\path{arXiv:0812.2710}},
  \href {http://dx.doi.org/10.1088/0004-637X/700/1/426}
  {\path{doi:10.1088/0004-637X/700/1/426}}.

\bibitem{Loewenstein:2009cm}
M.~Loewenstein, A.~Kusenko, {Dark Matter Search Using Chandra Observations of
  Willman 1, and a Spectral Feature Consistent with a Decay Line of a 5 keV
  Sterile Neutrino}, Astrophys. J. 714 (2010) 652--662.
\newblock \href {http://arxiv.org/abs/0912.0552} {\path{arXiv:0912.0552}},
  \href {http://dx.doi.org/10.1088/0004-637X/714/1/652}
  {\path{doi:10.1088/0004-637X/714/1/652}}.

\bibitem{Loewenstein:2012px}
M.~Loewenstein, A.~Kusenko, {Dark Matter Search Using XMM-Newton Observations
  of Willman 1}, Astrophys. J. 751 (2012) 82.
\newblock \href {http://arxiv.org/abs/1203.5229} {\path{arXiv:1203.5229}},
  \href {http://dx.doi.org/10.1088/0004-637X/751/2/82}
  {\path{doi:10.1088/0004-637X/751/2/82}}.

\bibitem{Riemer-Sorensen:2006fh}
S.~Riemer-Sorensen, S.~H. Hansen, K.~Pedersen, Sterile neutrinos in the milky
  way: Observational constraints, Astrophys. J. 644 (2006) L33--L36.
\newblock \href {http://arxiv.org/abs/astro-ph/0603661}
  {\path{arXiv:astro-ph/0603661}}.

\bibitem{malyshev:2014xqa}
D.~Malyshev, A.~Neronov, D.~Eckert, {Constraints on 3.55 keV line emission from
  stacked observations of dwarf spheroidal galaxies}, Phys.Rev. D90~(10) (2014)
  103506.
\newblock \href {http://arxiv.org/abs/1408.3531} {\path{arXiv:1408.3531}},
  \href {http://dx.doi.org/10.1103/PhysRevD.90.103506}
  {\path{doi:10.1103/PhysRevD.90.103506}}.

\bibitem{tamura:2014mta}
T.~Tamura, R.~Iizuka, Y.~Maeda, K.~Mitsuda, N.~Y. Yamasaki, {An X-ray
  Spectroscopic Search for Dark Matter in the Perseus Cluster with Suzaku},
  Publ.Astron.Soc.Jap. 67~(2) (2015) 23.
\newblock \href {http://arxiv.org/abs/1412.1869} {\path{arXiv:1412.1869}},
  \href {http://dx.doi.org/10.1093/pasj/psu156}
  {\path{doi:10.1093/pasj/psu156}}.

\bibitem{Ng:2015gfa}
K.~C.~Y. Ng, S.~Horiuchi, J.~M. Gaskins, M.~Smith, R.~Preece, {Improved Limits
  on Sterile Neutrino Dark Matter using Full-Sky Fermi Gamma-Ray Burst Monitor
  Data}, Phys. Rev. D92~(4) (2015) 043503.
\newblock \href {http://arxiv.org/abs/1504.04027} {\path{arXiv:1504.04027}},
  \href {http://dx.doi.org/10.1103/PhysRevD.92.043503}
  {\path{doi:10.1103/PhysRevD.92.043503}}.

\bibitem{Boyarsky:2007ge}
A.~Boyarsky, D.~Malyshev, A.~Neronov, O.~Ruchayskiy, {Constraining DM
  properties with SPI}, Mon. Not. Roy. Astron. Soc. 387 (2008) 1345.
\newblock \href {http://arxiv.org/abs/0710.4922} {\path{arXiv:0710.4922}},
  \href {http://dx.doi.org/10.1111/j.1365-2966.2008.13003.x}
  {\path{doi:10.1111/j.1365-2966.2008.13003.x}}.

\bibitem{Boyarsky:2006zi}
A.~Boyarsky, A.~Neronov, O.~Ruchayskiy, M.~Shaposhnikov, {Restrictions on
  parameters of sterile neutrino dark matter from observations of galaxy
  clusters}, Phys.Rev. D74 (2006) 103506.
\newblock \href {http://arxiv.org/abs/astro-ph/0603368}
  {\path{arXiv:astro-ph/0603368}}, \href
  {http://dx.doi.org/10.1103/PhysRevD.74.103506}
  {\path{doi:10.1103/PhysRevD.74.103506}}.

\bibitem{Neronov:2015kca}
A.~Neronov, D.~Malyshev, {Toward a full test of the $\nu$MSM sterile neutrino
  dark matter model with Athena}, Phys. Rev. D93~(6) (2016) 063518.
\newblock \href {http://arxiv.org/abs/1509.02758} {\path{arXiv:1509.02758}},
  \href {http://dx.doi.org/10.1103/PhysRevD.93.063518}
  {\path{doi:10.1103/PhysRevD.93.063518}}.

\bibitem{Boyarsky:2014ska}
A.~Boyarsky, J.~Franse, D.~Iakubovskyi, O.~Ruchayskiy, {Checking the Dark
  Matter Origin of a 3.53 keV Line with the Milky Way Center}, Phys. Rev. Lett.
  115 (2015) 161301.
\newblock \href {http://arxiv.org/abs/1408.2503} {\path{arXiv:1408.2503}},
  \href {http://dx.doi.org/10.1103/PhysRevLett.115.161301}
  {\path{doi:10.1103/PhysRevLett.115.161301}}.

\bibitem{Jeltema:2014qfa}
T.~E. Jeltema, S.~Profumo, {Discovery of a 3.5 keV line in the Galactic Centre
  and a critical look at the origin of the line across astronomical targets},
  Mon. Not. Roy. Astron. Soc. 450~(2) (2015) 2143--2152.
\newblock \href {http://arxiv.org/abs/1408.1699} {\path{arXiv:1408.1699}},
  \href {http://dx.doi.org/10.1093/mnras/stv768}
  {\path{doi:10.1093/mnras/stv768}}.

\bibitem{Urban:2014yda}
O.~Urban, N.~Werner, S.~W. Allen, A.~Simionescu, J.~S. Kaastra, L.~E. Strigari,
  {A Suzaku Search for Dark Matter Emission Lines in the X-ray Brightest Galaxy
  Clusters}, Mon. Not. Roy. Astron. Soc. 451~(3) (2015) 2447--2461.
\newblock \href {http://arxiv.org/abs/1411.0050} {\path{arXiv:1411.0050}},
  \href {http://dx.doi.org/10.1093/mnras/stv1142}
  {\path{doi:10.1093/mnras/stv1142}}.

\bibitem{Franse:2016dln}
J.~Franse, et~al., {Radial Profile of the 3.55 keV line out to $R_{200}$ in the
  Perseus Cluster}, Astrophys. J. 829~(2) (2016) 124.
\newblock \href {http://arxiv.org/abs/1604.01759} {\path{arXiv:1604.01759}},
  \href {http://dx.doi.org/10.3847/0004-637X/829/2/124}
  {\path{doi:10.3847/0004-637X/829/2/124}}.

\bibitem{Bulbul:2016yop}
E.~Bulbul, M.~Markevitch, A.~Foster, E.~Miller, M.~Bautz, M.~Loewenstein, S.~W.
  Randall, R.~K. Smith, {Searching for the 3.5 keV Line in the Stacked Suzaku
  Observations of Galaxy Clusters}, Astrophys. J. 831~(1) (2016) 55.
\newblock \href {http://arxiv.org/abs/1605.02034} {\path{arXiv:1605.02034}},
  \href {http://dx.doi.org/10.3847/0004-637X/831/1/55}
  {\path{doi:10.3847/0004-637X/831/1/55}}.

\bibitem{Iakubovskyi:2015dna}
D.~Iakubovskyi, E.~Bulbul, A.~R. Foster, D.~Savchenko, V.~Sadova, {Testing the
  origin of ~3.55 keV line in individual galaxy clusters observed with
  XMM-Newton}\href {http://arxiv.org/abs/1508.05186} {\path{arXiv:1508.05186}}.

\bibitem{Cappelluti:2017ywp}
N.~Cappelluti, E.~Bulbul, A.~Foster, P.~Natarajan, M.~C. Urry, M.~W. Bautz,
  F.~Civano, E.~Miller, R.~K. Smith, {Searching for the 3.5 keV Line in the
  Deep Fields with Chandra: the 10 Ms observations}\href
  {http://arxiv.org/abs/1701.07932} {\path{arXiv:1701.07932}}.

\bibitem{Anderson:2014tza}
M.~E. Anderson, E.~Churazov, J.~N. Bregman, {Non-Detection of X-Ray Emission
  From Sterile Neutrinos in Stacked Galaxy Spectra}, Mon. Not. Roy. Astron.
  Soc. 452~(4) (2015) 3905--3923.
\newblock \href {http://arxiv.org/abs/1408.4115} {\path{arXiv:1408.4115}},
  \href {http://dx.doi.org/10.1093/mnras/stv1559}
  {\path{doi:10.1093/mnras/stv1559}}.

\bibitem{Bulbul:2014ala}
E.~Bulbul, M.~Markevitch, A.~R. Foster, R.~K. Smith, M.~Loewenstein, S.~W.
  Randall, {Comment on ``Dark matter searches going bananas: the contribution
  of Potassium (and Chlorine) to the 3.5 keV line''}\href
  {http://arxiv.org/abs/1409.4143} {\path{arXiv:1409.4143}}.

\bibitem{Boyarsky:2014paa}
A.~Boyarsky, J.~Franse, D.~Iakubovskyi, O.~Ruchayskiy, {Comment on the paper
  ``Dark matter searches going bananas: the contribution of Potassium (and
  Chlorine) to the 3.5 keV line'' by T. Jeltema and S. Profumo}\href
  {http://arxiv.org/abs/1408.4388} {\path{arXiv:1408.4388}}.

\bibitem{Carlson:2014lla}
E.~Carlson, T.~Jeltema, S.~Profumo, {Where do the 3.5 keV photons come from? A
  morphological study of the Galactic Center and of Perseus}, JCAP 1502~(02)
  (2015) 009.
\newblock \href {http://arxiv.org/abs/1411.1758} {\path{arXiv:1411.1758}},
  \href {http://dx.doi.org/10.1088/1475-7516/2015/02/009}
  {\path{doi:10.1088/1475-7516/2015/02/009}}.

\bibitem{Muno:2004wh}
M.~P. Muno, J.~S. Arabadjis, F.~K. Baganoff, M.~W. Bautz, W.~N. Brandt, P.~S.
  Broos, E.~D. Feigelson, G.~P. Garmire, M.~R. Morris, G.~R. Ricker, {The
  Spectra and variability of x-ray sources in a deep Chandra observation of the
  Galactic Center}, Astrophys. J. 613 (2004) 1179--1201.
\newblock \href {http://arxiv.org/abs/astro-ph/0403463}
  {\path{arXiv:astro-ph/0403463}}, \href {http://dx.doi.org/10.1086/423164}
  {\path{doi:10.1086/423164}}.

\bibitem{Aharonian:2016gzq}
F.~A. Aharonian, et~al., {Hitomi constraints on the 3.5 keV line in the Perseus
  galaxy cluster}, Astrophys. J. 837~(1) (2017) L15.
\newblock \href {http://arxiv.org/abs/1607.07420} {\path{arXiv:1607.07420}},
  \href {http://dx.doi.org/10.3847/2041-8213/aa61fa}
  {\path{doi:10.3847/2041-8213/aa61fa}}.

\bibitem{Aharonian:2016pyf}
F.~Aharonian, et~al., {The Quiescent Intracluster Medium in the Core of the
  Perseus Cluster}, Nature 535 (2016) 117--121.
\newblock \href {http://arxiv.org/abs/1607.04487} {\path{arXiv:1607.04487}},
  \href {http://dx.doi.org/10.1038/nature18627}
  {\path{doi:10.1038/nature18627}}.

\bibitem{Neronov:2016wdd}
A.~Neronov, D.~Malyshev, D.~Eckert, {Decaying dark matter search with NuSTAR
  deep sky observations}, Phys. Rev. D94~(12) (2016) 123504.
\newblock \href {http://arxiv.org/abs/1607.07328} {\path{arXiv:1607.07328}},
  \href {http://dx.doi.org/10.1103/PhysRevD.94.123504}
  {\path{doi:10.1103/PhysRevD.94.123504}}.

\bibitem{Perez:2016tcq}
K.~Perez, K.~C.~Y. Ng, J.~F. Beacom, C.~Hersh, S.~Horiuchi, R.~Krivonos,
  {(Almost) Closing the $\nu$MSM Sterile Neutrino Dark Matter Window with
  NuSTAR}\href {http://arxiv.org/abs/1609.00667} {\path{arXiv:1609.00667}}.

\bibitem{abazajian:2006jc}
K.~N. Abazajian, M.~Markevitch, S.~M. Koushiappas, R.~C. Hickox, {Limits on the
  radiative decay of sterile neutrino dark matter from the unresolved cosmic
  and soft X-ray backgrounds}, Phys. Rev. D75 (2007) 063511.
\newblock \href {http://arxiv.org/abs/astro-ph/0611144}
  {\path{arXiv:astro-ph/0611144}}, \href
  {http://dx.doi.org/10.1103/PhysRevD.75.063511}
  {\path{doi:10.1103/PhysRevD.75.063511}}.

\bibitem{Bartalucci:2014mta}
I.~Bartalucci, P.~Mazzotta, H.~Bourdin, A.~Vikhlinin, {Chandra ACIS-I particle
  background: an analytical model}, Astron. Astrophys. 566 (2014) A25.
\newblock \href {http://arxiv.org/abs/1404.3587} {\path{arXiv:1404.3587}},
  \href {http://dx.doi.org/10.1051/0004-6361/201423443}
  {\path{doi:10.1051/0004-6361/201423443}}.

\bibitem{Gu:2015gqm}
L.~Gu, J.~Kaastra, A.~J.~J. Raassen, P.~D. Mullen, R.~S. Cumbee, D.~Lyons,
  P.~C. Stancil, {A novel scenario for the possible X-ray line feature at ~3.5
  keV: Charge exchange with bare sulfur ions}, Astron. Astrophys. 584 (2015)
  L11.
\newblock \href {http://arxiv.org/abs/1511.06557} {\path{arXiv:1511.06557}},
  \href {http://dx.doi.org/10.1051/0004-6361/201527634}
  {\path{doi:10.1051/0004-6361/201527634}}.

\bibitem{Shah:2016efh}
C.~Shah, S.~Dobrodey, S.~Bernitt, R.~Steinbrügge, J.~R.~C. López-Urrutia,
  L.~Gu, J.~Kaastra, {Laboratory measurements compellingly support
  charge-exchange mechanism for the 'dark matter' $\sim$3.5 keV X-ray line},
  Astrophys. J. 833~(1) (2016) 52.
\newblock \href {http://arxiv.org/abs/1608.04751} {\path{arXiv:1608.04751}},
  \href {http://dx.doi.org/10.3847/1538-4357/833/1/52}
  {\path{doi:10.3847/1538-4357/833/1/52}}.

\bibitem{mccammon:2002gb}
D.~McCammon, et~al., A high spectral resolution observation of the soft x-ray
  diffuse background with thermal detectors, Astrophys. J. 576 (2002) 188--203.
\newblock \href {http://arxiv.org/abs/astro-ph/0205012}
  {\path{arXiv:astro-ph/0205012}}.

\bibitem{Figueroa-Feliciano:2015gwa}
E.~Figueroa-Feliciano, et~al., {Searching for keV Sterile Neutrino Dark Matter
  with X-ray Microcalorimeter Sounding Rockets}, Astrophys. J. 814~(1) (2015)
  82.
\newblock \href {http://arxiv.org/abs/1506.05519} {\path{arXiv:1506.05519}},
  \href {http://dx.doi.org/10.1088/0004-637X/814/1/82}
  {\path{doi:10.1088/0004-637X/814/1/82}}.

\bibitem{XARM}
J.~Foust,
  \href{\url{http://spacenews.com/nasa-and-jaxa-to-develop-replacement-x-ray-astronomy-telescope/}}{Nasa
  and jaxa to develop replacement x-ray astronomy telescope}, [Online; posted
  April 1, 2017] (April 2017).
\newline\urlprefix\url{\url{http://spacenews.com/nasa-and-jaxa-to-develop-replacement-x-ray-astronomy-telescope/}}

\bibitem{Speckhard:2015eva}
E.~G. Speckhard, K.~C.~Y. Ng, J.~F. Beacom, R.~Laha, {Dark Matter Velocity
  Spectroscopy}, Phys. Rev. Lett. 116~(3) (2016) 031301.
\newblock \href {http://arxiv.org/abs/1507.04744} {\path{arXiv:1507.04744}},
  \href {http://dx.doi.org/10.1103/PhysRevLett.116.031301}
  {\path{doi:10.1103/PhysRevLett.116.031301}}.

\bibitem{Powell:2016zbo}
D.~Powell, R.~Laha, K.~C.~Y. Ng, T.~Abel, {Doppler effect on indirect detection
  of dark matter using dark matter only simulations}, Phys. Rev. D95~(6) (2017)
  063012.
\newblock \href {http://arxiv.org/abs/1611.02714} {\path{arXiv:1611.02714}},
  \href {http://dx.doi.org/10.1103/PhysRevD.95.063012}
  {\path{doi:10.1103/PhysRevD.95.063012}}.

\bibitem{Merloni:2012uf}
A.~Merloni, et~al., {eROSITA Science Book: Mapping the Structure of the
  Energetic Universe}\href {http://arxiv.org/abs/1209.3114}
  {\path{arXiv:1209.3114}}.

\bibitem{Zandanel:2015xca}
F.~Zandanel, C.~Weniger, S.~Ando, {The role of the eROSITA all-sky survey in
  searches for sterile neutrino dark matter}, JCAP 1509~(09) (2015) 060.
\newblock \href {http://arxiv.org/abs/1505.07829} {\path{arXiv:1505.07829}},
  \href {http://dx.doi.org/10.1088/1475-7516/2015/09/060}
  {\path{doi:10.1088/1475-7516/2015/09/060}}.

\bibitem{Nandra:2013shg}
K.~Nandra, et~al., {The Hot and Energetic Universe: A White Paper presenting
  the science theme motivating the Athena+ mission}\href
  {http://arxiv.org/abs/1306.2307} {\path{arXiv:1306.2307}}.

\end{thebibliography}

\end{document}